\pdfoutput=1
\documentclass[a4paper,oneside,openleft,12pt]{amsbook}

\usepackage[backend=bibtex,
            bibstyle=numeric,
            citestyle=numeric-comp,
            giveninits=true
            ]{biblatex}
\DefineBibliographyStrings{english}{andothers={\mbox{\&\textit{al}\adddot}}}
\DefineBibliographyStrings{english}{and={\&}}

\usepackage[]{geometry}
\newgeometry{%
            vscale=0.8,vcentering,
            hscale=0.75
            }
\savegeometry{default-layout}
\newgeometry{hscale=0.7}
\savegeometry{title-layout}

\usepackage{xfrac} 
\DeclareCollectionInstance{plainmath}{xfrac}{mathdefault}{math}{scale-factor = 0.7}
\UseCollection{xfrac}{plainmath}

\usepackage{amssymb,amsthm,amsxtra,mathrsfs} 

\usepackage[titletoc]{appendix} 

\usepackage{afterpage,booktabs,longtable,multirow} 

\usepackage{graphicx,subfig}    
\graphicspath{{chapter1/}{chapter2/}{chapter3/}{chapter4/}{chapter5/}{appendixA/}{appendixB/}}

\usepackage{lscape}             

\usepackage{exscale,relsize}    

\usepackage[parfill]{parskip}   

\usepackage{codelist} 

\usepackage{ifthen,xparse,xstring,expl3}
\usepackage{xspace}
\usepackage[normalem]{ulem}
\usepackage[usenames]{xcolor}
\usepackage[oldetc]{macrolist} 

\usepackage[colorlinks=true]{hyperref}

\usepackage[sort&compress,capitalise,nameinlink]{cleveref}
\Crefname{figure}{\text{Fig.}}{\text{Fig.}}
\Crefname{table}{\text{Table}}{\text{Tables}}
\Crefname{section}{\text{\S}}{\text{\S}}
\Crefname{subsection}{\text{\S}}{\text{\S}}
\numberwithin{section}{chapter}
\numberwithin{equation}{chapter}
\numberwithin{figure}{chapter}
\numberwithin{table}{chapter}

\clearpage{\pagestyle{empty}\cleardoublepage}

\makeatletter
\def\@starttoc#1#2{%
  \begingroup
  \setTrue{#1}%
  \let\secdef\@gobbletwo \chapter
  \let\@secnumber\@empty 
  \ifx\contentsname#2%
  \else \@tocwrite{chapter}{#2}\fi
  \typeout{#2}\@xp\chaptermark\@xp{#2}%
  \@makeschapterhead{#2}\@afterheading
  \parskip\z@skip
  \makeatletter
  \@input{\jobname.#1}%
  \if@filesw
    \@xp\newwrite\csname tf@#1\endcsname
    \immediate\@xp\openout\csname tf@#1\endcsname \jobname.#1\relax
  \fi
  \global\@nobreakfalse \endgroup
  \newpage
}
\def\contentsname{\normalfont\scshape{Contents}}
\def\tableofcontents{%
  \@starttoc{toc}\contentsname
}
\makeatother

\makeatletter
  \def\l@chapter{
      \@tocline{0}{8pt plus1pt}{0pt}{}{\bfseries}
  }
  \def\l@section{
     \@tocline{1}{1pt}{10pt}{}{}
  }
  \def\l@subsection{
     \@tocline{2}{1pt}{20pt}{}{}
  }
  \def\@tocline#1#2#3#4#5#6#7{\relax
  \ifnum #1>2 
  \else
    \par \addpenalty\@secpenalty\addvspace{#2}%
    \begingroup \hyphenpenalty\@M
      \@tempdima\csname r@tocindent\number#1\endcsname\relax
    \parindent\z@ \leftskip#3\relax \advance\leftskip\@tempdima\relax
    \rightskip\@pnumwidth plus4em \parfillskip-\@pnumwidth
    {#5\leavevmode\hskip-\@tempdima #6}\nobreak\relax
    \ifnum #1< 1 \hfill 	
    \else \dotfill 				
    \fi \hbox to\@pnumwidth{\@tocpagenum{#5{#7}}}\par
    \nobreak
    \endgroup
  \fi
  }
\makeatother

\makeatletter
\def\@makechapterhead#1
{\global\topskip 7.5pc\relax
  \begingroup
  \fontsize{\@xivpt}{18}\bfseries\centering
    \ifnum\c@secnumdepth>\m@ne
      \leavevmode \hskip-\leftskip
      \rlap{\vbox to\z@{\vss
          \centerline{\normalsize\mdseries
           \@xp{ \textsc{\chaptername} } 
          \enspace\thechapter} 
          \vskip 3pc}}\hskip\leftskip\fi
     #1\par \endgroup
  \skip@34\p@ \advance\skip@-\normalbaselineskip
  \vskip\skip@ 
}
\makeatother

\makeatletter
\def\partrunhead#1#2#3{%
  \@ifnotempty{#2}{\ignorespaces#1 #2\unskip\@ifnotempty{#3}{. }}%
  \textsc{#3}
}
\makeatother

\let\SMStheoremfont=\bfseries

\theoremstyle{remark}

\makeatletter
\def\@thm#1#2#3{%
  \ifhmode\unskip\unskip\par\fi
  \normalfont
  \trivlist
  \let\thmheadnl\relax
  \let\thm@swap\@gobble
  \let\thm@indent\noindent
  \thm@headfont{\SMStheoremfont}
  \thm@notefont{\fontseries\mddefault\upshape}%
  \thm@headpunct{:}
  \thm@headsep 5\p@ plus\p@ minus\p@\relax
  \thm@space@setup
  #1
  \@topsep \thm@preskip               
  \@topsepadd \thm@postskip           
  \def\@tempa{#2}\ifx\@empty\@tempa
    \def\@tempa{\@oparg{\@begintheorem{#3}{}}[]}%
  \else
    \refstepcounter{#2}%
    \def\@tempa{\@oparg{\@begintheorem{#3}{\csname the#2\endcsname}}[]}%
  \fi
  \@tempa
}
\makeatother



\def\Today{\ifcase\month\or January\or February\or March\or
  April\or May\or June\or July\or August\or September\or
  October\or November\or December\fi\space\number\year}

\makeatletter
\DeclareRobustCommand*\thesistype[1]{\gdef\@thesistype{#1}}
\DeclareRobustCommand*\degree[1]{\gdef\@degree{#1}}
\DeclareRobustCommand*\department[1]{\gdef\@department{#1}}
\DeclareRobustCommand*\university[1]{\gdef\@university{#1}}
\makeatother

\makeatletter
\def\@maketitle{
  \cleardoublepage\thispagestyle{empty}%
  \begingroup \topskip\z@skip
    \null\vfil
    \begingroup
    \LARGE\bfseries \centering
    \openup\medskipamount
     \@title\par
	\vspace{30pt}%
    \centering\mdseries\authors\par\bigskip
    \endgroup
    \vfil\vfil\vfil
    \begin{center}
        \@thesistype\ submitted in fulfillment of\\
        the requirements for the degree of\\
        \@degree
    \vfil\vfil
      \large\@department\par 
      \large\@university\par 
      \vskip18mm
      \includegraphics[width=50mm]{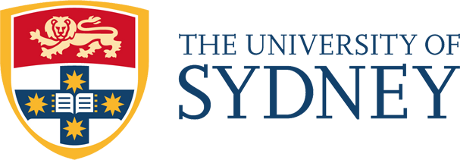}
    \vskip6mm
      \normalsize\Today
    \end{center}
    \vfil
  \endgroup
  \cleardoublepage
}
\makeatother

\author{Francesca von Braun-Bates}
\title{Gravitational lensing of pulsars as a probe of dark matter halos}
\thesistype{Masters by Research}
\university{University of Sydney}
\department{School of Physics}
\degree{Master of Science (Research)}
\date{26/03/2014} 
\makeatletter
\newcommand{\printdate}{\@date}
\makeatother

\makeatletter
\providecommand\phantomcaption{\caption@refstepcounter\@captype}
\makeatother

\newcommand{\HidefromToc}[1]{
	{\renewcommand{\addtocontents}[2]{}{#1}}
}

\newcommand{\vect}[1]{\vec{#1}} 
\newcommand{\ten}[1]{\mathfrak{#1}} 
\newcommand{\unit}[1]{\hat{#1}} 
\newcommand{\basis}[1]{\,\unit{e}_{#1}} 

\begin{document}    

\loadgeometry{title-layout} 
\maketitle          

\loadgeometry{default-layout} 

\pagenumbering{roman} 

\chapter*{Abstract}
A key question in cosmology is the properties of dark matter.  A particular open problem is whether dark matter on small scales is clumpy, forming gravitationally-bound halos distributed within the Galaxy.  The practical difficulties inherent in testing this hypothesis stem from the fact that, on astrophysical scales, dark matter is solely observable via its gravitational interaction with other objects.

This thesis presents a gravitational-lensing-based solution for the mapping and characterisation of low-mass dark matter halos via their signature in \msp{} observations.  This involves numerical calculations in three stages: first, determining the time delay and magnification surfaces generated in the frame of reference of the halo; second, obtaining the corresponding pulsar signature in the reference frame of the observer; and last, generalising the method to multiple halos at varying distances.  In both the single-lens and multiple-lens cases, we discuss whether the delay is observationally detectable.

Dark matter halos act as gravitational lenses which produce a variable flux and induce additional time delays in (tangent) bundles of photons passing near or through the halo.  The key dependency of the mass estimate is the density profile adopted for the halo.  I utilise a variety of proposed halo mass profiles --- namely the elliptical model of \textcite{ring-cycle}, the axially symmetric \swz{} and \hdisc{} lenses (both \cite{sef}) and the \nfw{} density profile \cite{nfw1,nfw2} --- which are applicable over a broad range of halo masses.  The pulsar simulations use the most realistic and sophisticated of these, the empirically-derived profile of \citeauthor{nfw1}.  I justify the adoption of a radially-symmetric density profile by showing that this greatly simplifies the calculation of the lens convergence.  Moreover, I demonstrate that the use of Hankel transforms is a novel way to increase the efficiency of the relativistic time delay.

The observational signatures of such halos are best identified using \msp{s}.  This remarkable subset of the pulsar population has both the highest rotational frequencies and the most period stability of all known pulsars.  Furthermore, the potential for gravitational wave detection using \msp{s} will result in an abundance of new data from pulsar surveys.  I propose that observational techniques do not require major adjustments when searching for signs of \gl{ing}, thus it is unnecessary to implement specialist data reduction pipelines, which enable the data from existing and future surveys to be examined for lensing with relative ease.

This thesis provides a practical method to search for \dmh{s} within our Galaxy and is readily extensible to nearby globular clusters and galaxies, pending the discovery of \msp{s} in these more distant systems. 


\HidefromToc{\chapter*{Statement of Originality}}

The work in this thesis is entirely my own, with supervision by Prof. Geraint Lewis.  It has not been reproduced in publications.  The chapter on dark matter \cref{ch:dm} is a literature review.  The mathematics in \cref{sec:propagation}-\cref{sec:tdpot} can be found elsewhere in the literature (as cited), apart from \cref{sec:hankel}, \cref{sec:potential}.  Otherwise, all of the thesis is my own work, including the code in \cref{ch:code}.  The \textsc{matlab} code used is all of my own design using existing functions (as of R2012a), except for \texttt{bessel\_zeros} used in the Hankel convolution code in \cref{sec:hankel} and \texttt{cb*}, \texttt{mtit} and \texttt{polyrev}, which are used for plotting.  Geraint Lewis assisted in debugging the code.

\vspace{3cm}
\begin{center}
\begin{minipage}[c]{.6\linewidth}
\textit{I certify that this report contains work carried out 
 by myself except where otherwise acknowledged. 
It has not been submitted to any other institution for the award of a degree.}
\end{minipage}
\end{center}
\vspace{8mm}

\begin{minipage}[c]{.1\linewidth}
\centering
Signed:
\end{minipage}%
\begin{minipage}[c]{.5\linewidth}
\bgroup\markoverwith{{\rule[+.7\baselineskip]{2pt}{1pt}}}
\ULon{\hfill
\hfill}
\end{minipage}
\hfill
\begin{minipage}[c]{.25\linewidth}
Date: \hspace{2mm}\printdate 
\end{minipage}%

\HidefromToc{\chapter*{Related material}}

\HidefromToc{\section*{Code repository}}
\url{https://github.com/vonbraunbates/pulsar-lensing}
\HidefromToc{\section*{Presentations Given}}
2014 cosmology seminar University of Oxford 

\tableofcontents


\newpage\setcounter{page}{1}\pagenumbering{arabic}
\chapter{Introduction}\label{ch:introduction}

The following section \cref{sec:purpose} examines the purpose of the thesis and the importance of the method which is proposed; \cref{sec:context} discusses the work undertaken by my predecessors; finally \cref{sec:structure} outlines the structure of the rest of the thesis and any notational conventions.

\section{Purpose}\label{sec:purpose}
This thesis proposes a method to detect \dmh{s} on galactic scales using the principles of \gl{ing}.  \dm{} comprises a diverse class of objects unified by the property that they are non-luminous.  The detection of \dmh{s} of small mass is possible by examining their gravitational interaction with signals from \msp{s}.  Such interaction may appear observationally in multiple ways; notably reception of multiple signals from a single source, changes in the amplitude of the signals and time delays imprinted upon the pulsar's period.  These effects form a part of a greater phenomenon termed gravitational lensing, which describes the relativistic interactions between matter and photons.  This corollary of general relativity permits the detection of \dm{} in an astrophysical (as opposed to a particle physics) context.

There are three underlying principles which form the core of this method.  Dark matter interacts with its surroundings purely gravitationally, which limits its ability to be detected on galactic scales.  One of the most-examined methods is \gl{ing}, which makes precise and observationally measurable predictions about the effect of (normal or dark) matter on photons which pass through the surrounding area.  The main obstacle to extracting information embedded in lensed signals is that this requires information about the source which emitted the signals \cite{mpa-notes}.  Resolution of this problem is provided by \msp{s} which act as very regular, point-like emitters: they are ideal candidates for lensing because their signals are emitted on short time scales ($\sim 1\, \textrm{ms}$) with short, non-cumulative errors ($\sim 1\mu \, \textrm{s}$; \cite{siegel1}). Thus, the small perturbations generated by the lensing effect are (relatively) easy to observe.  With such a source, it is possible for \dm{} to be readily detected.

The examples provided by this thesis show that the phenomenon is observable on human time scales, even considering a realistic rather than an idealised model for the \dmh{s}.  Moreover, a distribution of \dmh{s} at various distances can also be detected.  This forms an important extension to the single-lens, fixed-distance models previously examined (e.g. \cite{siegel1}).

Diverse generalisations of this method are possible.  Its flexibility enables the inclusion of any axially symmetric lens model.  This is particularly interesting because several modifications to the lens models used here (\cref{sec:models}) have been proposed (including \cite{nfw-improved}), which may be readily compared with the calculations here.  Increasing the accuracy of the multiple lens construction is achieved by introducing interaction between the lenses \cite{sef}.  This requires the so-called \mpl formalism, which is briefly examined in \cref{ch:mpl}.  

\section{Context}\label{sec:context}

This thesis extends the current literature in three areas: the lens profile, the inclusion of multiple lenses and the scale of the problem.  The first proposal to use pulsars as gravitational probes of dark matter arose from Siegel, Hertzberg and Fry \cite{siegel1}.  They utilised a single lens at a fixed radial distance, discussing three possible lens profiles and their effect on the observability of a signal lensed by the presence of a \dmh.  This project presents a more practical approach not only in the calculation of the lensing potential, but also in the inclusion of more than one lens between the sample pulsar and the Earth. 

The complexity of a suitable lens profile for \dmh{s} is a topic of some debate.  Most papers on solar-mass halos \eg \cite{siegel2} use a point mass (\swz{}) lens because it is analytically tractable.  Notably, \citeauthor{siegel1} examine three profiles: the \swz{} profile, a sphere of constant density and the radially-dependent NFW model \cite{siegel1}.  Of these lens models, I chose the most probable: the Navarro-Frenk-White model, which was hailed as a \lq\lq{}universal dark matter profile\rq\rq{} due to its good fit in N-body simulations across several decades of mass \cite{bartelmann-arcs}.  (The other two models I retained as analytical checks to my numerical simulations.  A further model with an elliptical potential was also used.  All the models are summarised in \cref{sec:models}.)  

A number of observational projects have detected lensing due to \dmh{s} in the Milky Way \cite{ml-summary}.  Collaborations including OGLE \cite{ogle-iii-lmc,ogle-iii-smc} and EROS \cite{eros1,eros} have used the technique of astrometric microlensing to limit the mass in \dmh{s}.  In contrast, the photometric microlensing technique utilised in this project has not been widely-implemented because the lensing signal is harder to detect \cite{siegel2,ml-summary}.

\section{Structure and remarks on notation}\label{sec:structure}
In \cref{ch:dm} I examine the astrophysical evidence for \dm{}, its distribution on a variety of cosmological to galactic scales and discuss possible candidates.  The main content of the thesis is in \cref{ch:gl}: the lens models are introduced in \cref{sec:models}; subsequent sections form a brief introduction to the mathematics of \gl{} theory; finally the numerical construction of the multiple-lens model is described in \cref{sec:multiple}. The main results of the thesis are described in \cref{ch:results}.  The final chapter \cref{ch:conclusion} outlines the main conclusions of my research, possible avenues for exploration and open questions in the field.  The first appendix \cref{ch:mpl} extends the material in \cref{ch:gl} to the case of multiple lenses.  The details are quite complex and under most circumstances it is sufficient to model multiple lenses as a superposition of their single lens behaviour \cite{sef}.  The second appendix \cref{ch:code} contains the exact procedures which are only outlined in pseudo-code in \cref{ch:gl}.

Physical constants set to unity are the speed of light \textit{in vacuo} $c \approx 3.0 \times 10^{8} \,\text{m}\,\text{s}^{-1}$ and Newton\rq{}s gravitational constant $G \approx 6.67 \times 10^{-11} \,\text{s}^{-1}\,\text{m}^2 \,\text{kg}^{-2}$.  Astronomical distances are measured in parsecs:~$1 \,\text{pc} \approx 3.09 \times 10^{16} \,\text{m}$; distances on the lens and source planes are measured in term of a scaling radius which depends upon the lens model.  Masses are given in units of the solar mass $M_{\odot} \approx 1.99 \times 10^{30} \,\text{kg}$.  Cosmological densities $\Omega_i$ are dimensionless fractions of the critical density $\rho_{\text{crit;}0} \equiv \sfrac{3H_0^2}{8\pi G} 
\approx 9.15 \times 10^{-33} \,\text{kg}\,\text{m}^{-3}$ where the Hubble constant is $H_0 \approx 72 \,\text{km}\,\text{s}^{-1}\,\text{Mpc}^{-1}$.

The mathematical convention chosen is to denote vectors by an over-arrow, except in the case of unit vectors, which are circumflexed.  The length (2-norm) of a vector is denoted with double vertical bars, as opposed to the modulus of a complex number, denoted by single bars.  Thus $\vect{x} = \norm{x}\unit{x}$ and $z \in \mathbb{C}$ has $\abs{z}^2 = z^*z$.  The vector differential operators in $\mathbb{R}^3$ are denoted by a nabla: the gradient and Laplacian are $\grad$ and $\grad^2$ respectively.  The co-ordinate systems used are Cartesian $\{ (x,y) \pmb{:} x \in \mathbb{R}, y\in \mathbb{R} \}$ and modified polar $ \{ (\rho,\phi) \pmb{:} \rho \in (-\infty, \infty), \phi \in [0,\pi) \}$.

The notation used in \gl{} literature is not widely standardised (for reasons listed in \cite{sef}).  Where a convention does exist, I have used it (\eg, $\kappa$ for the convergence and $\gamma$ for the shear of a lens).  There are some cases where this causes the symbols to overlap with standard mathematical notation (\eg $\phi$ for both the Fermat potential and the 2d polar co-ordinate) but the meaning should be clarified by the context.
\chapter{Dark matter halos}\label{ch:dm}

The disparity between the amount of luminous matter and the total matter present in the universe has remained an open problem in astronomy for three-quarters of a century \cite{Zwicky-1933}.  This has led to the hypothesis that some matter must be \lq\lq{}dark,\rq\rq{} \ie unable to be observed directly using the electromagnetic spectrum.  This chapter recounts the evidence for dark matter in \cref{sec:evidence}.  An overview of possible candidates follows, divided broadly into baryonic \cref{sec:baryonic} and non-baryonic \cref{sec:nbaryonic} classes.  

\section{Evidence for dark matter}\label{sec:evidence}

The historical development of the case for \dm{} is naturally fragmented.  In lieu of a chronological treatment, this section explains the cosmological motivation for \dm{}, before concentrating on the estimation of \dm{} on the sub-galactic scale probed by the method in this thesis.  We shall see that, despite the severity of the problem on cosmological scales, the situation is much reduced within individual galaxies.

\begin{figure}
\begin{longtable}{ lllll }
\toprule 
\multicolumn{2}{l}{Baryon form} & Max. likelihood & Upper bound & Lower bound\\
\midrule
\endfirsthead 
\bottomrule \\[-.1in]
\endlastfoot 
\multirow{3}{*}{Stars} & spheroids &$ 0.0026 $&  $0.0043 $
      &   $0.0014 $  \\
& discs of spiral and S0 galaxies & $0.00086 $&  $0.00129 $ 
&  $0.00051 $ \\
& irregular galaxies & $0.000069 $ & $0.000116 $
& $0.000033 $  \\
\midrule
\multirow{2}{*}{Gas} & neutral atomic & $0.00033 $& $0.00041 $ 
&  $0.00025 $ \\
& molecular & $0.00030 $& $0.00037 $ 
& $0.00023 $ \\
\midrule
\multirow{7}{*}{Hot gas} & in clusters & $0.0026 $& $0.0044 $
  & $0.0014 $ \\
& warm in groups & \multirow{2}{*}{$0.0056$} & \multirow{2}{*}{$0.0115$} & \multirow{2}{*}{$0.0029$}  \\
& (X-ray detection) &  &  &  \\
& cool in groups & \multirow{2}{*}{$0.002$} & \multirow{2}{*}{$0.003$} & \multirow{2}{*}{$0.0007$}  \\
& (Lyman-$\alpha$ absorption)& &  &   \\
& total in groups & \multirow{2}{*}{$0.014$} & \multirow{2}{*}{$0.030$} & \multirow{2}{*}{$0.0072$}  \\
& (scaled from clusters) &  & &  \\
\midrule
Sum\footnote{The various methods of estimating the hot gas in groups of galaxies are \lq\lq{}quite uncertain\rq\rq{}, with the possibility that the value for cool gas is under-estimated.  Thus, the total uses the sum of the warm and cool estimates for calculating the lower bound and the more reliable cluster extrapolation for the upper bound and best estimate.} & &  0.021   &  0.041   &  0.007   \\
\bottomrule
\end{longtable}
\captionof{table}{The fractional density of visible (baryonic) matter estimated at $z \simeq 0$.  (For details refer to \S 2.5 of \cite{baryons}.) \label{tab:baryons}}
\end{figure}
 
The existence of dark matter is necessary only if the amount of luminous matter in the Universe is less than the total amount.  Consequently, we must determine the quantity of luminous matter $\Omega_{\star}$ in the Universe.  A comprehensive treatment is given by \cite{baryons}, whose estimates%
\footnote{The estimates assume a Hubble constant of $H_0 = 70 \,\mathrm{kms^{-1}Mpc^{-1}}$: this is sufficiently close to $H_0 = 72 \,\mathrm{kms^{-1}Mpc^{-1}}$ that I have omitted the factors of $h_{70}^{-1}$ which appear in the original paper.  }
 are duplicated in \cref{tab:baryons} (with some simplification).  The total listed in \cref{tab:baryons} can be further constrained by nucleosynthesis from the Big Bang (BBN) \cite{dm-bartelmann}.  The primordial abundances of light elements (namely ${}^{2}$H, ${}^{4}$He, ${}^{7}$Li and isotopes ${}^{3}$He and D) are related to the ratio between the photon $n_{\gamma}$ and baryon $n_B$ number densities \eqref{eq:density-ratio} :
\begin{equation}\label{eq:density-ratio}
\eta = \frac{n_B}{n_{\gamma}} = 10^{-10}\eta_{10}, \quad \eta_{10} \equiv 273\Omega_B h^2
\end{equation}
Given $n_{\gamma}$ from the temperature of the CMB, it is possible to combine abundance estimates to calculate $\eta$ and thus find $\Omega_B$. Theoretical prediction of abundance estimates is possible by solving a coupled set of ODEs describing the element production and destruction in the radiation-dominated era \cref{tab:bbn} to find the initial abundances as a function of $\eta_{10}$ \cite{bbn-summary}.  

\begin{figure}
\includegraphics[width=0.75\textwidth]{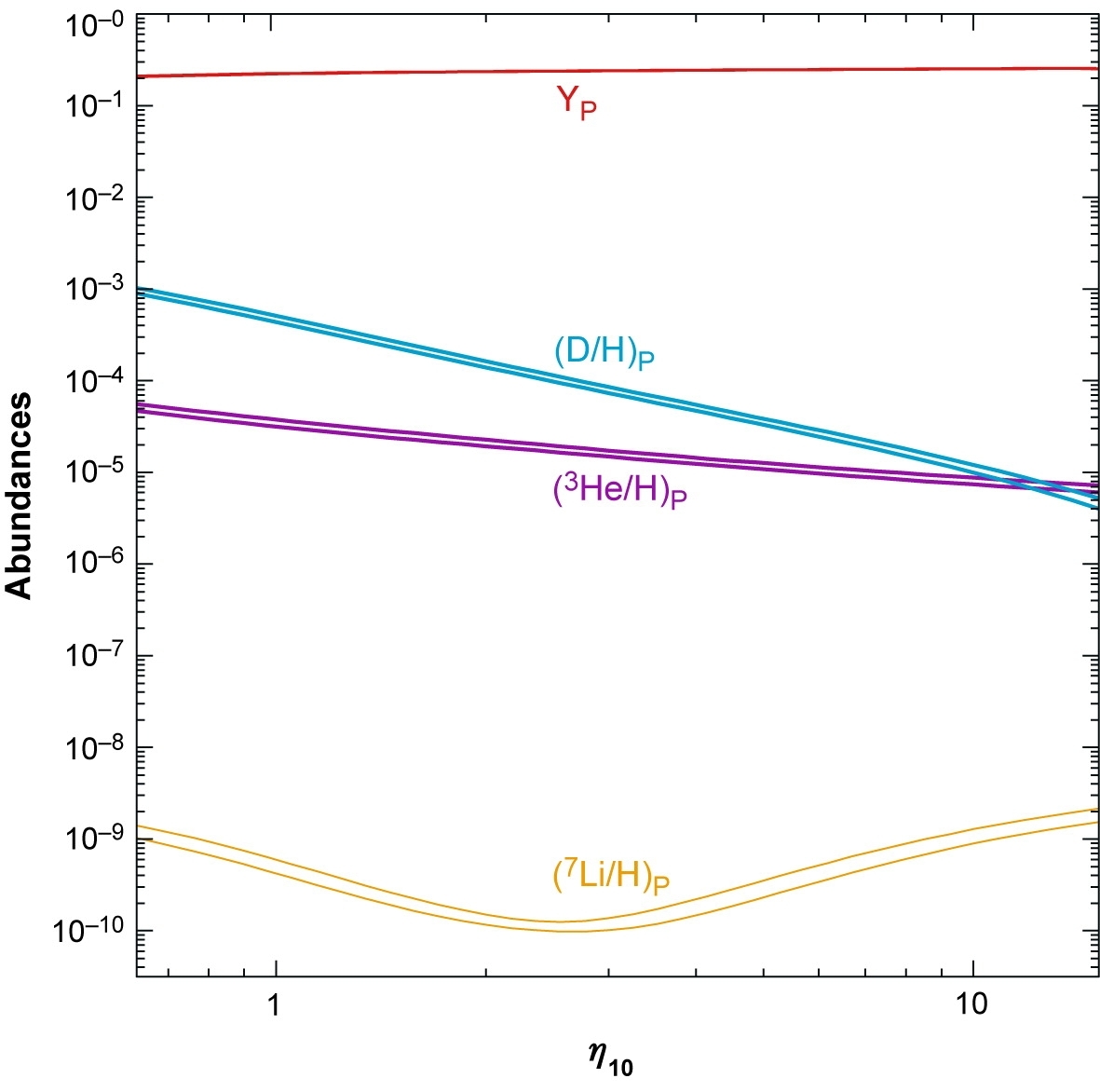} 
\caption{Theoretically expected abundances (relative to hydrogen) of the light elements deuterium ($D$), helium ($^{3}$He and $Y = ^{4}$He) and lithium($^{7}$Li) as functions of the abundance ratio $\eta_{10}$. The widths of each band are caused by the uncertainties in the nuclear and weak-interaction rates. (Fig.~5 in \cite{bbn-summary}) \label{fig:bbn}}
\end{figure}

Comparison of these results with observation is difficult due to possible depletion from the primordial abundances.  Bartelmann asserts in \cite{dm-bartelmann} that such depletion is unlikely in the case of the deuterium abundance measured in high-redshift QSOs.  Given the strong dependence of the deuterium abundance on $\eta_{10}$, this makes it an ideal estimator of the ratio $\eta_{10}$.  (Other elements, such as ${}^7$Li from low-metallicity halo stars in the Galaxy, can be used to confirm a consistent estimate.)  We thus find that Big-Bang nucleosynthesis alone implies:
\begin{equation}
0.0207 \leq \Omega_B h^2 \leq 0.0234 \quad\text{or}\quad 0.0399 \leq \Omega_B \leq 0.045
\end{equation}
based on the deuterium abundance in high-redshift absorption systems and assuming the Standard Model of particle physics \cite{dm-bartelmann}.  Thus, a reasonable estimate of the baryon content of the universe is $\Omega_B \sim 0.04$ from both BBN and astrophysical estimators, of which $\Omega_{\star} \sim 0.02$ is luminous.

\newsavebox{\mybox}
\sbox{\mybox}{
\begin{minipage}{0.48\linewidth}
\begin{align*}
\begin{aligned}
 p + n &\rightarrow d + \gamma \;
\end{aligned} &\text{fusion of deuterium} \\
\left.\begin{aligned}
d + p &\rightarrow \text{$^3$He} + \gamma \\
d + d &\rightarrow \text{$^3$He} + n \\
d + d &\rightarrow t + p \\
\text{$^3$He} + n &\rightarrow t + p \\
\end{aligned}
\;\right\}\;
&\text{\begin{minipage}{3.2cm}production of $^3$He \\ and tritium $t$ \end{minipage}} 
\\
\left.\begin{aligned}
\text{$^3$He} + d &\rightarrow \text{$^4$He} + p \\
t + d &\rightarrow \text{$^4$He} + n \\
\end{aligned}
\;\right\}\;
&\text{conversion to $^4$He} \\
\left.\begin{aligned}
t + \text{$^4$He} &\rightarrow \text{$^7$Li} + \gamma \\
\text{$^3$He} + \text{$^4$He} &\rightarrow \text{$^7$Be} + \gamma \\
\text{$^7$Be} + e^- &\rightarrow \text{$^7$Li} + \nu_e \\
\end{aligned}
\;\right\}\;
&\text{Lithium production}
\end{align*}
\end{minipage}
}

\begin{figure}[b!]
\begin{longtable}{p{0.5\linewidth} l} 
\toprule 
Reaction & Element production \\
\midrule  \\[-.25in]
\endfirsthead 
\bottomrule 
\endlastfoot 
\multicolumn{2}{c}{\usebox{\mybox}} \\
\bottomrule
\end{longtable}
\captionof{table}{Reactions involved in primordial neucleosynthesis.  \label{tab:bbn}} 
\end{figure}

\subsection{Cosmological mass fraction}\label{sec: cosmological}

The missing mass problem is greatest on cosmological scales.  The total matter(-energy) content in the Universe must be inferred from its geometry, as implied by the field equation of general relativity:
\begin{equation}
\ten{R} - \dfrac{1}{2}\ten{g}R = \dfrac{8\pi G}{c^4}\ten{T}
\end{equation}
This equation supplies a relation between the energy distributed within spacetime and the deformation of the spacetime caused by the presence of the energy \cite{Hobson}.  Spacetime is described by a pseudo-Riemannian manifold of dimension 4, with metric $\ten{g}$ determining the choice of inner product on the metric \cite{doCarmo}
\begin{equation}\label{eq:field-eqn}
\ten{g}_{\mu\nu} = \langle x_{\mu} , x_{\nu} \rangle
\quad \text{or in terms of the line segment} \quad
\dx{s}^2 = \ten{g}_{\mu\nu}\dx{x^{\mu}}\dx{x^{\nu}}
\end{equation}
The curvature of the manifold is described by the Riemann curvature tensor $\ten{R}$, of which the first- and second-order contractions appearing in \eqref{eq:field-eqn} are the Ricci tensor $\ten{R}$ and scalar $R$ respectively \cite{Hobson}.  The corresponding energy is given by the stress-energy-momentum tensor $\ten{T}$ \cite{Hobson}.

\begin{figure}
\includegraphics[width=0.75\textwidth]{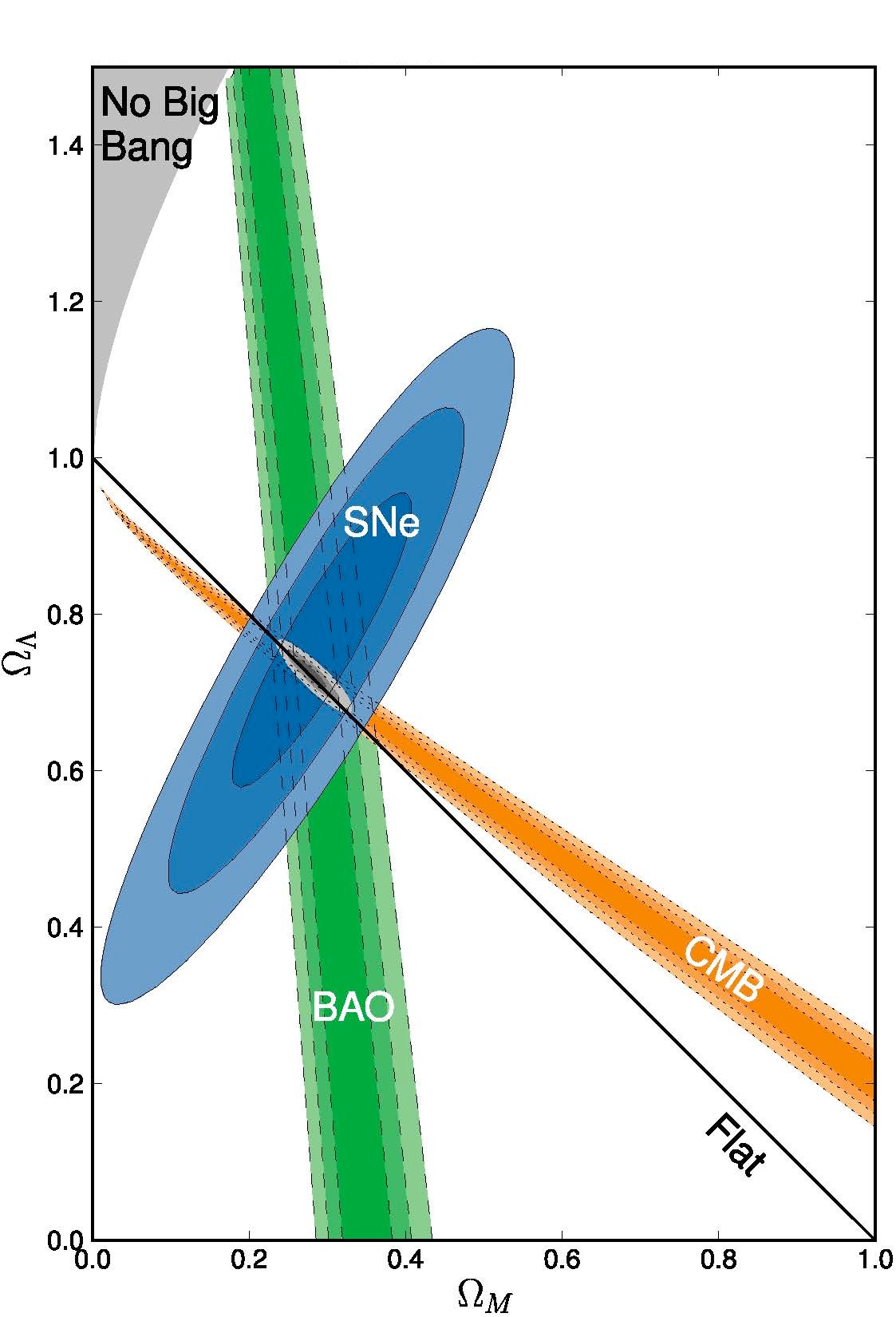}
\caption{Observational constraints on the total fraction of matter $\Omega_M$ and dark energy (as a cosmological constant) $\Omega_{\Lambda}$, where a value of $\Omega_i = 1$ represents a density equal to the critical density of the universe (hence the line showing no spatial curvature at $\Omega_M + \Omega_{\Lambda} = 1$).  The contours show the posterior probabilities, with darker contours representing 1, 2 and 3 $\sigma$ credible regions respectively.  The blue contours are from the Union Supernova Project; green from baryon acoustic measurements from the Sloan Digital Sky Survey and yellow from WMAP measurements of the CMB.  The combined posterior is shown in grey. \cite{union2} \label{fig:omega}}
\end{figure}

The assumptions of isotropy and homogeneity diagonalise the left-hand side of \eqref{eq:field-eqn} by necessitating a geometry invariant under both rotation and translation \cite{doCarmo}.  Consequently, the large-scale contents of the Universe are limited to perfect fluids, which diagonalise $\ten{T}$: lacking both heat conduction and bulk and shear viscosity, perfect fluids are entirely characterised by their energy density $\rho = \sfrac{\ten{T}^{00}}{c^2}$ and energy pressure  $p = \ten{T}^{ii}$.  In cosmological units, we prefer to express $\rho$ as a fraction of the critical density $\rho_{\textrm{crit}}$ and define $\Omega_i \equiv \sfrac{\rho_i}{\rho_{\textrm{crit}}}$.  Similarly, we express the energy pressure via the equation of state: $w(z) \equiv \sfrac{p(z)}{\rho(z)}$.  The stress-energy-momentum tensor is the sum of the contribution from each fluid, so a specification of $\Omega_{\sigma}$ and $w_{\sigma}(z)$ is sufficient to determine $\ten{T}$. \cite{Hobson}

These assumptions, termed the cosmological principle, reduce the 20 possible equations of \eqref{eq:field-eqn} to three \cite{Hobson}.  These form a coupled set of \textsc{ode}s known as the Friedmann equations \cite{Blau}:
\begin{subequations}\label{Friedmann}
\begin{align}
-3\frac{\dot{a}}{a} 
& =  4\pi G \sum_{\sigma}(\rho_{\sigma}(t)+3p_{\sigma}(t)) 
& \implies & \sum_{\sigma} \Omega_{\sigma}(z) 
= 1
\label{expansion}\\
\frac{\ddot{a}}{a(t)} + 2 \frac{\dot{a}^{2} +K}{a^2(t)} 
& =  4\pi G \sum_{\sigma}(\rho_{\sigma}(t) - p_{\sigma}(t))
& \implies & \Omega_k(z) 
= \frac{-3K/\rho_{\textrm{crit}}}{8\pi Ga^2(z)} \;\text{and}\; w_k(z) = -\frac{1}{3}
\label{Hubble} \\
\frac{\dx{\rho}_{\sigma}}{\dx{t}}  
& = -3 (\rho_{\sigma}(t) + p_{\sigma}(t)) \frac{\ddot{a}}{a}
& \implies & q(z) 
= \frac{1}{2} \sum_{\sigma} (1 + 3w_{\sigma}(z)) \Omega_{\sigma}(z)
\label{adiabatic} 
\end{align}
\end{subequations}
These correspond to the time-time and space-space components of \eqref{eq:field-eqn} and a third equation which prescribes local conservation of energy (derivable from the other two) \cite{Blau}.  We have reformulated them (following \cite{Blau}) into redshift-dependent equations, introducing the \lq\lq{}generalised deceleration parameter\rq\rq{} $q(z)$ and explicitly including the curvature contribution $k \in \{-1,0,1\}$ as a perfect fluid of density $\Omega_k$ and equations of state $w_k$.  As expected from the Bianchi relations, the evolution of the scale factor $a(t)$ and the different cosmological fluid densities $\Omega(z)$ are not independent \cite{Blau}.  The equations can then be solved numerically for any desired number of cosmological fluids to find the scale factor $a(t)$, which is the key element of distance in the Universe.

Thus, the fractional content of the universe affects inner products on the metric, which are used to measure distances.  Inversely, distance-redshift measurements of standard candles (\eg Type Ia SNe \cite{union2}), standard rulers (\eg baryon acoustic oscillations \cite{bao}) and the cosmic microwave background \cite{wmap7}, allow the present-day value of $\Omega_{i0}$ to be estimated via Bayesian inference \cref{fig:omega}.  The different observations show varying correlations between the fraction of matter $\Omega_M$ and dark energy/cosmological constant $\Omega_{\Lambda}$: in combination they give strict limits on $\Omega_M$.  Thus, best estimates of the fractional matter content of the Universe on cosmological scales are $\Omega_Mh^2 = 0.1352 \pm 0.0036$ \cite{wmap7}, or $\Omega_M \sim 0.3$ (\cf the visible mass fraction $\Omega_{\star} \sim 0.02$ on the same scale). 

\subsection{Mass in galaxies}
An estimate of the \dm{} fraction within galaxies is given by comparison of the galactic and stellar mass-to-light ratio.  Galaxy masses are extrapolated from their luminosities \cite{dm-bartelmann}.  By observation, the distribution of galaxy luminosities is the Schechter function:
where the penultimate approximation uses $\alpha \approx 1$ and $\Gamma$ is the gamma function.  Using the same mass-to-light ratio as the previous subsection, we obtain for the galaxy population a corresponding mass density of:
\begin{equation}
\mathcal{M}_{\text{gal}} = \langle \frac{m}{l} \rangle \mathcal{L}_{\text{gal}}
\approx 1.1 \times 10^{-4} M_{\odot} \, \text{Mpc}^{-3}
\end{equation}
This forms an upper bound on the mass contained in galaxies due to our choice of mass-to-light ratio: a more conservative estimate of $\langle \tfrac{m}{l} \rangle = 30$ would give a value one-fifth of this.  Combining this result with the critical density, the cosmological matter density of galaxies is $\Omega_{\text{gal}} \approx 0.08$.

Justification for the mass-to-light ratios in the previous paragraph follow from the rotation curves of spiral galaxies and the Tully-Fisher luminosity relation.  The variation in tangential velocity \wrt radius from the galactic centre can be measured spectroscopically using stars and (further out) neutral hydrogen lines \cite{dm-bartelmann}.  These rotation curves trace the mass enclosed within a given radius: assuming an axisymmetric mass distribution (which by necessity causes circular orbits) we find that
\begin{equation}\label{eq:v-rot}
v^2_{\text{rot}} = \frac{GM(r)}{r} \implies M(r) = \frac{v^2_{\text{rot}}r}{G}
\end{equation}
The observations of $v_{\text{rot}}(r)$ show that it increases rapidly, but becomes constant at some radius (beyond which stars can still be observed) and remains so even at radii at which stars are not visible.  These flat rotation curves require a mass profile of $M(r) \propto r$, equivalent to a density profile of $\rho(r) \propto r^{-2}$.  Since this mass profile diverges as $r \rightarrow \infty$, it is necessary to define the cutoff radius $R$ for the profile which is chosen such that the galaxy has mean overdensity of 200:  
\begin{equation}\label{eq:r-rot}
\frac{M(R)}{v_{\text{rot}}(R)} = \frac{3M(R)}{4\pi R^3} = 200\rho
\end{equation}
Given that typical values for $\rho$ and $v_{\text{rot}}$ are known, we can solve \eqref{eq:v-rot} and \eqref{eq:r-rot} for $M$ to give:
\begin{equation}
M = \frac{v_{\text{rot}}^2 R}{G} = 2.7 \times 10^{12}M_{\odot} \left( \frac{v_{\text{rot}}}{200\,\text{km}\,\text{s}^{-1}} \right)
\end{equation}
We have an equation for the typical mass of a spiral galaxy; we require one for the typical luminosity.  This is provided by the Tully-Fisher relation
\begin{equation}
L = L_{\ast} \left( \frac{v_{\text{rot}}}{220\,\text{km}\,\text{s}^{-1}} \right)^{\alpha} \quad \text{for $L_{\ast} \approx 2.4 \times 10^{10} \text{L}_{\odot}$ and $\alpha \in (3,4)$}
\end{equation}
Judicious choice of the typical rotation velocity and the virial radius gives a mass-to-light ratio of $\langle \tfrac{m}{l} \rangle \approx 150$.  This, although greatly simplified, does provide an upper bound on the mass-to-light ratios for typical  galaxies (both spiral and elliptical).  This justifies the upper bound for the mass density fractions in the previous two sections.
 
\subsection{Mass in stars}\label{sec:stars}
The stellar mass-to-light ratio will be far less than that for the entire galaxy.  As before, we can use the luminosity of a \lq\lq{}typical\rq\rq{} stellar population%
\footnote{We nominate \lq\lq{}typical\rq\rq{} to mean a stellar population visible at optical and near-infrared wavelengths.  Stars in this regime have peaks in their blackbody curves at $\lambda \lessapprox 10^{-4}\,\text{cm}$, \ie an effective temperature of $T \lessapprox 2900\,\text{K} \approx 0.5 T_{\odot}$.}
 to estimate its mass.  The mass distribution of stellar populations is defined as the number of stars $N$ formed per unit mass $dm$ per unit time.  Normalising the mass distribution to unity (\ie $m_0 \leq m < \infty$) and expressing the mass in solar units ($m = M/M_{\odot}$) gives the frequently-used Salpeter distribution \cite{Salpeter}:
\begin{equation}\label{eq:salpeter}
\ddx{\ln M}{N}{} \propto M^{-1.35} 
\implies \ddx{m}{N}{} =  \frac{1.35}{m_0} \left( \frac{m_0}{m} \right)^{2.35} \quad \text{where $m = \sfrac{M}{M_{\odot}}$}
\end{equation}
where we choose a lower bound of $m_0 = 0.25 M_{\odot}$ to ensure that the stars produce measurable luminosities (which we can translate into masses).  Consideration of a star as an ideal gas in hydrostatic equilibrium and obeying mass conservation and the energy transport equation shows that luminosity and stellar mass are related by $L \sim M^3$.  We can use this to estimate the mass-to-light ratio from \eqref{eq:salpeter}:
\begin{equation}
\langle \frac{m}{l} \rangle 
= \integ{m}{m_0}{\infty}{\frac{m}{l} \ddx{m}{N}{}}
= \integ{m}{m_0}{\infty}{\frac{1}{m^2} \ddx{m}{N}{}}
\approx 6.4
\end{equation}
Although we have neglected to include more complex physics, (\eg{} spectral energy distributions, non-main-sequence stars), we may justifiably assume that our calculation represents the correct order-of-magnitude result for the stellar mass-to-light ratio.  Our result shows that the mean stellar mass-to-light ratio differs by orders of magnitude from the mean galactic one.  

This demonstrates that the case for dark matter within galaxies (rather than in their surrounding \dmh{s}) is a valid one.  It is this scale which is addressed by this thesis.

\section{Baryonic dark matter}\label{sec:baryonic}
The first choice for a \dm{} candidate is one that we know to exist: baryonic matter.  At galactic scales it is difficult to hide baryonic matter \cite{olive}, which limits the possibilities.  The major candidates in the literature have been gas, brown-dwarf-like objects and stellar remnants \cite{olive}.

\subsection{Primordial hydrogen}
The exclusion of sublimed or gaseous hydrogen can be made by X-ray observations.    Under the assumption that concentrations of primordial H (with some He) still exist today, we conclude that either they are electrostatically bound \lq\lq{}snowballs\rq\rq{} of frozen H or gravitationally bound clouds of gaseous H (since H sublimes).  

In the first instance, a lower bound can be placed on their age by assuming that the snowballs are collisionless,%
\footnote{It is possible to compare the binding energy of a sample halo to its kinetic energy to show that it must be collisionless in order to have survived.  The details are not particularly relevant, as we shall see that this is an unsuitable candidate for \dm{}.}
which implies that they can only form when the average density of the universe is equal to the density $\rho_H$ of the halo.  This was at $z = 2.5$, corresponding to a microwave background temperature of 9.5 K \cite{olive}.  At this temperature, the hydrogen would sublimate and we need only consider the gaseous case.  

In the second instance, a halo of H gas formed at $z = 2.5$ would now be in hydrostatic equilibrium, since the age of its host galaxy is greater than the collapse time for the halo to form.  Then we can find the equilibrium temperature $T$ by simultaneously solving: 
\begin{align*}
P(r)  &= \frac{2\rho(r)}{m_P}kT 
\qquad\text{and}\qquad
\ddx{r}{P(r)}{}  = -\frac{GM(r)\rho(r)}{r^2} 
\intertext{(where $\rho$ is the halo density, $m_P$ the proton mass and $M$ the halo mass enclosed at radius $r$) to find}
T  &= \frac{Gm_pM(r)}{4kr} 
\quad \sim 1.3 \times 10^6 \mathrm{K}
\end{align*}
Gas at this temperature would give off X-rays, which conflicts with observations \cite{olive}.  Consequently, we may rule out hydrogen as a \dm{} candidate.

\begin{landscape}
\renewcommand{\arraystretch}{1.5}
\vspace*{\fill}
\begin{longtable}{l c l c c l c}
\toprule 
\multirow{2}{*}{Survey}
& Survey 
& \multirow{2}{*}{Field observed}
& Candidate 
& Microlensing 
& Mass fraction (\%) 
& Average MACHO  \\
& time (yr)
& 
&  source stars
&  events
& (MACHO mass $M_{\odot}$)
& mass ($M_{\odot}$) \\
\midrule
\endhead
\bottomrule
\endfoot
\multirow{3}{*}{EROS-2 \cite{eros}}
& \multirow{3}{*}{6.7}
& LMC 
& \multirow{3}{*}{$7 \times 10^6$}
& \multirow{3}{*}{1}
& \multirow{3}{*}{$\begin{aligned}
 &< 4 & &(M \in [10^{-3},\, 10^{-1}]) \\
 &< 10 & &(M \in [10^{-6},\, 1])
 \end{aligned}$ }
& \multirow{3}{*}{---------} \\
& & SMC & & & & \\
& & Bulge & & & & \\
\midrule 
MACHO \cite{macho-i} 
& 2.3 
& LMC
& $8.5 \times 10^6 $
& 8
& $ < 20 \; (M \in [10^{-6}, \, 2 \times 10^{-2}]) $
& 0.3 - 0.8 \\
\midrule 
MACHO \cite{macho-ii}
& 5.7
& LMC
& $ 12 \times 10^{6} $
& 13 - 17
& $\begin{cases}
20 &\text{max. likelihood} \\
8-50 &\text{95\& CL}
\end{cases} $
& 0.15 - 0.9 \\
\midrule 
OGLE-III \cite{ogle-iii-lmc}
& 8
& LMC
& $ 35 \times 10^6 $
& 2-4
& $\begin{cases}
3 &\text{max. likelihood} \\
1-5 &\text{95\& CL}
\end{cases} $
& 0.2 \\
\midrule 
OGLE-III \cite{ogle-iii-smc}
& 8
& SMC
& $ 5.5 \times 10^6 $
& 1
& $ \begin{aligned}[c] 
  &<4 & &(M < 10^{-1} \\
  &<6 & &(M \in [10^{-1},\, 4 \ \times 10^{-1}]) \\
  &<9 & &(M = 1) \\
  &<20 & &(M = 20)
 \end{aligned} $
& ------ \\
\end{longtable}
\captionof{table}{Summary of \textsc{MACHO} data inferred from astrometric microlensing experiments.  \lq\lq{}Field observed\rq\rq{} refers to whether the source stars were in the Large or Small Magellanic Clouds (LMC, SMC) or the Galactic Bulge.  \lq\lq{}Mass fraction\rq\rq{} denotes the percentage of the galactic dark matter budget comprised of \textsc{MACHO}s of various masses. The number of microlensing events is so small that is some cases no estimate for the typical mass of the compact dark matter halos was (attempted to be) determined.
\label{tab:macho}}
\vspace*{\fill}
\end{landscape}

\subsection{Massive compact halo objects} \label{sec:machos}
The next step is to look for solid objects which do not (observably) radiate.  These are known alternately as \lq\lq{}Jupiter-like objects\rq\rq{} or \lq\lq{}massive compact halo objects\rq\rq{} (\textsc{MACHO}s).  As the name suggests, this describes any object which is massive enough to avoid fragmentation ($M > 0.007 M_{\odot}$) but insufficiently massive for nuclear fusion ($M < 0.08 M_{\odot}$) \cite{olive}.
The contribution of MACHOs to the \dm{} budget is determined by the initial mass function of the stellar population in the galaxy.  Recall from \cref{sec:stars} that this follows a logarithmic law:
\begin{equation}\label{eq:imf}
\ddx{\;\ln m}{N}{} \propto m^{-(1+x)}
\end{equation}
where the slope $x$ is determined empirically.  Unlike the case for main-sequence stars (for which the Salpeter form $x = -0.35$ is a reasonable fit), the \textsc{MACHO}s obey a distribution which is not well-known: constraints from infrared observations in the galactic disc only constrain $x$ from below to $x > 1.7$ \cite{olive}.  The issue is further complicated by the fact that the disc and halo have different stellar populations, hence different IMFs.  
Thus, the possible contribution of \textsc{MACHOs} to the baryonic \dm{} budget of the galaxy must be inferred from observation rather than derived from theory.

To this end, various collaborations such as \textsc{eros} \cite{eros} and \textsc{ogle} \cite{ogle-iii-lmc,ogle-iii-smc} have calculated estimates via astrometric microlensing experiments.  
Astrometric microlensing is based upon the principle that observation of a rich field of background stars will counteract the low optical depth of potential foreground lenses \cite{macho-i}.  Accordingly, the experiments involve the collection of stellar fluxes (in the Large and Small Magellanic Clouds, sometimes the Galactic bulge) over a protracted length of time and the subsequent reduction of the data into light curves \cite{macho-ii}.  The light curves are then searched for transient events, of which a subset are extracted as variations due to gravitational lensing \cite{macho-ii}.  By examining millions of stars over several years, a very few microlensing events may be detected (\cref{tab:macho}).  Two major cuts are then performed \cite{ogle-iii-smc}: removing both extragalactic and already-known lenses.  In the first instance, a star in the Large Magellanic Cloud may be lensed by a halo in the Milky Way, or by a halo in the LMC itself \cite{macho-ii}.  The spatial distribution of lensing events is used to determine whether each event is consistent with \lq\lq{}self-lensing\rq\rq{} (\eg{}LMC-LMC lensing) or \lq\lq{}galaxy lensing\rq\rq{} \cite{ogle-iii-lmc}.  In the second instance, there may be evidence for the lens candidate to be an already-identified object, whether visible (\eg{} a foreground star) or not (\eg{} a black hole), in which case the lens cannot be a dark matter halo \cite{eros,ogle-iii-smc}.  The time-scales and spatial and temporal frequency of each event can then be used to place bounds upon the Galactic \textsc{macho} budget.  From \cref{tab:macho}, we see that this fraction varies considerably from survey to survey.  This should not be surprising, given the differences between the SMC and LMC environments (which creates data reduction biases \cite{eros}) and the small number of events (which is problematic for statistical analysis).  Despite the variations, the consensus is that \textsc{macho}s are not the main contribution to the Galactic \dm{} budget.

\subsection{Stellar Remnants}
The final possibility for baryonic \dm{} on sub-galactic scales is stellar remnants.  Black holes of both intermediate mass $M \sim 10^6 M_{\odot}$ and stellar mass have extremely small abundances which are limited by a lack of microlensing events and stellar dynamics arguments respectively \cite{dm-bartelmann}.  Cold white dwarfs are similarly unrealistic choices, since the ejecta produced during supernovae would (over-)contaminate the galactic disc with heavy elements: a large population would contradict the observed existence of low metallicity objects \cite{olive}.  This rules out dead stellar remnants as a significant contributor to the interstellar \dm{} budget.

\setlength{\LTcapwidth}{\textwidth}

\begin{landscape}
\begin{centering}

\begin{longtable}{*{2}{l} *{5}{p{.2\textwidth}}} 
\\\\\\\\
\toprule 
	& 	& WIMPs & SuperWIMPs & Hidden DM & Neutrinos & Axions  \\
\midrule
\endfirsthead 
\endfoot
\bottomrule \\[.1in]
\caption{Table of high energy physics dark matter candidates based on \cite{dm-particles}. \\ [\baselineskip]
The motivations for proposing each particle are: the gauge hierarchy problem (GHP); the new physics flavour problem (NPFP); the finite mass of neutrinos; the violation of invariance under parity-charge conjugation operations in the strong nuclear force (strong CP).  
\\ [\baselineskip]
The various detection methods are: direct detection from colliders; scattering off normal matter in lab experiments; dark-dark interactions resulting in annihilation or decay of the tabulated particles; possible signals from the early Universe such as effects on the CMB.  The other factors are discussed in the text.}
\endlastfoot 
\multicolumn{2}{l}{Motivation}
	& GHP &  GHP &  GHP + NPFP &  $\nu$ mass &  Strong CP \\
\midrule
\multicolumn{2}{l}{Temperature} 
	& Cold &  Cold / Warm &  Cold / Warm &  Warm &  Cold \\
\multicolumn{2}{l}{Mass range} 
	&GeV -- TeV &  GeV -- TeV &  GeV -- TeV &  keV &  $\mu$eV -- meV \\
\multicolumn{2}{l}{Naturally correct relic density}
	& Yes &  Yes &  Perhaps &  No &  No \\
\multicolumn{2}{l}{Production mechanism} 
	& Freeze-out &  Decay &  Various &  Various &  Various \\
\midrule 
\multirow{4}{*}{Detection}
	& Colliders &  Yes &  Yes &  Perhaps &  No &  No \\
	& Laboratory scattering &  Yes &  No &  Perhaps &  No &  Yes \\
	& Annihilation/decay &  Yes &  Perhaps &  Perhaps &  Yes &  No \\
	& Early Universe &  No &  Yes &  Perhaps &  No &  No
\end{longtable}

\end{centering}
\end{landscape} 

\section{Non-baryonic candidates}\label{sec:nbaryonic}
The following paragraphs constitute a brief review of non-baryonic \dm{} candidates.  The majority of these are \lq\lq{}thermal relics,\rq\rq{} so-called because they were produced by thermally efficient interactions in the early Universe \cite{axions}.  As the equilibrium temperature in the Universe decreased, the thermal relics were \lq\lq{}frozen out\rq\rq{} of the background plasma via phase transitions, preserving their primordial abundances \cite{olive}.  The non-baryonic \dm{} candidates can be divided broadly into neutrinos (\cref{sec:neutrinos}), \textsc{wimp}S (\cref{sec:wimps}) and hypothetical low mass relics (\cref{sec:axions}, \cref{sec:hidden}).

\subsection{Neutrinos and Sterile Neutrinos}\label{sec:neutrinos}
The simplest case is to consider particles which already exist in the Standard Model, namely neutrinos.  There are a multitude of different techniques for investigating the neutrino mass fraction (described in detail in \cite{olive}).  Of these, the technique most accessible to cosmologists is the measurement of CMB temperature fluctuations, which constrain the neutrino mass fraction to:
\begin{equation}
0.0005 \leq \Omega_{\nu}h^2 \equiv \sfrac{m_{\nu}} {94 \,\textrm{eV}} \leq 0.09
\end{equation}
That the (three left-handed) neutrinos have mass at all is (the strongest) evidence for the incompleteness of the Standard Model.  Sterile neutrinos are necessary to explain the non--zero neutrino mass.  

The ($N \geq 2$) sterile neutrinos are a right--handed analogue of the left-handed \lq\lq{}ordinary\rq\rq{} neutrinos.  Their existence enables the addition of left-- and right--handed neutrino coupling terms to the Standard Model Lagrangian \cite{dm-particles}.  In this way, the coupling terms bestow masses upon the corresponding neutrinos, whereas the lack of these terms in the Standard Model forces all neutrinos to be massless.

\begin{figure}
\centering
\includegraphics[width=\textwidth]{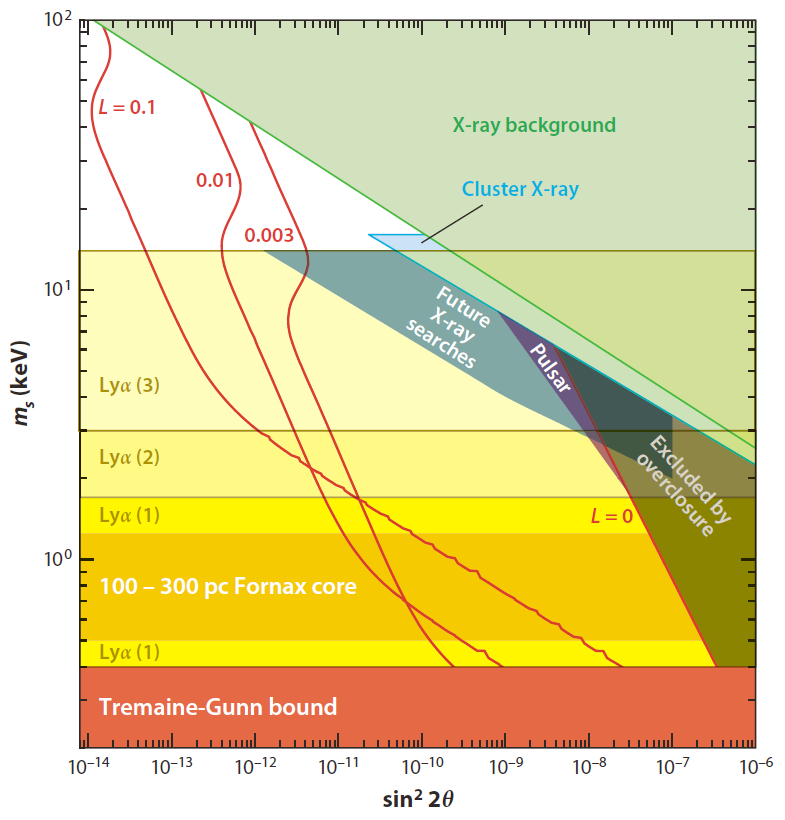} 
\caption{Parameter space for the sterile neutrino mass $m_{\nu}$ and mixing angle $\theta$.  Regions which are yellow are \textit{favoured} by astrophysical observations (darker regions have higher likelihood).  Regions shaded in other colours are \textit{excluded} by their respective observations while the red region is excluded on theoretical grounds. (Fig.~20 in \cite{dm-particles}) \label{fig:sterile}}
\end{figure}

The masses of (both active and sterile) neutrinos are determined by the eigenstates of the $(3+N) \times (3+N)$ neutrino mass matrix \cite{dm-particles}.  In practice, the sterile neutrino mass(es) $m_s$ and mixing angle $\sin^2 2\theta$ are degenerate in parameter space, so estimates \cref{fig:sterile} of the sterile neutrino relic density $\Omega_{\nu_s} = 2 \times 10^7 \,\sin^2 2\theta \left(\sfrac{m_s}{3\,\textrm{keV}} \right)^{1.8}$ are difficult to obtain, despite a number of constraint--imposing searches \cite{dm-particles}.  

Nevertheless, the consensus is that the neutrino mass fraction contributes insufficiently for neutrinos to be a candidate for the majority of (non--baryonic) \dm.  A corollary to this is that non-baryonic \dm{} necessitates a major extension to the current Standard Model of particle physics.

\begin{figure}
\centering
\includegraphics[width=0.8\textwidth]{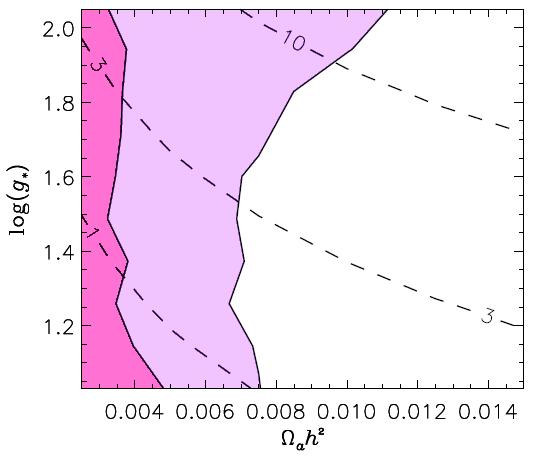} 
\caption{Credible regions for the axion parameters $g_{\ast}$ (thermal degrees of freedom) and $\Omega_a$ (cosmological density) from cosmological sources.  The axion mass in eV is indicated by the dashed lines.  Regions which are unshaded are \textit{excluded} at the 95\% CL; regions which are lightly shaded are also excluded at the 68\% CL.  (Fig.~5 in \cite{axions}) \label{fig:axions}} 
\end{figure}

\subsection{Axions}\label{sec:axions}
The next step is to consider a minor extension to the Standard Model.  This extension solves the \lq\lq{}strong CP problem\rq\rq{} extant in Standard Model quantum chromodynamics (QCD) via the introduction of a massive particle known as an axion.  

The \lq\lq{}strong CP problem\rq\rq{} is a conflict between the predicted and observed value of the neutron electric dipole moment $d_e$ \cite{axions}.  Experimentally, the dipole moment has not yet been observed, which restricts its value to $d_e < 2.9 \times 10^{-26}\,e\textrm{cm}$ \cite{dm-particles}.  Theoretically, its value is determined by the Lagrangian term%
\footnote{This term is $\sfrac{g^2_3 \theta_3}{32 \pi^2} \epsilon^{\mu\nu\rho\sigma} G^{\alpha}_{\mu\nu} G^{\alpha}_{\rho\sigma}$, where $g_3$ is the coupling of the strong interactions, $\theta_3$ is an angle parameter, $\epsilon$ is the totally antisymmetric 4-index tensor, and $G$ is the gluon field strength.  Given the values of $g_3$ and setting $\theta_3$ to unity gives the estimate of $d_e$.}
describing 
interactions via the strong nuclear force: this is expected to be $d_e \sim 10^{-16}\,e\,\textrm{cm}$ \cite{dm-particles}.  The cause of the apparently contradictory results is the nature of QCD as a CP-violating force \cite{dm-particles,axions}: to avoid a fine-tuning problem it is necessary to restore CP-conservation.

The most elegant way of achieving this is to introduce another symmetry which becomes broken at some large energy scale $f_a$.  As with all gauge symmetries, there exists a corresponding boson which is termed an axion \cite{axions}.  The symmetry breaking gives the axion a small mass \cite{axions}.  Both the mass and the number density of axions depend upon $f_a$ (up to a constant which varies with axion production model) \cite{dm-particles}
 and both can be constrained cosmologically \cite{axions}.  The relic density $\Omega_a$  takes different forms depending on whether the axion production mechanism is thermal or non-thermal and in the latter case, upon whether the symmetry-breaking phase transition occurs before or after inflation \cite{dm-particles}.\footnote{%
The exact details are not relevant to this thesis: the interested reader may find \S7 of \cite{dm-particles} illuminating.  Cosmological constraints on the axion mass and energy scale  are discussed in \cite{axions}.}

\subsection{WIMPs and SuperWIMPs}\label{sec:wimps}
The weakly-interacting massive particles (\textsc{wimp}s) are the most-studied candidates for dark matter because they have several appealing features, namely a naturally correct relic density (cosmological mass fraction); (particle-) model-independent properties.   
Indeed, Bertschinger asserts in \cite{wimp-decoupling} that \textsc{wimp}s are the leading candidate to comprise the majority of non-baryonic, non-relativistic \dm{}.

Motivation for the existence of \textsc{wimp}s is provided by the gauge hierarchy problem: the question of why the Higgs boson mass is finite but small.  
The \lq\lq{}natural\rq\rq{} value expected for dimensionful quantities is either zero or the combination of fundamental constants which has the same dimension: for the Higgs boson mass, this is the Planck mass $M_{Pl} = \sfrac{\sqrt{hc}}{G} = 1.2 \times 10^{19}\,\textrm{GeV}$.  In contrast, the physical mass of the Higgs boson is $\sim 125 \,\textrm{GeV}$.  Generation of the physical mass $m_b$ from the natural one is achieved by \lq\lq{}correcting\rq\rq{} the tree-level mass $m_{b0}$ with the loop-level adjustment $\Delta m_{b}$ \cite{dm-particles}:
\begin{equation}
m_b^2 
= m_{b0}^2 + \Delta m_b^2 
= M_{Pl}^2 - \frac{\lambda^2}{16 \pi^2} \Lambda^2 
\approx M_{Pl}^2 \left( 1 - \frac{\lambda^2}{16 \pi^2} \right) 
\implies 1 - \frac{\lambda^2}{16 \pi^2} 
\approx \left(\frac{m_b}{M_{Pl}}\right)^2
\end{equation}
a fine-tuning problem of one part in $10^{36}$!  The solution is to modify $\Lambda$, the energy scale at which the Standard Model is no longer valid \cite{dm-particles}.  Similarly to the case of axions, this introduces a new family of particles, \textsc{wimp}s, whose properties are associated with a symmetry-breaking field which resolves the fine-tuning problem.

The behaviour of \textsc{wimp}s is similar to nucleons.  Both types of particles \lq\lq{}froze out\rq\rq{} of the plasma in the early Universe, but remained coupled to it via scattering interactions until recombination (or its \textsc{wimp} analogue, kinetic decoupling); consequently both particles left acoustic oscillation signals on cosmic structure (BAOs on Mpc scales and \textsc{wimp} acoustic oscillations on pc scales) \cite{wimp-decoupling}.  These similarities produce two favourable properties for \textsc{wimp}s: a model-independent relic density of $\Omega_{\textsc{wimp}} \sim \Omega_{DM}$ \cite{dm-particles} and the possibility of an astrophysical detection method \cite{wimp-decoupling}.

It is possible that \textsc{wimp}s decay into particles with extra-weak interactions, denoted super\textsc{wimp}s.  The super\textsc{wimp} theory requires a second phase transition caused by \textsc{wimp} decay, but if the masses of \textsc{wimp}s and super\textsc{wimp}s are of the same order, then the super\textsc{wimp} relic density retains the correct order of magnitude to comprise the majority of \dm{}.

\subsection{Hidden dark matter}\label{sec:hidden}
The final strong possibility for \dm{} is a type which has no Standard Model interactions: hidden \dm{}.  The existence of hidden \dm{} addresses the issue that \lq\lq{}all solid evidence for dark matter is gravitational\rq\rq{} by suggesting that \dm{} should not be given gauge interaction properties when there is no firm evidence that it has them \cite{dm-particles}.  

The increased freedom implied by the lack of strong and (electro-)weak interactions produces that drawback that hidden \dm{} encompasses a wide range of particles: correspondingly, its interaction characteristics are difficult to define.  A corollary of this is that it is difficult for a single detection method to be effective.  Nevertheless, a variety of detection methods are proposed in \S6 of \cite{dm-particles}.  Of particular interest is the possibility that astrophysical methods can be used to limit the hidden \dm{} particle mass and \lq\lq{}fine structure constant,\rq\rq{} as demonstrated by \eg \cite{bullet-cluster,ellipticity}.  Ultimately, the definition of hidden \dm{} as a family of particles with no Standard Model interactions has the direct implication that its only effect on ordinary matter is gravitational.

\section{Conclusions}
There is considerable evidence for the existence of \dm{} on all astronomical scales.  Moreover, the directly-observable matter in galactic discs and bulges interacts gravitationally with this dark matter.  The baryonic sources of \dm{} are difficult to quantify: although primordial gas clouds and stellar remnants  have been ruled out as major contributors to the \dm{} budget, it is difficult to obtain estimates for \textsc{jlo}s and \textsc{macho}s.  Furthermore, the range of proposed candidates for non-baryonic \dm{} precludes a single detection method from particle physics techniques.  Consequently, the only detection method which is sensitive to all possible forms of \dm{} must be purely gravitational.  This suggests that one should examine the gravitational interaction of matter (dark or otherwise) with photons.  This is a well-observed, theoretically sound phenomenon called gravitational lensing.
\chapter{Simulations of gravitational lensing}\label{ch:gl}

General Relativity predicts a quantifiable interaction between matter and photons.  The spacetime in a neighbourhood of some matter distribution will have its geometry changed by the presence of the matter.  Consequently, any photons passing through this region will be affected.  It follows that we can reverse this process and use the changes in the photons to infer details about the matter distribution.  

Despite the long history of \gl{ing} in theoretical papers, the field is considered relatively new \cite{sef}.  The first quantitative paper \cite{Soldner} suggesting that the path of light rays would be perturbed by the presence of matter is due to Solder in 1804.  He calculated that the deflection of light at the solar limb would be $0."87$, in contrast to the relativistic calculation by Einstein%
\footnote{In 1911, Einstein recast Soldner's calculations in the framework of special relativity, deriving the same value.  Only after finishing the theory of \gr in 1915 did Einstein publish his correct value for light deflection at the solar limb.  In the meantime, the expedition sent from Potsdam to test Einstein's first calculation during the solar eclipse of September 1914 had had their equipment confiscated and returned from the Crimean Peninsula without a result.  That Einstein's second calculation agreed with observations was not confirmed until Eddington's measurements of the next solar eclipse in 1920 \cite{gl-bartelmann}.}
of $1."68$.  A few lensing papers were published in the 1920s and 1930s, including papers by Chowlson and Einstein on the formation of a circular image when source and lens were perfectly aligned (now termed an \lq\lq{}Einstein ring.\rq\rq{}) \cite{Einstein-1936,Chowlson}.  The next theoretical advances did not occur until the 1960s: in particular the formulae used in modern gravitational lensing were mostly derived by Refsdal in 1964 \cite{ml-summary}.  The next major theoretical development occurred in 1986, with the suggestion by Pacz\`{y}nski that a collection of unresolvable micro-images might moderate the intensity of the macroscopic lensing image in observable ways \cite{ml-summary}.  This technique, called \lq\lq{}microlensing,\rq\rq{} has been applied widely to search for dark matter.  Establishing the existence of lensing as a useful observational technique did not occur until the 1980s \cite{sef}.  The discovery of a gravitationally-lensed quasar in 1979 \cite{walsh} was the first example of this phenomenon outside the solar environment.  This triggered the development of further searches for examples of lensing on galactic and cosmological scales.  Thus, the theory behind lensing was already well-defined before its observational application.

This chapter explains the simulation of a single \gl{}, including an introduction to the theory of (single-)\gl{} models.  Firstly in \cref{sec:models} we introduce the various lens models used.  A brief explanation of light propagation in the single-lens setup follows in \cref{sec:propagation}.  The next section uses each model to examine the key effects of \gl{ing}, which fall into three categories: multiple images from a single source; a Shapiro-like time delay induced in the photons and magnification effects due to flux conservation.  A relativistic version of Fermat\rq{}s principle, derived in \cref{sec:fermat} determines a general formula for the time delays (\cref{sec:lenseq}) and image locations (\cref{sec:images}).  Magnification effects are discussed in \cref{sec:flux}.  Each subsection describes a key theory and new approaches to its efficient calculation before the lens models are used to demonstrate the numerical accuracy of the code.  This completes the theoretical basis of the single-lens simulations.  Finally, the transition from a single to multiple lens system is described in \cref{sec:multiple}.  This is divided into three sections: a co-ordinate translation so that the lens rather then the source moves; appropriate superposition of the time delays onto the puslar signal; lastly a useful choice of iteration scheme to model the lens\rq{} progress between the pulsar and the Earth.  This completes the method used in the full simulation.

\section{Lens models}\label{sec:models}
Four lens models were chosen for the project.  Of those, three possess axial symmetry.  The simplest lens is the Schwarzschild lens: a point mass, one-parameter lens which models a compact and dense lensing object akin to a black hole.  The homogeneous disc is the simplest model with finite radius, comprising a flat disc of constant density.  The most realistic model is the spherically-symmetric \nfw{} profile.  The remaining model --- the elliptical lens --- lacks axial symmmetry.  For this reason, it is usually defined in terms of the lensing potential rather than the density profile.  The introduction of each model is discussed in the next section.  A summary of the key properties of each lens model is in \cref*{tab:lens-models}.

\begin{figure}[b]
\includegraphics[height=.37\textheight]{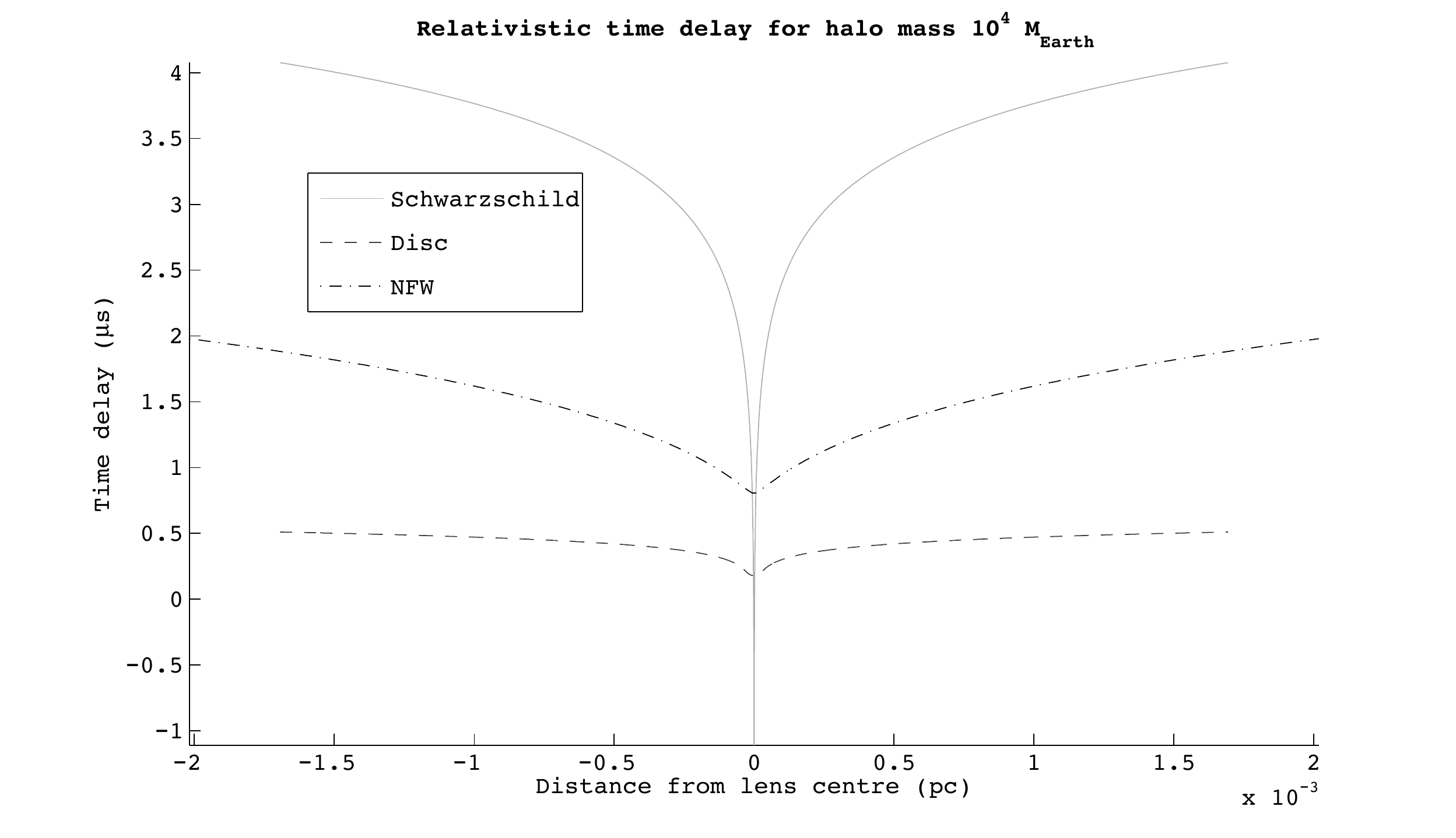}
\caption{Comparison of the relativistic components of the time delays from the three axisymmetric models.}
\end{figure}

\subsection{\swz{} lens}\label{sec:swz}
The Schwarzschild lens is the simplest possible lensing geometry.  It represents a point-mass: its only free parameter is the total mass of the lens $M$.  Of particular interest is the natural length scale defined by this lens:
\begin{equation}
R_E = \sqrt{ R_S \frac{D_{\textrm{d}} D_{\textrm{ds}}}{D_{\textrm{s}}} } \; \text{where $R_S = \dfrac{2GM}{c^2}$ }
\end{equation}
The radii $R_E$ and $R_S$ are the Einstein and \swz{} radii respectively.  The Einstein radius is defined by the \swz{} lens thus: a collinear lens and source produce a ring-shaped image with radius equal to the Einstein radius \cite{mpa-notes}.  The \swz{} radius acts as a condition for a lens with finite physical radius to be modelled as a point mass, namely that the impact parameter is $\norm{\vect{\xi}} \gnsim R_S$. The representation of the convergence is Dirac\rq{}s delta function.    As a consequence, the time delay, magnification factor and lens equation are all analytically tractable \cref{tab:lens-models}.  It is this property which makes the \swz{} lens a useful tool for numerical analysis.

\subsection{The homogeneous disc model}\label{sec:hdisc}
The next step is to add a finite radius to the lens.  The \hdisc{} model is exactly what it suggests, namely a disc-shaped lens with constant density.    Given a lens of total mass $M$ and physical radius $\rho_0$, its surface mass density is $\Sigma(\xi) = \sfrac{M}{\pi\rho_0^2}$.  Using the scaling relation defined in \eqref{eq:convergence}, the convergence is $\kappa(x) = \sfrac{1}{x_0^2}$ inside the lens (and zero outside).  As may be expected from such a simple density profile, the lens equation is easily (albeit piecewise-)invertible and the magnification factor has an analytical form.  The potential time delay is not analytical (\cf the \swz{} case) but it is sufficiently simple to be a suitable test for the integral transform used for numerical calculation of the time delay (as we shall see in \cref{ch:results}).

\subsection{The Navarro-Frenk-White profile}\label{sec:nfw}
The last radial lens model is the most realistic.  The \nfw{} model was developed via numerical simulation of \dmh{s} using the standard (cold dark matter) cosmology \cite{nfw1,nfw2}.  They concluded that dark matter halos in four decades of mass showed a \lq\lq{}universal\rq\rq{} density profile of the form:%
\footnote{This is the notation used in \cite{bartelmann-arcs} rather than that of Navarro, Frenk \& White\rq{}s original papers \cite{nfw1,nfw2}.}
\begin{equation}
\rho({\vect{r}}) = \dfrac{\rho_s}{\sfrac{\norm{r}}{r_s} \left( 1 + \sfrac{\norm{r}}{r_s} \right)^2} \; \text{where $\vect{r} \in \mathbb{R}^3$ and $\rho_s$, $r_s \in \mathbb{R}^+$}
\end{equation}
This lens does not fulfil the thin-lens approximation because it is extended in the radial direction: it is necessary to project this density profile onto the lens plane.  Setting the natural length scale to be $r_s$, let $x = \sfrac{\norm{r}}{r_s}$.  Scaling and applying the convergence definition \eqref{eq:convergence} implies:
 \begin{subequations} \label{eq:nfw}
\begin{align}
\kappa(x) &= \dfrac{2\kappa_s}{x^2 - 1} \left(1 - g(x) \right)  \\
\intertext{Similarly, the enclosed-mass integral \eqref{eq:mass} implies:}
m(x) &= 4\kappa_s \left( \ln \dfrac{x}{2} + g(x) \right)
\intertext{where $\kappa_s = \rho_s r_s \Sigma_{cr}^{-1}$ is a constant and $g(x)$ is the continuous function:}
g(x) &= 
\begin{cases}
\dfrac{2}{ \sqrt{x^2 - 1} } \arctan \sqrt{ \dfrac{x-1} {x+1} }  
&\text{for $x > 1$} \\[1em]
\dfrac{2}{ \sqrt{1 - x^2} } \operatorname{arctanh} \sqrt{ \dfrac{1-x} {1+x} } &\text{for $x < 1$} \\[1em]
1 &\text{for $x = 1$}
\end{cases} 
  \end{align}
\end{subequations} 
This lens profile has three free parameters: the lens mass $M$ which contributes to the critical mass density $\Sigma_{cr}$, the scale radius $r_s$ chosen such that the lens has unit turnover radius and the scaled density $\rho_s$ which is related to the concentration of the lens \cite{bartelmann-arcs,nfw1,nfw2}.  

\subsection{Elliptical lens}\label{sec:elliptical}
The elliptical lens is a purely empirical model.  It was suggested by Kochanek\etal{} to model the shape of a lensed radio source \cite{ring-cycle}. 
The name originates from the elliptical potential for the lens plane:
\begin{equation}\label{eq:elliptical}
\Psi(x,y) = b\sqrt{s^2 + (1-e)(x - x_0)^2 + (1+e)(y - y_0)^2} 
\end{equation}
which is centred on $(x_0, y_0)$ and has ellipticity $0 \leq e \leq 1$.  The two other free parameters are the core radius $s$, which is the radius of the circular lens (\ie the corresponding lens with $e=0$) and the lens strength $b$, which scales the potential \wrt{} the geometric delay term.  Since the time delay is explicitly defined, it is unnecessary to calculate the convergence (which obeys the Poisson equation) and the lens equation (which in 2d can be found more easily by searching for extrema of the time delay surface).  For our purposes, the elliptical lens merely serves as a test parametrisation of the relativistic time delay.

\setlength{\LTcapwidth}{\textwidth}
\renewcommand{\arraystretch}{1.75} 

\begin{landscape}
\begin{longtable}{ lllll }
\toprule 
Lens
& Free Parameters
& Convergence $\kappa(x)$
& Lens equation $ x - \sfrac{m(x)}{x}$
& Magnification factor $\mu(x)$
 \\
\midrule
\endfirsthead 

\bottomrule
\endlastfoot 


Schwarzschild
& $M$ Lens mass
& $ \delta(x)$
& $ x - \dfrac{1}{x} $  
& $\left( 1 - \dfrac{1}{x^4} \right)^{-1}$ 
\\[.1in]
 \midrule 
\\[-.2in]

\multirow{2}{*}{Homogeneous disc} 
& $M$ Lens mass 
& \multirow{2}{*}{ $ \kappa_0 \equiv 
\begin{cases}
\dfrac{1}{x_0^2} &\text{for $x \leq 1$} \\ 
0  \vphantom{\frac{a}{b}}  &\text{for $x > 1$}
\end{cases} $}
& \multirow{2}{*}{ $
 \begin{cases}
x - \dfrac{x}{x_0^2} &\text{for $x \leq 1$} \\
x - \dfrac{1}{x} &\text{for $x > 1$} 
\end{cases} $ }
& \multirow{2}{*}{ $
\begin{cases}
\dfrac{1}{(1 - \kappa_0)^{2}} 
&\text{for $x \leq 1$} \\
\dfrac{1}{(1 - \kappa_0)^{2}} + 
&\text{for $x > 1$} 
\end{cases} $ } 
\\ 
& $x_0$ Lens radius & & & \\
\\[-.2in]
 \midrule 
\multirow{3}{*}{\begin{minipage}{2.7cm}Navarro--Frenk \\ --White profile \end{minipage}}
& $M$ Lens mass
& \multirow{3}{*}{ $ \dfrac{2\kappa_s}{x^2 - 1} \left( 1 - g(x) \right) $ }
& \multirow{3}{*}{ $ x - \dfrac{4\kappa_s}{x} \left( \ln \dfrac{x}{2} +  g(x) \right) $ }
& \multirow{3}{*}{ $ 
\dfrac{1}{
\vphantom{\Biggl( \Biggr)} 
 \left( 1 - \dfrac{m(x)}{x^2} \right) \left( 1 + \dfrac{m(x)}{x^2} - 2\kappa(x) \right)  }
$ } 
\\ 
& $x_0$ Lens radius & & & \\
& $\rho_s$ Density scale & & & \\
\midrule

\multirow{3}{*}{Elliptical}
& $e$ Ellipticity
& \multirow{3}{*}{---------} 
& \multirow{3}{*}{---------} 
& \multirow{3}{*}{---------} 
 \\
& $b$ Lens strength & & &\\
& $s$ Core size & & & \\

\end{longtable}
\captionof{table}{Summary of the key properties of the four lens models. The elliptical, \swz{} and \hdisc{} lenses are scaled by the Einstein radius $x = \sfrac{\xi}{r_E}$ whereas the \nfw lens is scaled by the scale radius $r_s \gg r_E$. The functions $g(x)$ are defined in the text. The elliptical lens model does not show the convergence, lens equation nor magnification factor, since these were not used in the lens modelling (and due to the lack of radial symmetry, these expressions are not particularly enlightening in radial co-ordinates). \label{tab:lens-models}}
\vspace*{\fill}
\end{landscape}
\renewcommand{\arraystretch}{1.0} 

\section{Light propagation in gravitational lens systems}\label{sec:propagation}
\begin{figure}[hb]
\includegraphics[width=\textwidth]{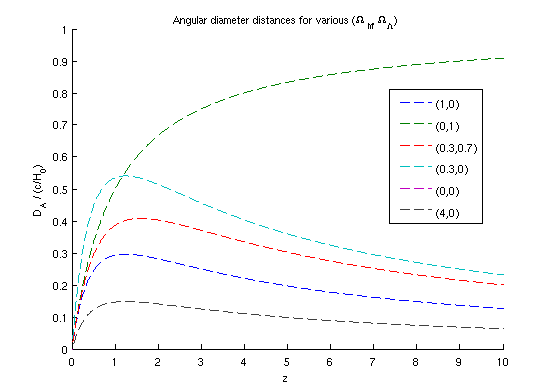} 
\caption{Angular diameter distances in an FLRW universe for various fractional densities ($\Omega_M$, $\Omega_{\Lambda}$) of matter and dark energy respectively.  The dark energy in this case is the cosmological constant, with $w(z) = -1$.  At nearby redshifts $z \ll 1$, the distances converge regardless of the cosmology.}
\label{fig:angular-diameter-dist}
\end{figure}

\begin{figure}
\includegraphics[width=0.6\textwidth,clip]{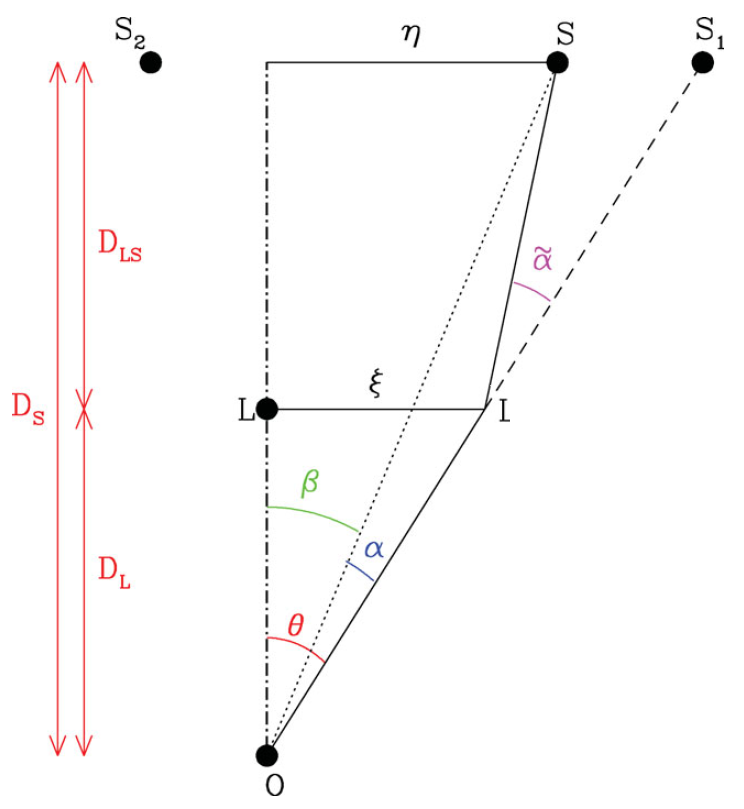}
\caption{Geometry of a typical gravitational lens system. The positions of the observer, source and lens are represented by \lq{}$O$\rq{},  \lq{}$L$\rq{} and \lq{}$S$\rq{} respectively.  The two apparent image locations are denoted \lq{}$S_1$\rq{} and \lq{}$S_2$.\rq{}  The angular diameter distances $D_{\text{L}}$, $D_{\text{S}}$ and $D_{\text{LS}}$ are between observer-lens, observer-source, and source-lens.  
Image credit: Fig.~3 in \cite{gl-bartelmann}
\label{fig:lens-setup}} 
\end{figure}

The propagation of photons according to \gr can be separated into global and local parts.%
\footnote{Due to the non-linearity of \gr{}, this is only possible if the local perturbations are sufficiently small that a linear approximation is appropriate.  The exception to this case is when the metric of the local perturbations is not a form of the weak-field (Newtonian) metric.  If the object is a black hole, for example, the \swz{} metric must be applied.}
The \lq\lq{}local\rq\rq{} part describes the perturbation due to the presence of the gravitational lens \cite{sef}.  The key details are derived in \cref{sec:fermat}.  The \lq\lq{}global\rq\rq{} part describes the distance along the un-lensed geodesic \cite{sef} prescribed by the metric (\cf \cref{sec: cosmological}).  
   We still require a definition of observable (rather than co-ordinate) distance on the manifold.  In general, different (practical) measurement methods on a metric will give different distance quantities: it is necessary to \textit{define} a distance by the method by which it is calculated.  The geometry of a lens system \cref{fig:lens-setup} suggests that we need to relate the physical cross-section $\delta A$ of an object at redshift $z_2$ and the solid angle $\delta\omega$ that it subtends for an observer at $z_1$ \cite{mpa-notes}: thus we define
\begin{equation}\label{dist-ang}
D_{\textrm{ang}}(z_1,z_2) 
\equiv \Bigl(\frac{\delta A|_{z_2}}{\delta\omega|_{z_2}}\Bigr)^{1/2}
= \frac{1}{1+z} \frac{c}{H_0} \integ{z^{\mathsmaller\prime}}{0}{z}{\sqrt{\mathsmaller{\sum}_{\sigma}\Omega_{\sigma}(z^{\mathsmaller\prime}) (1+z^{\mathsmaller\prime})^{3(1+w_{\sigma})}}} 
\end{equation}
This is the angular diameter distance $D_{\textrm{ang}}$ \eqref{dist-ang}.  Calculation of $D_{\textrm{ang}}$ depends upon the values of the cosmological mass fractions $\Omega_{\sigma}$ and the equations of state $w_{\sigma}$ of the contents of the universe, as well as the value of the Hubble constant $H_0$ \cref{fig:angular-diameter-dist}.  At the redshifts used to test the lens scaling in \cref{sec:scaling}, the non-Euclidean form of the angular-diameter distance becomes important: in particular it is not linear, \ie $D_{\textrm{ang}}(z_1,z_2) \neq D_{\textrm{ang}}(z_0,z_2) - D_{\textrm{ang}}(z_0,z_1)$ unless $z_{1,2} \approx 0$.  At the redshifts at which pulsars are detected, it converges regardless of the contents of the Universe \cref{fig:angular-diameter-dist}.  This justifies the use of Euclidean distances throughout the pulsar lensing calculations.  Our definition of distance enables the calculation of the path length of the geodesics which the photons trace from source to observer.  This is explained in the next section.

\section{Fermat's principle}\label{sec:fermat}
The geodesic linking the source and the observer is perturbed by the presence of the intervening lens.  As in classical optics, the path length of the geodesic is an extremum, following Fermat's Principle.  This causes the apparent (observed) position of the source to differ from its true (physical) position.  This section demonstrates a geometric argument for the relationship between the lens and the true and observed locations of the source in \cref{sec:lenseq}.  Then \cref{sec:fermat} shows that a relativistic version of Fermat's Principle can be used to calculate the corresponding image locations \cref{sec:images} and time delays \cref{sec:tdtot}.

\subsection{The lens equation}\label{sec:lenseq}
The relationship between the observed images of the source and its true position is called the lens equation.  A typical lensing geometry is shown in \cref{fig:lens-setup}.  The optical axis is chosen such that the observer $O$ and lens $L$ is centred upon it.  The two spheres of radius $D_{\textrm{s}}$ and $D_{\textrm{d}}$ mark the radial (angular diameter) distance to the source and lens respectively.  Adopting angular co-ordinates, the true position of the source subtends an angle $\vect{\beta}$.  Without the presence of the lens, the observed position of the source $S$ would subtend the same angle at the lens plane.  With the deflection of the lens, the source appears at (possibly more than one) image location $S'$.  The point mass lens in the example diagram causes two images on either side of the lens.  The corresponding angular deflection $\vect{\alpha}$ is a(n as-yet arbitrary) function.  The angles involved must be sufficiently small to replace the spherical geometry with two planes tangent to their respective spheres at $L$ and $S$: the lens and source planes.  By definition of angular diameter distance, we can establish
\clearpage
 a co-ordinate chart on the planes using length instead of angles:
\begin{subequations} \begin{align}\label{eq:Cartesian}
\vect{\theta} &\equiv \frac{\vect{\xi}}{D_{\textrm{d}}} \iff \vect{\xi} = D_{\textrm{d}}\vect{\theta} \\
\vect{\beta} &\equiv \frac{\vect{\eta}}{D_{\textrm{s}}} \iff \vect{\eta} = D_{\textrm{s}}\vect{\beta} = \frac{D_{\textrm{s}}}{D_{\textrm{d}}}\vect{\theta}
\end{align} \end{subequations}

We require one further condition: the thin-lens approximation.  This states that the radial extent of the lens is much less than the distance to it (or between it and the source plane).  Provided that the lens is geometrically thin, the geodesics between the two planes and the observer can be approximated by the piecewise-straight line $SIO$ (recall that our three-space metric is Euclidean).  The actual (curved) 
path of the light is represented by the deflection angle $\tilde\alpha(\theta)$ which links the two asymptotes of the real geodesic \cite{sef}.  The lens equation follows directly:
\begin{subequations}\label{eq:lens-eqn}
\begin{align}
\vect{\beta} &= \vect{\theta} - \vect{\alpha}(\vect{\theta}) 
&\text{(angular co-ordinates)} \\
\vect{\eta} &= \frac{D_{\textrm{d}}}{D_{\textrm{s}}} \vect{\xi} - D_{\textrm{ds}} \vect{\tilde{\alpha}}(\vect{\xi})
&\text{(linear co-ordinates)} 
\end{align}
\end{subequations} 
This is a mapping from the set of image vectors $\vect{\xi}$ to the source vector $\vect{\eta}$.  

\subsection{Image locations}\label{sec:images}
Our aim is to invert the mapping: given the images by observation, we require the source position.  The surjectivity of the equation renders this analytically insoluble for all but a few lens configurations \cite{sef}.  The aim of this section is to rewrite \eqref{eq:lens-eqn} into a variational problem.  

Scaling is necessary to avoid potential numerical errors in the modelling process, such as catastrophic cancellation.  For this reason, we introduce a fiducial scale parameter $\xi_0$ in the lens plane and a corresponding parameter $\eta_0$ in the source plane.  These define new Cartesian co-ordinates $(x,y)$:
\begin{equation}\label{eq:scaling}
\vect{x} = \frac{\vect{\xi}}{\xi_0} \quad \vect{y} = \frac{\vect{\eta}}{\eta_0} \quad \text{where } \eta_0 = \xi_0 \frac{D_{\textrm{s}}}{D_{\textrm{d}}} 
\end{equation}
The lens equation is now dimensionless, so it can be rewritten as a gradient:
\begin{equation}
\vect{\eta} = \frac{D_{\textrm{d}}}{D_{\textrm{s}}} \vect{\xi} - D_{\textrm{ds}} \vect{\alpha}(\vect{\xi})
\implies \vect{y} = \vect{x} - \vect{\alpha}(\vect{x}) 
\end{equation}
Rearranging:
\begin{equation}
0 = (\vect{y} - \vect{x}) - \alpha(\vect{x}) = \nabla \left( \frac{1}{2}(\vect{y} - \vect{x})^2 \right) - \alpha(\vect{x}) = \nabla \left( \frac{1}{2}(\vect{y} - \vect{x})^2 - \Psi(\vect{x}) \right)
\end{equation}
where we have introduced the lensing potential $\Psi(\vect{x})$,  whose gradient is the deflection angle $\vect{\alpha}$.  The gradient is taken with respect to $\vect{x}$: this is the independent variable because we aim to find the image locations in the lens plane (hence using its co-ordinate system).  Thus, we have reduced the problem of inverting the map \eqref{eq:lens-eqn} to that of finding the zeros of the gradient function:
\begin{subequations}\label{eq:fermat-potential}
\begin{align}
\nabla \left( \frac{1}{2} \left(\vect{x} - \vect{y} \right)^2 - \Psi(\vect{x}) \right) &= \nabla\phi(\vect{x})
\intertext{Under the assumption of an axisymmetric lensing potential, this simplifies to:}
\frac{\partial}{\partial\rho} \left( \frac{1}{2} \left(\rho - \norm{y} \right)^2 - \Psi(\rho) \right) &= 
\frac{\partial\phi}{\partial\rho} 
\end{align}
\end{subequations}
The new lens equation is analogous to Fermat's principle: zeros $\{ x_0 \boldsymbol{:} \nabla\phi(x_0) = 0 \}$ correspond to extrema of the (total) potential $\phi$.  

The identification of image locations is a root-finding problem.  In two dimensions, this can be done by using ray-tracing in the case of a general lens potential.  This proves to be an unnecessary complication for the lens models used here.  For the elliptical lens, an analytical formula for the derivative in the $x-$ and $y-$direction exists, so the problem reduces to a set of one-dimensional equations.  The other three lenses have radial symmetry which fixes the angular co-ordinates of source and image to be equal (for proof see \cref{sec:hankel}).  Thus we require only the zeros of the radial equations.  

The one-dimensional root-finding procedure is shown in detail in \cref{code:rootsearch}.  The key problem is that we can only search for intervals within which the function changes sign: this will erroneously include solutions which diverge and exclude double (quadruple \etc) roots which touch but do not cross the axis.  Consequently, we can only find all roots by searching for roots of higher derivatives (since a double root of $f(x)$ is a single root of $\ddx{x}{f}{}$).  Then it is necessary to evalyate $f(x)$ at all the \lq\lq{}roots\rq\rq{} to confirm that they are zeros of the function rather than divergent points where $f(x)$ is infinite.  

\begin{samepage}A simplified algorithm is as follows:
\begin{center}
\begin{minipage}[][][c]{.8\textwidth}
\begin{enumerate}
  \item{Generate an array of $x \in [a,b]$.}
  \item{Given $f(x)$, calculate the analytical forms of $\ddx{x}{f}{}$ and $\ddx[2]{x}{f}{}$.}
  \item{Find the $x_i^{(2)} \in [a,b]$ where $\ddx[2]{x}{f}{}$ changes sign.  Set $x_0^{(2)}=a$ and $x_{end}^{(2)}=b$.}
  \item{For each interval $[x_i^{(2)}, x_{i+1}^{(2)}]$:}
    \begin{enumerate}
      \item{Find the $x_j^{(1)} \in [x_i^{(2)}, x_{i+1}^{(2)}]$ 
where $\ddx{x}{f}{}$ changes sign.}
      \item{For each interval $[x_j^{(1)}, x_{j+1}^{(1)}]$:}
        \begin{enumerate}
           \item{Find the $x_k^{(0)} \in [x_j^{(1)}, x_{j+1}^{(1)}]$ 
where $f(x)$ changes sign.}
          \end{enumerate}
    \end{enumerate}
    \item{Combine the arrays $x^{(0)}, x^{(1)},x^{(2)}$.}
    \item{Evaluate $f(x)$ at each value in the array to check that it is a root.}
\end{enumerate}
\end{minipage}
\end{center}
\end{samepage}
The weakness of such a scheme is that the derivatives of $f(x)$ have to be calculated, but unlike the case for integrals, they will always exist in closed form.  We have now shown that, given a lens equation, it is possible to find the co-ordinates which satisfy extremisation of the path from source to observer.

\begin{figure}
\includegraphics[width=\textwidth,clip,trim=30 95 30 85]{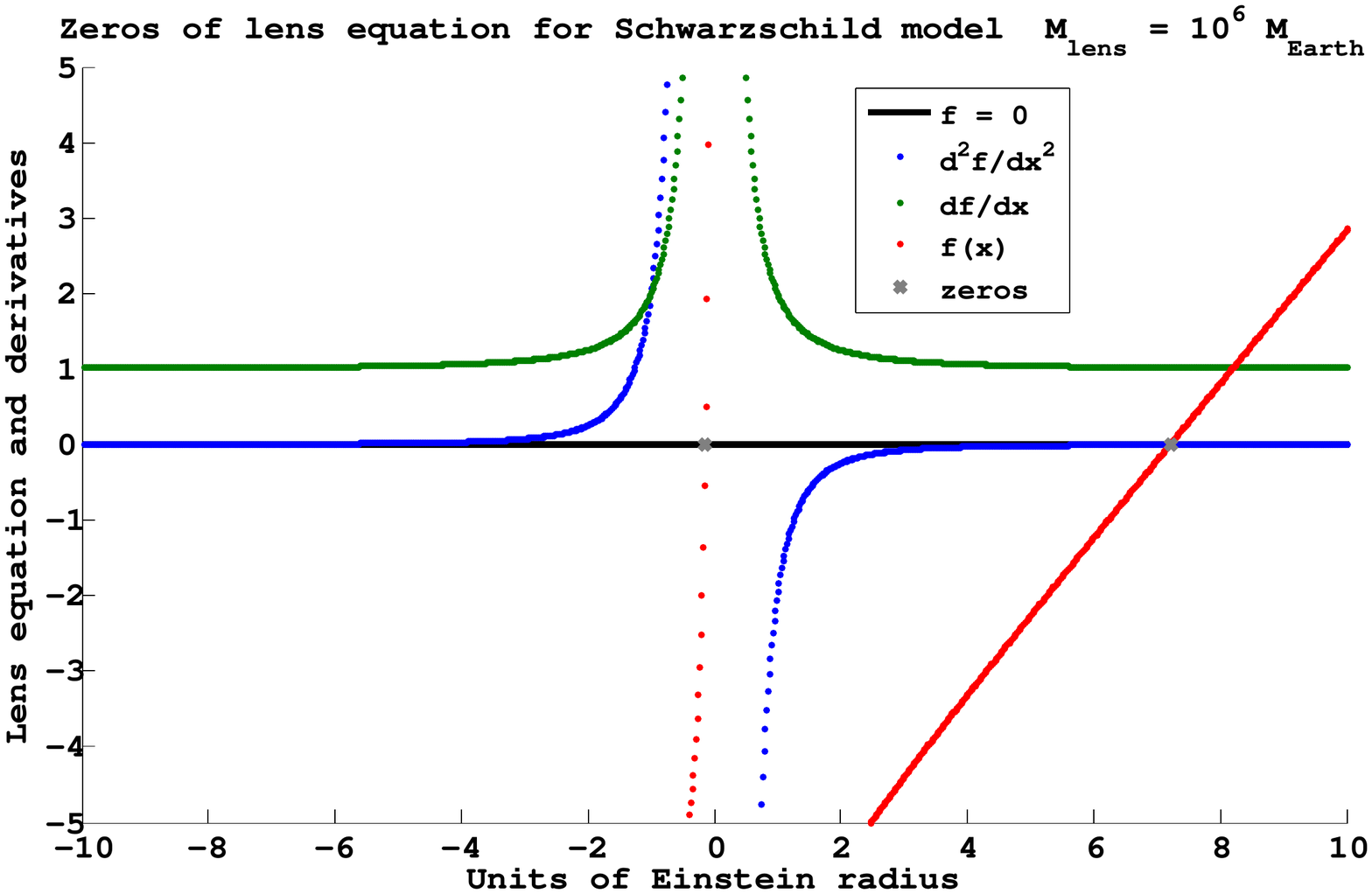}
\caption{Root-finding plot for the \swz{} lens with realistic parameters.  The two zeros are marked by a black cross.  The lens equation's zeroth, first and second derivatives are shown in red, green and blue respectively. \label{fig:nfw-zeros}}
\end{figure}

\subsection{Time delay}\label{sec:tdtot}
The final step is the proof that the Fermat potential $\phi(\vect{x})$, whose gradient is the lens equation, is precisely the (scaled) time delay caused by the presence of the lens.   First we concern ourselves only with the distance added by the new path length.\footnote{Granted that the speed of light \textit{in vacuo} is constant, recall that we have set $c = 1$.}
The lensing diagram \cref{fig:lens-setup} shows that the geodesic of the unperturbed system is the straight line $\overrightarrow{SO}$.  The geometric term of the time delay is the path length difference $\overrightarrow{SIO} - \overrightarrow{SO}$ calculable from Pythagoras\rq{} Theorem.  
Thanks to the thin-lens approximation, the relativistic perturbation of the geodesic occurs only in the infinitesimally thin section of the geodesic which intersects the lens plane.  Thus, the relativistic time delay is the solution to Poisson\rq{}s equation in the lens plane (\qv \eqref{eq:poisson}), namely the lensing potential.  The time delay is formed from the geometric and potential terms by redshifting from the source plane.  The resulting time delay is:
\begin{equation}\label{eq:td-scaling-ch3}
\tau 
\equiv \tau_{\text{geom}} + \tau_{\text{pot}} 
= (1 + z_d) \left( 
\frac{D_{\textrm{d}}D_{\textrm{s}}}{2D_{\textrm{ds}}}\left( \vect{\theta} - \vect{\beta} \right)^2 - \Psi(\vect{\theta})
\right) 
= (1 + z_d)\frac{D_{\textrm{s}} \xi_0^2}{D_{\textrm{ds}}D_{\textrm{d}}} \left( 
\frac{1}{2} \left( \vect{x} - \vect{y} \right)^2 - \Psi( \vect{x} )
\right) 
\end{equation}
Thus, we have shown that the scaled time delay and the Fermat potential are identical.

\section{The lensing potential}\label{sec:tdpot}
We now require a relation between the lensing potential $\Psi(\vect{x})$ and the properties of the lens itself.  A general expression is given in \cref{sec:smd}, which can be simplified, as shown in \cref{sec:hankel}, if we assume that the lens is axisymmetric. 

\subsection{Surface mass density}\label{sec:smd}
This section introduces the convergence $\kappa$ of the lens.  Dividing the lens into infinitesimal volumes $\dx{V}$ with mass $\dx{m}$, let a light ray pass the element at $(\xi_1',\xi_2',r_3')$.  Recall that by definition of our co-ordinate system, the impact vector of the light ray at $(\xi_1,\xi_2,r_3)$ relative to the mass element is $\norm{\vect{\xi} - \vect{\xi}'}$, independent of $r_3$.  Now we can calculate the deflection caused by the mass element \cite{mpa-notes}:
\begin{subequations}
\begin{align}
\dx{\vect{\alpha}}(\xi_1,\xi_2,r_3) 
&= \frac{4G}{c^2} \, \dx{m}(\xi_1',\xi_2',r_3') \, \frac{\vect{\xi} - \vect{\xi}'} {\norm{\vect{\xi} - \vect{\xi}'}^2}
\intertext{Integrating, we find that the deflection due to the total mass is:}
\vect{\alpha}(\vect{\xi}) 
&= \frac{4G}{c^2} \int \dx{m}(\xi_1',\xi_2',r_3') \frac{\vect{\xi} - \vect{\xi}'} {\norm{\vect{\xi} - \vect{\xi}'}^2} \\
&= \frac{4G}{c^2} \integ{\vect{\xi'}}{}{}{ \integ{\vect{r_3'}}{}{} {\rho(\xi_1',\xi_2',r_3')}} \frac{\vect{\xi} - \vect{\xi}'} {\norm{\vect{\xi} - \vect{\xi}'}^2} \\
&= \frac{4G}{c^2} \integ{\vect{\xi'}}{}{}{ \frac{\vect{\xi} - \vect{\xi}'} {\norm{\vect{\xi} - \vect{\xi}'}^2} \integ{\vect{r_3'}}{}{} {\rho(\xi_1',\xi_2',r_3')}} \\
&= \frac{4G}{c^2} \integ{\vect{\xi'}}{}{}{ \frac{\vect{\xi} - \vect{\xi}'} {\norm{\vect{\xi} - \vect{\xi}'}^2} \Sigma(\vect{\xi}) }
\end{align} 
\end{subequations}
defining the surface mass density $\Sigma(\vect{\xi}) = \integ{\vect{r_3}}{}{} {\rho(\xi_1',\xi_2',r_3')}$.  Scaling this to the $\vect{x}$ co-ordinate system defines a corresponding factor, the convergence $\kappa \left( \vect{x} \right)$:
\begin{equation}
\vect{\alpha}(\vect{x}) = \frac{1}{\pi} \integ{\vect{x'}}{}{}{ \frac{\vect{x} - \vect{x}'} {\norm{\vect{x} - \vect{x}'}^2} \kappa(\vect{x}) } 
\quad \text{defining $\kappa(\vect{x}) \equiv \dfrac{\Sigma(\xi_0\vect{x})}{\Sigma_{cr}}$ and 
$ \Sigma_{cr} = \dfrac{c^2}{4\pi G}\dfrac{D_{\textrm{s}}}{D_{\textrm{d}}D_{\textrm{ds}}}$ } \label{eq:convergence} 
\end{equation}
The convergence $\kappa \left( \vect{x} \right)$ is related to the surface mass density $\Sigma({\vect{\xi}})$ by the critical surface mass density $\Sigma_{cr}$ which quantifies the strength of the lens: any lens which has $\Sigma({\vect{\xi}}) > \Sigma_{cr}$ (equivalently $\kappa \left( \vect{x} \right) > 1$) will generate multiple images \cite{mpa-notes}.  This completes the expression of the deflection angle in terms of the scaled surface mass density.

\subsection{Simplifications due to axisymmetry}\label{sec:hankel}
The integration performed in \eqref{eq:convergence} is not numerically simple.  The aims of this section are twofold: first, to show that an equivalent integral can be reduced to a Fourier convolution or, with additional symmetry constraints, a Hankel convolution; second, to simplify the lensing equation.

The equivalent problem to finding the deflection angle is to calculate the lensing potential.  The lensing potential satisfies Poisson\rq{}s equation with respect to the convergence.  
\begin{equation}\label{eq:poisson}
\nabla^2\Psi(\vect{x}) = 2\kappa(\vect{x})
\end{equation}
This implies, by comparison with \eqref{eq:convergence}, that the lensing potential takes the form:
\begin{equation}\label{eq:integral}
\Psi(\vect{x})
\equiv \frac{1}{\pi} \integ{\vect{x}}{\mathbb{R}^2}{}{ \kappa(\vect{x}) 
\ln|\vect{x} - \vect{x_0}|} 
\quad\text{using $\nabla \ln|\vect{x}| = \frac{\unit{x}}{\norm{\vect{x}}}$}
\end{equation}
Thus the relativistic portion of the time delay reduces to the convolution of the convergence with the logarithm of the radial distance \cite{mpa-notes}.  

The next paragraph is entirely routine and shows that the above result \eqref{eq:integral} is also expressible as a product of integral transforms.  
Consider two scalar-valued functions $f(\vect{x})$ and $g(\vect{z})$, where the vectors $\vect{x}$ and $\vect{z}$ refer to the same physical quantity (but are denoted differently because the variable appears in two different roles).  The convolution of these functions is defined to be:
\begin{equation}\label{eq:convolution}
f \ast g \equiv h\left(\vect{z}\right) = \integ{\vect{x}}{-\infty}{\infty} {f \left( \vect{x} \right) g \left( \vect{z} - \vect{x} \right) }
\end{equation}
Each point in the region of integration contributes \textit{twice} to the value of the integral: the value within $x$ and $x+\dx{x}$ is mapped to $f(\vect{x})$, translated by an amount $\vect{z} - \vect{x}$ into a region of width $\dx{z}$, then mapped to $g(\vect{z} - \vect{x})$.  Thus, calculation of a convolution is computationally expensive if we integrate using its definition.  Conversely, if an integral can be represented as a convolution, then its calculation can be simplified.  The simplification is due to the convolution theorem \eqref{eq:fourier}.  The (Fourier) convolution theorem states that:
\begin{subequations} \begin{align}\label{eq:fourier}
f \ast g 
&=\mathscr{F}^{-1}\bigl\{ 2\pi \mathscr{F}\bigl\{ f \bigr\} 
\mathscr{F}\bigl\{ g \bigr\}\bigr\} 
\quad \begin{cases}
\mathscr{F}\bigl\{ f(\vect{x}) \bigr\} &=
\dfrac{1}{(\sqrt{2\pi})^{{2}}} \mathlarger{\int} \mathrm{d}^{2}{x} \; f(\vect{x}) 
\exp{-i\vect{k} \cdot \vect{x}} 
\\[1ex]
\mathscr{F}^{-1}\bigl\{ F(\vect{k}) \bigr\} &= 
\dfrac{1}{(\sqrt{2\pi})^{2}} \mathlarger{\int} \mathrm{d}^{2}k \; F(\vect{k}) 
\exp{i\vect{k} \cdot \vect{x}}
\end{cases}
\intertext{
where we have explicitly defined the two-dimensional Fourier transform $\mathscr{F}$ and its inverse.  Under the assumption of axial symmetry, $f(\vect{x}) = f(\rho,\phi) = 
f(\rho)$ in (plane) polar co-ordinates.  Then the 2-d Fourier transform  reduces to a 1d Hankel transform:
}\label{eq:hankel}
f \ast g 
&=\mathscr{H}^{-1}\bigl\{ 2\pi \mathscr{H}\bigl\{ f \bigr\} 
\mathscr{H}\bigl\{ g \bigr\}\bigr\} 
\quad \begin{cases}
\mathscr{H}\bigl\{ f(\rho) \bigr\} &=
\int \mathrm{d}^{}{\rho} \; \rho f(\rho) J_0(k\rho)
\\[1ex]
\mathscr{H}^{-1}\bigl\{ F(k) \bigr\} &= 
\int \mathrm{d}^{}k \; k F(k) J_0(k\rho)
\end{cases}
\end{align} \end{subequations}
where $J_0$ denotes the zeroth-order Bessel function of the first kind.  By comparison of \eqref{eq:convolution} with \eqref{eq:integral}, we identify $f(\vect{x})$ with $\kappa(\vect{x}) \equiv \kappa(\rho)$ and $g(\vect{z} - \vect{x})$ with $\ln\norm{\vect{z} - \vect{x}} \equiv \ln(\rho - \rho_0)$.  This completes the proof that the potential term of the time delay is a Hankel convolution.

The assumption of radial symmetry also facilitates calculation of the lens equation.  For the remainder of this section we use plane polar co-ordinates $(\rho,\phi)$ such that $x = \rho\cos\phi$ and $y = \rho\sin\phi$.  For clarity, we express the vectors explicitly in terms of the (orthonormal) basis functions:
\begin{align}
&\basis{\rho} = \cos\phi\basis{x} - \sin\phi\basis{y} 
&\basis{\phi} &= \sin\phi\basis{x} + \cos\phi\basis{y}
\end{align} 
\Wlog we can orient the co-ordinate system such that $\phi = 0$ \ie $\rho$ is aligned with the x-axis.%
\footnote{In curvilinear co-ordinates the direction of the basis vectors is a function of position.  Thus $\basis{x} = \basis{\rho} \neq \basis{\rho'}$ and similarly $\basis{\phi} \neq \basis{\phi'}$ for the vectors $(\rho,\phi)$ and $(\rho',\phi')$.}
Then we have the lemmata: 
\begin{subequations} \begin{align}
\vect{\rho} - \vect{\rho}^{\;\prime} 
&= \left( \rho - \rho'\cos\phi' \right)\basis{\rho}
+ \left( -\rho'\sin\phi' \right)\basis{\phi} \\
\norm{\vect{\rho} - \vect{\rho}^{\;\prime}} &= \sqrt{\rho^2 + \rho'^2 - 2\rho\rho'\cos\phi'}
\end{align}   
\end{subequations}
Substitution of an axisymmetric convergence $\kappa(\norm{\vect{x}}) = \kappa(\rho)$ into the plane polar form of the deflection angle \eqref{eq:convergence} gives:
\begin{subequations} \begin{align}
\vect{\alpha} \cdot \basis{\rho}
&= \frac{1}{\pi} \integ{\rho'}{0}{\infty}{ \rho \kappa(\rho) 
\integ{\phi'}{0}{2\pi}{\frac{(\vect{\rho} - \vect{\rho}^{\;\prime}) \cdot \basis{\rho} } 
{\norm{\vect{\rho} - \vect{\rho}^{\,\prime}}^2}   
}}
&= \frac{1}{\pi} \integ{\rho'}{0}{\infty}{ \rho \kappa(\rho) 
\integ{\phi'}{0}{2\pi}{\frac{ \rho - \rho'\cos\phi' } 
{\rho^2 + \rho'^2 - 2\rho\rho'\cos\phi'} 
}} \label{eq:a-rho}\\
\vect{\alpha} \cdot \basis{\phi}
&= \frac{1}{\pi} \integ{\rho'}{0}{\infty}{ \rho \kappa(\rho) 
\integ{\phi'}{0}{2\pi}{\frac{(\vect{\rho} - \vect{\rho}^{\;\prime}) \cdot \basis{\phi} } 
{\norm{\vect{\rho} - \vect{\rho}^{\,\prime}}^2}
}}
&= \frac{1}{\pi} \integ{\rho'}{0}{\infty}{ \rho \kappa(\rho) 
\integ{\phi}{0}{2\pi}{\frac{-\rho'\sin\phi'} 
{\rho^2 + \rho'^2 - 2\rho\rho'\cos\phi'}
}} \label{eq:a-phi}
\end{align}  \end{subequations}
The inner integral in \eqref{eq:a-phi} vanishes: $\vect{\alpha} \cdot \basis{\phi} = 0$.  Thus we see that the deflection angle is parallel to the radial basis vector.  The inner integral in \eqref{eq:a-rho} vanishes for $\rho' > \rho$ whereas for  $\rho' \leq \rho$ it evaluates to $\sfrac{2\pi}{x}$.  Hence the only contribution to the deflection angle is:
\begin{equation}\label{eq:mass}
\vect{\alpha}(\vect{\rho}) 
= \left( \vect{\alpha} \cdot \basis{\rho} \right) \basis{\rho}
  + \left( \vect{\alpha} \cdot \basis{\phi} \right) \basis{\phi}
= \basis{\rho} \left( \frac{1}{\pi}\integ{\rho'}{0}{\infty}{ 
\frac{2\pi}{\rho}\rho' \kappa(\rho') } \right)
= \basis{\rho} \frac{m(\rho)}{\rho}
\end{equation}
where the last line defines the mass within a circle of radius $\rho$.  The interpretation of the result is as follows: at a radius $\rho$ from the centre of the lens, the matter within that radius contributes as if it were a point mass at the origin and the matter without does not contribute.  Using the scaling introduced in \eqref{eq:scaling}, the corresponding lens equation is:
\begin{equation}\label{eq:radial-lens-eq}
\vect{y} = \vect{x} - \vect{\alpha}(\vect{x}) \implies y = x - \frac{m(x)}{x}
\end{equation}
The imposition of axisymmetry on the convergence of the lens has simplified the solution of the lens equation to a one-dimensional problem.

\subsection{Calculating the potential}\label{sec:potential}
Recall from \cref{sec:hankel} that the relativistic part of the time delay can be expressed in multiple ways.  Accordingly, we can evaluate any of these equivalent expressions for the potential term:
\begin{itemize}
\item Gaussian quadrature methods to calculate the axisymmetric integral 
\eqref{eq:integral}
\item Convolution of 2d Fourier transforms \eqref{eq:fourier}
\item Convolution of 1d Hankel transforms \eqref{eq:hankel}
\end{itemize}
It is necessary to select the method which balances computational efficiency with accuracy, bearing in mind that extra time would be required if the routine were not pre-existing.  

The first attempt used Fourier convolution.  This proved unsatisfactory due to limitations on the grid fineness.  The grid size is governed by two opposing factors: fitting in the source and image locations and accurately modelling the radial density profile of the lens.  
On the one hand, the grid must be sufficiently fine to represent the density profile smoothly.  In the case of the \nfw{} lens, sensible values for the turnover radius and the physical extent of the lens are $10^{-4.5}\,\text{pc}$ and $10^{-3}\,\text{pc}$ respectively: since we require at least one point inside the turnover radius to approximate the piecewise-smooth density profile, modelling the lens radius alone uses at least $(2 \times 31.6)^2 \sim 2^{6} \times 2^{6}$ grid points.  
On the other hand, the grid must cover a sufficiently large area to contain both the source and the resulting images.
  Due to the relative motion between the halo and the pulsar transiting behind that \dmh{}, the furthest extent of the source-lens distance depends upon the time taken for observations.  Using typical values (\cref{sec:parameters}), as well as taking into account the impact factor of the lens, \ie{} that at closest approach the lens and source may be offset, an order-of-magnitude value for the transit radius is $10^{-2}\,\text{pc}$.  Scaling to \lq\lq{}lens plane units\rq\rq{} of the turnover radius, the length of the grid edge is $\sfrac{2 \times 10^{-2}} {10^{-4.5}}\,\text{pc}$ or 632 units.  Furthermore, \matlab{} requires the Fast Fourier Transform to be performed on a square matrix of $2^{2n}$ elements for efficiency reasons.  Thus we see that accommodating both requirements necessitates at least $2^{10}$ points along the grid edge.  (Recall that this includes only a single pixel within the turnover radius of the \nfw{} lens --- hardly a smooth approximation to the density profile!)  
However, for more than $2^{11}$ points on a square edge, \matlab{} encountered \lq\lq{}out-of-memory\rq\rq{} errors.  Consequently this method had to be abandoned.  It became clear that the 2d integration was untenable, so a 1d method had to be used.

The second attempt was a custom routine based upon Gaussian quadrature of Bessel functions.  Gaussian quadrature is well-established as an efficient method for 1d numerical integration \cite{RHB}.  The choice of polynomial for the approximation gives accurate, fast-converging results for integrands which are close in form; correspondingly it is inaccurate for integrands which are not.  The difficulty in this method is that Bessel functions are not conventional polynomials.  The oscillatory nature of Bessel functions \cite{oscillatory} and their infinite roots \cite{RHB} are highly non-trivial problems \cite{oscillatory}.  While Gaussian schemes (e.g. \cite{bessel-quad}) for integrands of the form $f(x)J_{\nu}(kx)$ do exist, they are difficult to implement due to the restrictions on $f(x)$ if the integral is to converge.  My supervisor convinced me that there were more efficient and less complicated ways of solving my dilemma.  Thus, the quadrature scheme was abandoned.

The sole remaining approach was to solve the Hankel transforms \eqref{eq:hankel}.  Following the symmetric matrix algorithm of \cite{0qdht,pqdht}, writing the \matlab{} routine required an adjustment to the definition of the Hankel transform.  I derived the appropriate equations and implemented them numerically: the algorithm is shown in \cref{code:hankel}.  The success of this method was proven using the \hdisc{} lens profile, since it is an extended lens (\cf the delta function of the \swz{} lens) which is sufficiently simple to yield analytical forms for the image locations and magnification factors.

\section{The flux theorem}\label{sec:flux}
A corollary of the change in the geodesic is the effect of the lensing potential on bundles of light rays.  The cross-section of a given bundle will change as the direction of each individual ray is deflected slightly by the lens.  For a sufficiently small area, this change can be related to the Jacobean determinant of the lens equation.

Surface brightness is conserved by \gl{ing}.  The gravitational effects do not trigger the emission of absorption of photons, thus their total number is conserved \cite{gl-bartelmann}.  It follows that specific intensity $I$ is constant as the light propagates.  Conversely, the flux from the source is not conserved.  The flux is defined as the product of surface brightness with the area of emission: accordingly it changes with the area distortion caused by the lens.  Let $(\Delta\omega)_0$ be an infinitesimal, undeflected area and $(\Delta\omega)$ its corresponding lensed area.  Then the ratio between the corresponding fluxes is as follows \cite{sef}:
\begin{equation}\label{eq:mu-defn}
\mu = \frac{\int_{(\Delta\omega)} \dx{\vect{\omega}} \cdot I} {\int_{(\Delta\omega)_0} \dx{\vect{\omega}} \cdot I} \\
= \frac{I \int_{(\Delta\omega)} \dx{\vect{\omega}}} {I \int_{(\Delta\omega)_0} \dx{\vect{\omega}} } \\
= \Biggl(\frac{ (\Delta\omega)_0 } { (\Delta\omega) } \Biggr)^{-1} \\
= \Biggl( \mathrm{det} \Biggl( \frac{ \partial{\vect{\beta}} } { \partial{\vect{\theta}}  }\Biggr) \Biggr)^{-1}
\end{equation}
The area is much smaller than the scales upon which the source properties change, in which case, the lens equation is (locally) linearisable \cite{mpa-notes}.  Hence the magnification factor $\mu$ is the inverse of the Jacobean of the lens mapping.  

Accordingly, it is possible to relate the magnification factor to the lensing potential.  Since the lens mapping is the gradient of the Fermat potential
\begin{equation}\label{eqn:jacobean}
\mathscr{J}_{ij} 
= \frac{\partial \vect{y}_i}{\partial \vect{x}_j}
= \frac{\partial}{\partial x_j} \left( \vect{x} - \nabla \Psi(\vect{x}) \right)_i
= \delta_{ij} - \frac{\partial^2 \Psi}{\partial x_i \partial x_j}
\end{equation}
As in the previous section, denote the convergence $\kappa$ and introduce the complex shear $\gamma$, which is related to the lens potential $\Psi$:
\begin{equation}\label{eq:shear}
\Re{\gamma}
= \frac{1}{2} \left(\frac{\partial^2 \Psi_1}{\partial \vect{x}_1} - 
\frac{\partial^2 \Psi_2}{\partial \vect{x}_2} \right) \quad
\Im{\gamma}
= \frac{\partial^2 \Psi_1}{\partial \vect{x}_2} 
= \frac{\partial^2 \Psi_2}{\partial \vect{x}_1} 
\end{equation}
The convergence satisfies Poisson's equation \eqref{eq:poisson}.  Rewriting the Jacobean using \eqref{eq:poisson} and \eqref{eq:shear}:
\begin{subequations} \begin{align}
\mathscr{J} &= 
\begin{pmatrix}
1 - \kappa - \Re{\gamma} & - \Im{\gamma} \\ 
- \Im{\gamma} & 1 - \kappa + \Im{\gamma}
\end{pmatrix}
\intertext{The determinant follows:}
\mathrm{det} \mathscr{J}  
&= \left(1 - \kappa - \Re{\gamma} \right) \left( 1 - \kappa + \Re{\gamma} \right)  - \left( - \Im{\gamma} \right)^2 \\
&= (1 - \kappa)^2 - \abs{\gamma}^2
\end{align} \end{subequations}
The symmetry argument presented in \cref{sec:hankel} allows further simplification.  Recalling that $m(x)$ is the lens mass enclosed by a radius $x$, we can apply \eqref{eq:poisson} once more to relate $m(x)$ to $\gamma$:
\begin{equation}
\ddx{x}{m}{} = 2x\kappa(x) 
\implies \gamma^2 = \left( \frac{m(x)}{x^2} - \kappa^2 \right)^2
\end{equation}
The final form of the magnification factor is thus:
\begin{equation}\label{eq:magnification}
\mu = \frac{1}{\mathrm{det} \mathscr{J} } 
= \left( \left( 1 - \frac{m(x)}{x^2} \right) \left( 1 + \frac{m(x)}{x^2} - 2\kappa(x) \right)  \right)^{-1}
\end{equation}
Note that this value is (always) larger than one.  The fact that this does not violate energy conservation reveals a subtlety in the definition.  The magnification factor defined by \eqref{eq:mu-defn} is the magnification relative to an empty universe rather than relative to a \lq\lq{}smoothed out\rq\rq{} universe (with matter) \cite{mpa-notes}.  The magnification factor is the last phenomenon that we discuss in connection with the single-lens case.  In the next section \cref{sec:multiple} we will show how to apply the formulae of the single-lens geometry to multiple lenses.

\addtocontents{toc}{\newpage} 
\section{Multiple-lens algorithm}\label{sec:multiple}
\subsection{Testing the lens scale}\label{sec:scaling}
The first test of the numerical model was the reproduction of time delays observed in lensed QSOs.  The simple, elliptical model \cref{sec:elliptical} was unlikely to model the actual system accurately beyond first order.  Consequently, judging whether the time delay calculations were functioning correctly became rather subjective.  Furthermore, the number of systems available for testing was small: for those systems with a known relative time delay, the angular separation between the images with known delay as well as both source and lens redshift had to be known to obtain the angular diameter distance.  Fortunately \cite{sef} lists a set of lensing candidates which served as a master list from which test candidates were chosen.  The results \cref{tab:lens-table} show sufficient agreement with those described in \cite{sef} to reach two conclusions: first, the angular diameter distances to source and lens planes and between them were correctly calculated; second, the conversion from the natural length units of the Einstein radius to the physical units required to calculate the observed time delay was correct.  Thus results from single lenses at different distances can be combined in physical units, a prerequisite for the multiple lens system.
\begin{table}
\begin{tabular}{llllll}
\toprule
System & $z_d$ & $z_s$ & max. sep. $('')$ & Time delay \cite{sef} & Time delay (simulation) \\
\midrule
0957+561 & 0.36 & 1.41 & 6.1 & 415 days & 323 days\\
0142-100 & 0.49 & 2.72 & 2.2 & \lq\lq{}a few weeks\rq\rq{} & 53 days\\
2016+112 & 1.01 & 3.27 & 3.8 & $\sim$ 1 yr & 398 days\\
2237+0305 & 0.039 & 1.69 & 1.8 & $\sim$ 1 day & 1.7 days\\
\bottomrule
\end{tabular}
\caption{The gravitational lens systems listed in \cite{sef} for which the source and lens redshifts $z_s$, $z_d$, the maximum separation between images and the relative time delay between those images are known.  The rightmost column shows the numerically-calculated time delay found by modelling the system as Einstein rings.
\label{tab:lens-table}} 
\end{table}

\subsection{Analytical comparison}
\begin{figure}
\includegraphics[width=\textwidth]{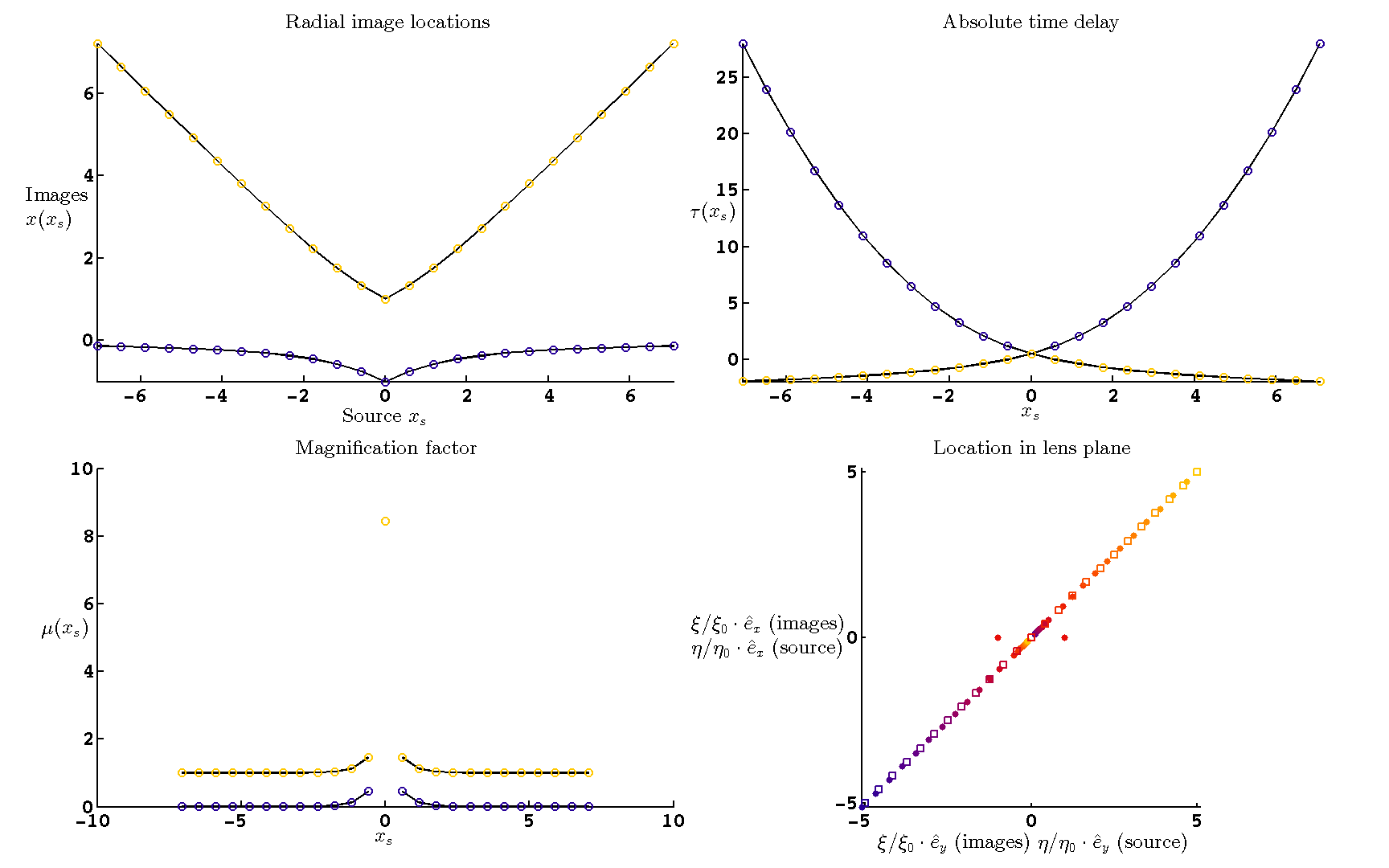}
\caption{Comparison of the theoretical (lines) and numerical (points) results for a $10^6 M_{\oplus}$ \swz{} lens.  There are two images (\textsc{top left}) for each source location, with a corresponding time delay \textsc{(top right)} and magnification factor \textsc{(bottom left)}.  The lens plane ($D_{\text{d}} = 5\,\textrm{kpc}$) and source plane ($D_{\text{s}} = 10\,\textrm{kpc}$) are shown in lens-centric co-ordinates \textsc{(bottom right)} \ie{} the lens appears fixed to the optical axis while the source moves.  At conjunction, the result is an Einstein ring -- a circle rather than two distinct images -- shown by the two markers not aligned with the source angle.
\label{fig:plot-summary}}
\end{figure}

The accuracy of the numerical routines was confirmed by comparison with analytical solutions \cref{fig:plot-summary}.  The \swz{} lens \cref{sec:swz} was used, as it is the only axially symmetric model for which there is an analytical form for both the time delay and the image locations \cite{sef}.  The magnification factor calculations and the root-finding algorithm \cref{code:rootsearch} used to find the lens locations  worked satisfactorily even for the Einstein ring case (when the source and deflector are aligned, a singular point in the lens mapping).  The original Fourier-transform-based code for the potential proved unsatisfactory even with zero-padding and was replaced by the faster and more accurate Hankel-transform-based code \cref{code:hankel} (\qv \cref{sec:potential}).  The \hdisc{} lens \cref{sec:hdisc} was then used to compare the analytical and numerical calculations for the magnification factor and the lens locations.  This served as a further check with a radially extended lens potential (rather than a point mass, which is analytically transformable) and a piecewise-invertible lens mapping.  The accuracy and performance of the simulation was sufficient to utilise the more complex \nfw{} model \cref{sec:nfw} for the full simulations.

\subsection{Moving lenses}
\begin{figure} 
\centering
\begin{minipage}[b]{.8\textwidth}
\centering
\includegraphics[width=.6\textwidth]{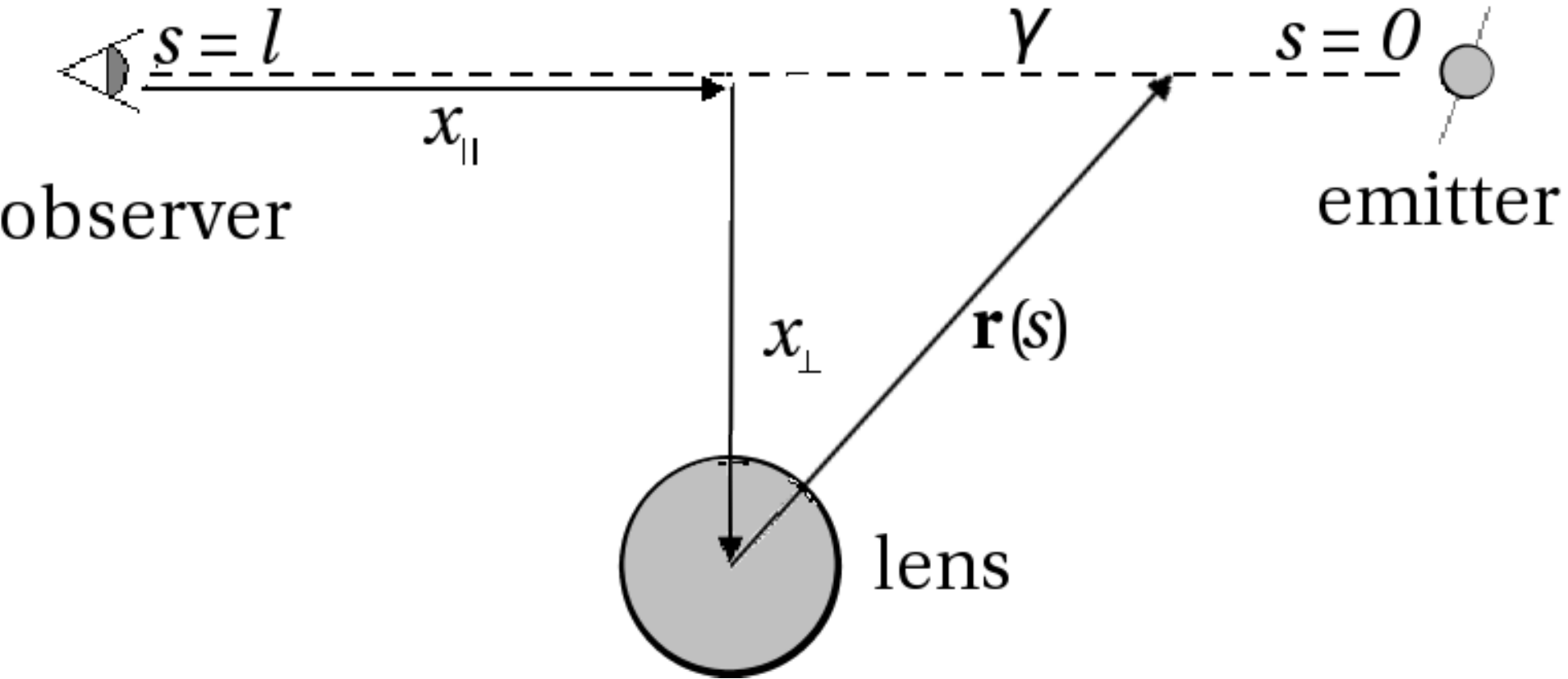}
\end{minipage}\\[-10pt]
\begin{minipage}[t]{\linewidth}
\caption{The geometry of a gravitational lens system with a transiting lens and a source fixed to the optical axis.  The geodesic $\gamma$ is parameterised by arc-length $s$; the vector from the lens to the geodesic is marked by $\textbf{r}(s) \in \mathbb{R}^3$; via the thin-lens approximation it can be decomposed into an impact factor $x_{\perp}$ and a lens-observer distance $x_{\parallel}$.
\label{fig:source-lens}}
\end{minipage}
\end{figure}

The last single-lens step is to simulate motion of the lens.  Until this stage, it is the lens which is fixed and the source which moves: this is necessary because the lens must be at the origin of the co-ordinate system to take advantage of axial symmetry.  Realistically \cref{fig:source-lens}, it is the lens which transits between the pulsar and the Earth: the source is fixed at the axis of this new co-ordinate system.  Hence, the calculations for each lens are done in lens-centric co-ordinates, then translated after calculating to source-centric co-ordinates (since the source lies along the optical axis, this is equivalent to centring the co-ordinate system on the observer).  

The key consequence of the translation is the addition of another term in the time delay.  Given two lenses at different (radial) locations $y$ (relative to the optical axis), the quantity of interest is the relative time delay between their respective images $x$.  Thus, as long as we measure all the time delays relative to the same geodesic (\ie the unlensed ray corresponding to the optical axis), the time delays from different lenses can be compared.  The geometric time delay component 
becomes $\tfrac{1}{2}(x-y)^2 + \tfrac{1}{2}y^2$ for each image, rather than $\tfrac{1}{2}(x-y)^2$ as it was previously.

The motion of the lens is approximated by assigning an array of $N \gg 1$ points (in this case 100) equally-spaced along the transit: given the velocity $v$ of the lens and the radius of the simulation $a$, the points represent a sample of the pulsar emission at times $n (\sfrac{a}{vN})$ with $n \in \{1, \ldots , N\}$.  The reason for this sampling is that the transit time is of a different order of magnitude to the pulsar period ($\sfrac{a}{v} \sim 3 \times 10^7\,\textrm{s}$ compared to $\Tres \sim 10^{-3}\,\textrm{s}$): it follows that it is impractical to build an array storing the effect of the \gl{} system on every signal emitted by the pulsar.  Instead, the aim is to take sufficient samples such that the data accurately represent the evolution of the pulsar signal(s).

\subsection{Multiple lenses}
In lieu of the complex \mpl{} algorithm 
 (\cref{ch:mpl}), a simple method of combining the effects of each lens is necessary.  The problems are twofold: creating a realistic distribution of lenses from which to draw lenses with appropriate free parameters; then combining the data from each lens into a compound signal from the pulsar.  These problems are discussed in the next two paragraphs. The parameter selection is left to \cref{sec:parameters}.

The lens distribution is inferred from the density of the Galaxy.  The radial density profile of the Milky Way is well-approximated by the \nfw{} profile: given the scale radius $r_s$ and the total galaxy mass $M$, the density profile as a function of radius from the galactic centre is completely determined.  The difficulty lies in ensuring that the distribution of lenses with radius has the correct number density to match the density profile.  Consider the number distribution of halos with volume.  There is no reason to assign any subspace of the cone with a higher density of halos than any other.  For the prior distribution of halos to obey maximum entropy, it follows that the number of halos scales according to the volume \ie $\dx{N} = \dx{V} = \pi\dx{r^2}$.  Accordingly, the array of halo distances could not be generated using a pre-existing routine: it was necessary to write and test a subfunction which created the correct probability density function.  

The signal which reaches the Earth is the superposition of each signal from all of the images produced by each lens.  Na{\"i}vely, this is represented by the array of magnification factors and arrival times ${\mu(t)}$; the situation is complicated by the fact that the pulsar emitting the signals has a non-zero timing residual $\Tres$ even in its un-lensed state.  This residual represents an uncertainty in the arrival time of the pulsar signal: any two signals arriving within $\Tres$ of each other appear to be a single signal, with an amplitude generated by the superposition of the individual pulses \cite{siegel2}.  It follows that the signals detected by the observer are not precisely the same as the signals generated by the simulations from the time delay equation.  

It is necessary to artificially combine the signals.  The na{\"i}ve method to do this is to bin the signals by their arrival time, combining any which arrive within $\Tres$ of each other.  However, there are two time-scales to the problem, which makes this method unsuitable: the observation time $T_{\text{obs}}$ and transit time of the halos are on the order of years, whereas the time delays are measured in microseconds.  A more sophisticated method, shown in \cref{code:tdelay}, is to create a \lq\lq{}comb\rq\rq{} of histograms, ignoring the times during which no signal was emitted, but binning those signals which arrive around the pulse emissions at $T_n = T_{\text{obs}} (\sfrac{n}{N}) \; n \in \left\{1, ... N \right\}$.  The usefulness of this method depends on the fine-tuning required for the definition of \lq\lq{}close\rq\rq{} (which I took as $10\Tres$).  Since \textit{a priori} neither the number of images per lens nor the delay induced in each image is known, the search interval must be sufficiently wide to trap all of the signals, yet sufficiently narrow that it is covered in a manageable number of bin widths (\ie a small number of $\Tres$).  Once this binning is complete, the amplitude of the composite signal must be calculated: this is the product of the magnification factors from each image (proof requires \mpl{$\!\!$}, \qv \cref{ch:mpl}).  This completes the transformation of the lens images from the output of the lensing program to an observable representation of the pulsar signal.

In this chapter, we have identified the three key consequences of \gl{ing}, namely creation of multiple images, addition of time delays to the arrival time of each image, and the (de)magnification of the images.  We have derived quantitative formulae for each property in the case of a single lens.  Finally we have seen how this can be applied to multiple lenses in a relatively simple fashion and formulated the signals observed in such a situation. In the next chapter, we will examine some examples of this method.
\chapter{Results}\label{ch:results}

\section{Simulation parameters}\label{sec:parameters}
The practical application of this lensing model determines the parameter choice for the simulations.  The parameters in the simulations ought to reflect the \dmh{} population in the Galaxy.  This requires a choice of reasonable predictions for the following (open) questions:
\begin{samepage}
\begin{center}
\begin{minipage}[][][c]{.9\textwidth}
\begin{enumerate}
\item \label{list:a} What is the area of influence between the pulsar and the Earth?
\item \label{list:n} How many halos are likely to intersect this region?
\item \label{list:t} For how long would they remain within it?
\item \label{list:m} What are suitable masses for these halos?
\item \label{list:d} What is their distribution along the \los?
\item \label{list:T} What is a sensible time period for the observations?
\end{enumerate}
\end{minipage}
\end{center}
\end{samepage}
The choices for \cref{list:a} and \cref{list:t} are linked, as are \cref{list:n} and \cref{list:m}, whereas selection of \cref{list:T} is largely arbitrary.  Although initially fixed, \cref{list:d} later became a free parameter.  The trajectory of the halo between the pulsar and the Earth introduced two more (free) parameters: the linear and angular displacement of (the centre of mass of) the lens.    The simulations also require sensible values for the source: the pulsar-Earth distance $D_{\textrm{s}}$ and the pulsar timing residual $\Tres$.

\begin{figure}
\includegraphics[width=0.8\textwidth]{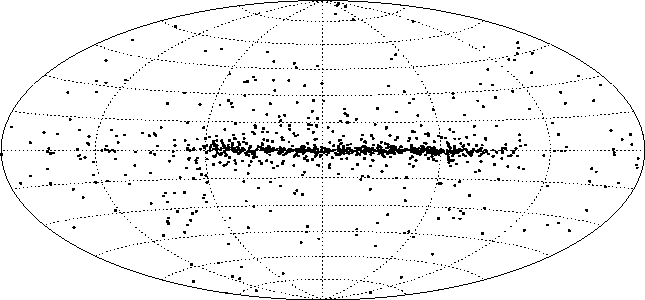}
\caption{The sky distribution of pulsars in the Milky Way.  1026 pulsars are shown projected onto galactic coordinates.  [Fig.~6 in \cite{psr-summary}]
\label{fig:pulsar-distribution-galaxy}}
\end{figure}

\begin{figure}
\includegraphics[width=\textwidth]{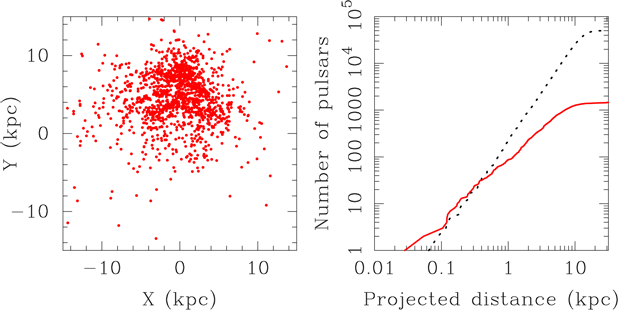}
\caption{\textsc{Left:} The Galactic pulsar sample (as of 2008) projected onto the Galactic plane.  The Galactic centre is at $(0,0)$ and the Sun at $(0,8.5)$. \textsc{Right:} The cumulative number count of Galactic pulsars as a function of distance from the Sun, showing the observed sample (solid line) and a model population after selection effects are accounted for (dashed line). [Fig.~11 in \cite{psr-summary}]
\label{fig:pulsar-distribution-plane}}
\end{figure}

First we examine \cref{list:a} and \cref{list:t}.  The distribution of pulsars within the Milky Way is shown in \cref{fig:pulsar-distribution-plane,fig:pulsar-distribution-galaxy}.  The vast majority of pulsars lies in the Galactic plane (\cref{fig:pulsar-distribution-galaxy}), whence we may assume that the \msp{} population also lies largely at small Galactic latitudes.  However, the numerical simulations in this project are only dependent upon the radial distance $D_\text{s}$ to the pulsar, rather than its angular position on the sky.  Accordingly, for simplicity we may project all pulsar locations to the plane of the Milky Way (\cref{fig:pulsar-distribution-plane}).
A sensible value for the pulsar distance was taken to be $D_{\textrm{s}} = 10\,\textrm{kpc}$.  This was selected by examining the pulsar catalogue of the Parkes Pulsar Timing Array \cite{ppta} for \msp{s} and choosing an order-of-magnitude estimate for their distance.  Examination of \cref{fig:pulsar-distribution-plane} confirms that there is a high density of pulsars at this distance, which --- assuming a correlation between the distribution of all pulsars and the millisecond subfamily --- affirms the sensibility of our fiducial Earth-pulsar distance.  The corresponding \lq\lq{}region of influence\rq\rq{} was a cone with its vertex at the Earth and base of radius $a \approx 10^{-2}\,\textrm{pc}$ centred on the pulsar.  This is a sufficiently small value for the cone radius that the Hankel transform could be used to calculate the time delay for a halo at any radial distance within it.  Conversely, it is sufficiently large to enclose a useful number of halos while maintaining a realistic density.  The transit time of each halo also depends upon its velocity $v$ and the lens-Earth distance $D_{\textrm{d}}$.  The halo velocity was set to be a constant $220 \,\textrm{km}\,\textrm{s}^{-1}$ for all halos.  The properties of the pulsars relevant to their role as lensing sources are now completely specified.

\begin{figure}
\includegraphics[width=\textwidth]{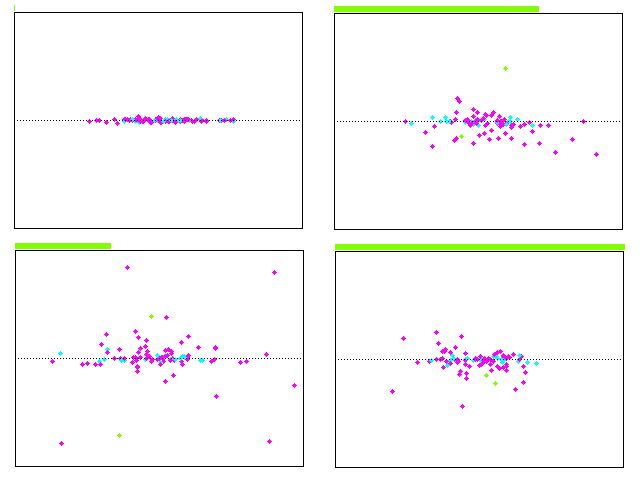}
\caption{Frames from an animation illustrating the effect of the Galactic potential on pulsars over $200 \, Myr$.  The dotted line indicates the Galactic plane, $30\,\text{kpc}$ across, while the height of the box is $\pm 10\,\text{kpc}$. The bar illustrates the length of time elapsed before each snapshot.  [The complete animation is Fig.~10 in \cite{psr-summary}.]
\label{fig:psr-potential}}
\end{figure}

Next we consider \cref{list:m}.  Suitable values for the lens masses are $M \mathrel{\substack{\textstyle\in\\[-0.1ex]\sim}} \{10^4,\,10^5,\,10^6\}\, M_{\oplus}$ (based upon \cite{siegel1,siegel2}).  Converting to solar masses ($1 M_{\oplus} \approx 3 \times 10^{-6} M_{\odot}$), we find that the $10^{6}\,M_{\oplus}$ mass falls within (and the $10^{5}\,M_{\oplus}$ mass slighly below) the range $0.05 - 1 M_{\odot}$ proposed by \cite{halo-cluster} to be the most likely bounds from survey data (\cf{} \cref{tab:macho}).  The $10^{5}\,M_{\oplus}$ is precisely within the mass range $0.02 - 0.08 M_{\odot}$ for which the \textsc{eros} and \textsc{macho} observng programs were designed \cite{eros}.  This leaves the $10^{4}\,M_{\oplus}$ mass as an order-of-magnitude lower bound on sensible \dmh{} masses; but it is also useful to test whether the pulsar-based method of lensing experiment outlined in this thesis is sensitive to such small halos ($ \sim 3 \times 10^{-3} M_{\odot}$), in contrast to the stellar microlensing surveys outlined in \cref{sec:machos}.  The small fiducial mass of the \dm{} lenses renders them sensitive to the potential of the Galaxy.  Specifically, if the lensing halos do not lie at the same Galactic latitude as the majority of source pulsars (as the Earth does), lensing images are unlikely to be visible from Earth.  Given the discussion of \dm{} candidates in \cref{sec:baryonic,sec:nbaryonic}, it is reasonable to assume that the halos are compact rather than diffuse objects.  By examining the effects of the Galactic potential on other compact objects whose dynamics have been examined in more detail, we may infer the effect on the \dmh{s}.  One such example is the trajectory of pulsars.  Due to their broad range of velocities ($1-10^3 \, \text{km}\,\text{s}^{-1}$), they form an especially useful example of the ability of the potential well of the Milky Way to restrain the motion of compact objects.  \cref{fig:psr-potential} demonstrates that the fate of the pulsars is highly sensitive to their initial velocity.  While those with the highest velocities do escape, a significant fraction have orbits whose amplitude decays over the lifetime of the simulation and most barely leave the Galactic plane at all.  Given that dark matter halos are expected to have much smaller velocities (\cf{} stellar proper velocities of $10-50\,\text{km}\,\text{s}^{-1}$), we may assume that \dmh{s} created within the plane of the Milky Way remain at low Galactic latitudes over Myr.  This reinforces the lack of concern on this topic in the initial papers on pulsar lensing by \dm{} \cite{siegel1,siegel2}.  This concludes the examination of suitable estimates for the \dmh{} masses.

Now \cref{list:n} can be derived from the assumptions made for \cref{list:m}.  The mass and number of halos is linked by the density of the galaxy.  The masses of all lenses were equal in each simulation.  Consequently, the number of halos per simulation is simply the mass within the cone divided by the mass of a single halo.  The mass within the cone was calculated by approximating the density profile of the galaxy as an \nfw{} profile ($r_s = 25\,\textrm{kpc}$, $M_{\textrm{gal}} = 1.2 \times 10^{12}M_{\odot}$), then performing a volume integral over the cone.  Since the number of halos must be a natural number, the ratio of enclosed mass to the individual lens mass was rounded to the nearest integer.  The cone radius $a$ and halo mass $M$ were adjusted so that the rounding was minimal: the values given in this chapter are given to one significant figure.

Sensible values for \cref{list:T} cover a significant range.  A lower bound is given by the time for a halo to traverse one Einstein radius, \ie
\begin{equation}
T_{\textrm{min}}
= \frac{1}{v}\sqrt{\frac{2GM}{c^2} \frac{D_{\textrm{d}} D_{\textrm{ds}}}{D_{\textrm{s}}} }
\approx \begin{cases}
5.3 \times 10^3 \,\textrm{s} \approx 1.7 \times 10^{-4}\,\textrm{yr}& \text{for $M = M_{\oplus}$} \\
3.1 \times 10^6 \,\textrm{s} \approx 9.7 \times 10^{-2}\,\textrm{yr}& \text{for $M = M_{\odot}$}
\end{cases}
\end{equation}
Similarly, an upper bound is given by the time taken for a halo to traverse the \lq\lq{}region of influence\rq\rq{} of the halo, \ie
\begin{equation}
T_{\textrm{max}}
= \frac{2a}{v} 
\approx 2.8 \times 10^9 \,\textrm{s}
\approx 89 \,\textrm{yr}
\end{equation}
The lower bound is considerably less than the time between subsequent observations of the same pulsar in a typical survey \cite{ipta}.  The upper bound, while within a human lifetime, is likewise impractical.  Despite this, representative values for the observing time must still fall within the two extremes.  Therefore the observing time was taken to be $T_{\textrm{obs}} \in \left[1, 25 \right]\,\textrm{yr}$ respectively.

The Earth-lens distance was at first fixed, then allowed to vary.  The fixed value of $D_{\textrm{d}}$ was halfway between the Earth and the pulsar, \ie $D_{\textrm{d}} = 5 \,\textrm{kpc}$.  Accordingly, the varied distances were initially chosen so that they were distributed in a Gaussian about $D_{\textrm{d}} = 5 \,\textrm{kpc}$.  This served as a temporary measure for two tests.  Firstly, it provided a check that a correct set of $D_{\text{d}}$ values was drawn from the Gaussian distribution.  Secondly, it provided a means of ensuring that the scaling of the time delay shown in \cref{eq:td-scaling} varied correctly with distance.  Unfortunately, this is not a realistic distribution for the lenses: to maintain a constant halo (number) density between the Earth and the pulsar, the number of halos has to scale as $N(d) \propto d^2$ (for $d = \sfrac{D_{\textrm{d}}}{D_{\textrm{s}}}$ as before).  Then the array of lens-Earth distances $D_{\textrm{d}}$ was drawn from this distribution appropriately.  

The transverse motion of the lens requires two further parameters.  Given a constant speed $v$ and an observation time $\Tobs$, the lens moves in a line $\vect{\ell}$ covering a distance $\Tobs v$.  This line has an angular displacement $\varphi$ by which it is rotated anti-clockwise from the x-axis of the lens plane (\ie $\ell \cdot \basis{x} = \norm{\ell} \cos \varphi $).  Perpendicular to $\varphi$, the vector $\vect{b}$ from the origin to $\vect{\ell}$ forms the shortest distance between the centre of the halo and the pulsar.  The length of this vector is the \emph{impact factor} $b$.  The angles were drawn at random from a uniform distribution $\varphi \in \left[-\pi ,\, \pi \right)$.  The impact factors were drawn from a uniform distribution $\left( 0, b_{\textrm{max}} \right)$ where the maximum impact factor (in lens plane units) $b_{\textrm{max}}$ was retained as a free parameter, $b_\textrm{max} \in \left[1, 10 \right]$.  Since the source plane is two-dimensional, the halo paths are now completely described.

The pulsar timing residual was chosen to be $\Tres = 1 \,\mu\textrm{s}$, reflecting an optimistic estimate of the uncertainty in the pulsar period \cite{siegel2}.  This completes the choice of realistic parameters for the lensing simulations.

Next, in \cref{sec:single-lens} we examine the single-lens model.  Then in \cref{sec:multi-lens} we compare the effect of the different lens masses and timing residuals on the observed signals for a model with multiple lenses at a fixed distance.  Then in \cref{sec:vardist-lens} we allow the lenses to be distributed along the \los{}.  Finally, in \cref{sec:discussion} we compare the results to the point-mass so frequently used in literature and discuss whether or not lensing effects have been observed.

\section{Single lens at a fixed distance } \label{sec:single-lens}

The single-lens case best emphasises the effects of the intrinsic parameters.  The parameters intrinsic to the lens are its total mass $M$, the impact factor $b$, the lens radius $r_\textrm{max}$, the turnover radius $r_0$ and the lens scale $r_s$, whereas the timing residual $\Tres$ is extrinsic (a property of the source).  Since the radii were fixed to $r_\textrm{max} = 10^{-3}\,\textrm{pc}$ and $r_0 = r_s = 10^{-4.5}\,\textrm{pc}$, there are only three free parameters: $b$, $M$ and $\Tobs$.  The two lens masses were set at $M \mathrel{\substack{\textstyle\in\\[-0.1ex]\sim}} \{10^5,\,10^6\}\, M_{\oplus}$.  The measurable effects of the lensing are the image locations of the pulsar, the \emph{change} in the period (corresponding to the change in the time delay) and the magnification and time delay on the pulsar signal itself.

Only one image is produced at each observation.  This shows that the source-lens mapping \cref{eq:lens-eqn} is an injective function for the two (maximum) impact parameter values considered, namely $b_\textrm{max} \in \left[1, 10 \right]$.  It is notable that $b \neq x - y$: the geometric term of the time delay $\sfrac{1}{2} \left(\vect{x} - \vect{y} \right)$ is \emph{not} quadratic in the \emph{lens}-source distance, but in the \emph{image}-source distance.  The actual position of the lens is not observable, nor can it be calculated directly from the lens mapping.  Thus the exact relationship between the impact factor and the resulting time delay signal is difficult to calculate.  

Instead, it is more useful to ask what relationship exists between the impact factor and the number of images.  Before discussing the results, we require the following lemmata \cite{sef}:
\begin{center}
\begin{minipage}[][][c]{.9\textwidth}
\begin{enumerate}
\item \label{thm:1}
Provided that the lens profile is axisymmetric, any image at $x > 0$ produced by a source at $y > 0$ lies at $x \geq y$.
\item \label{thm:2}
For piecewise-continuous convergence $\kappa(x)$, the enclosed mass $m(x)$ is also continuous.  Then $\kappa(x)$ is bounded from above and: 
\begin{align}
 \dfrac{\kappa(x)}{\abs{x}} < c && \text{and} &&
 \dfrac{m(x)}{\abs{x}} < d &&\text{for $c,d \in \mathbb{R}^+$}
\end{align}
\end{enumerate}
\end{minipage}
\end{center}
Proof of \cref{thm:1} is as follows: Due to axisymmetry, we need only consider sources at $y > 0$ in lensing-polar co-ordinates.  By definition, the mass enclosed within a radius $x$ is positive.  Substituting $m(x) > 0$ into the lens equation completes the proof.  

The second lemma \cref{thm:2} is somewhat more convoluted.  We begin by noting that a necessary condition for the convergence to be piecewise-continuous is that it is a well-defined function\footnote{Thus we exclude the \swz{} lens, since $\kappa(x) = \delta(x)$ which is, strictly speaking, the limit of a function.} at all radii.  Physical arguments require that:
\begin{itemize}
\item The lens itself must be finite in extent, so there is some $\xmax$ such that $\kappa(x > \xmax) = 0$.
\item The surface mass density does not diverge to infinity, in which case a real number can always be found that is larger than any value of the convergence, \ie $\kappa(x) < \kappa_{\textrm{max}} < \infty \;\forall x$.  
\item The finite total mass $M$ of the lens bounds $m(x)$.
\end{itemize}
Then we may take the limits:
\begin{align*}
\lim_{\abs{x} \rightarrow \infty} x\kappa(x) = 0 
&\implies \lim_{\abs{x} \rightarrow \infty} \kappa(x) 
< \lim_{\abs{x} \rightarrow \infty}cx 
\iff \sfrac{\kappa(x)}{\abs{x}} < c 
\\
\lim_{\abs{x} \rightarrow \infty} m(x) = M
&\iff M = \lim_{\abs{x} \rightarrow \infty} \frac{2}{x}\integ{\bar{x}}{0}{x}{\bar{x}\kappa(\bar{x})}
&&\text{by definition of $m(x)$}
\\
&\implies M < \lim_{x \rightarrow \infty} \frac{2}{x}\integ{\bar{x}}{0}{x}{\bar{x}\kappa_{\text{max}}} = \kappa_{\text{max}}x
&&\text{using $\kappa(x) \leq \kappa_{\textrm{max}} \;\forall x$} 
\\
&\implies \sfrac{m(x)}{\abs{x}} < d
&&\text{setting $d = \kappa_{\text{max}}$}
\end{align*}
This completes the proof.  We can now apply the lemmata to the two different impact factors in the simulations.

Consider the case where $b_\text{max} = 10$ \cref{fig:single-lens:a,fig:single-lens:c,fig:single-lens:e,fig:single-lens:g}.  The source is never blocked by the lens.  Using the lens co-ordinates in \cref{ch:gl,thm:1}, we have $x \geq y > \xmax$.  Consequently the radial density profile of the lens is not important and we can approximate it by an \hdisc{} lens of the same radius $\xmax$ and total mass $M$.  The corresponding lens equation is readily invertible:
\newsavebox{\xcases}\begin{lrbox}{\xcases}
\begin{minipage}{.4\linewidth}$
	\begin{cases}
	x - \dfrac{x}{\xmax^2} &\text{for $x \leq \xmax$} \\[1em]
	x - \dfrac{1}{x^2} &\text{for $x \geq \xmax$}
	\end{cases}
$\end{minipage}
\end{lrbox}
\newsavebox{\ycases}\begin{lrbox}{\ycases}
\begin{minipage}{.4\linewidth}$
	\begin{cases}
	y \dfrac{x}{\xmax^2 - 1}  &\text{for $y \leq \dfrac{\xmax^2 - 1}{\xmax}$} \\[1em]
	\dfrac{y}{2} + \sqrt{\dfrac{y^2}{4} + 1} &\text{otherwise}
	\end{cases}
$\end{minipage}
\end{lrbox}
\begin{align}
y &= \usebox{\xcases}
\intertext{and we are only interested in the latter case, which has solution:}
x &= \usebox{\ycases}
\end{align}
The single root of the lens equation produces a single image.  When the impact factor is larger than the physical radius of the lens, we see only the single image which is predicted.

The case where $b_\text{max} = 1$ in \cref{fig:single-lens:b,fig:single-lens:d,fig:single-lens:f,fig:single-lens:h} is more complex.  The convergence of the lens does become important, but the \nfw{} model does not have an analytically-invertible lens equation.  Under these circumstances, are there any limits to be placed on the number of images?  We now generalise to any axisymmetric lens with convergence $\kappa(x) \propto \abs{x}^{1-\epsilon}$ since this does not affect the complexity of the proof.  Consider the effect of axisymmetry on the deflection angle $\alpha(x)$:
\begin{equation}
-\alpha(-x) = \frac{-m(-x)}{-x} = \frac{m(-x)}{x} = \frac{m(x)}{x} = \alpha(x)
\end{equation}
We have shown that it is an odd function, but the linear combination of odd functions is also an odd function.  Using this, the lens equation $x - \alpha(x)$ is also odd.  Since odd functions have an odd number of roots, the number of images, if any, must be odd: $n = \left\{0,2m+1 | m \in \mathbb{N} \right\}$.  Furthermore, we can show that there must be at least one image:
\begin{align}
x - y = \frac{m(x)}{x} \leq d &&\text{by Lemma \cref{thm:2}} \\
f \leq x - y \leq d &&\text{substituting Lemma \cref{thm:1}} \\
\intertext{Similarly, for the gradient:}
\lim_{\abs{x} \rightarrow \infty} \frac{\dx{y}}{\dx{x}} 
= \lim_{\abs{x} \rightarrow \infty} \left(1 - \frac{\dx{}}{\dx{x}} \frac{m(x)}{x}\right) = 1 &&\text{by Lemma \cref{thm:2}}
\end{align}
Since, for sufficiently large $y$, the lens equation is linear in $x$, there is one and only one root.  Thus the zero-image possibility is discounted: $n = 2m+1,\, m \in \mathbb{N}$.  A generalisation of this is known as the Odd Number Theorem.

A more qualitative view of the role of the impact factor is provided by the time delay surface.  Recall that images occur where the time delay surface has an extremum.  The geometric contribution to the surface is quadratic in $x$, so there is a single minimum.  It is this \lq\lq{}geometric surface\rq\rq{} to which the time delay surface is asymptotic, far from the lens.  To obtain multiple images, the relativistic contribution must distort the time delay surface sufficiently to produce additional extrema.  This distortion depends upon the size of the lens (an extended lens will induce smaller gradients than a concentrated one) and its mass (a larger lens will increase the magnitude of the distortion).  When the impact factor is large, the lens and source are sufficiently separate that the asymptotic behaviour occurs, producing a single image near the source.  When the impact factor is small, even a massive lens will not produce multiple-image-creating perturbations in the surface, if the mass of the lens is distributed about a large radius, \ie if the value of the convergence is small (compared to the value of the geometric delay).  This \lq\lq{}intuition\rq\rq{} is confirmed by the above calculations.

We have now explained why there is only a single image shown in \cref{fig:single-lens}.  In the case of a small impact factor $b_\text{max} = 1$ this is due to the Odd Number Theorem, whereas in the case of a large impact factor $b_\text{max} = 10$ it follows directly from the lens equation when the lens does not obstruct the geodesic from source to observer.

The magnification produced by the lens is a useful indicator of a lensing event.   The evolution is smoothly-varying with time, with its peak width determined by the mass and its magnitude determined by the impact factor.  Unlike the time delay, there is a closed form for the \mf{}: 
\newsavebox{\casebox}\begin{lrbox}{\casebox}
\begin{minipage}{.4\linewidth}
	\begin{align*}
		 	\left[ \left( 1 - \frac{1}{\xmax^2} \right) \left( 1 + \frac{1}{\xmax^2} - 2\kappa_0 \right) \right]^{-1}
			&= \left( 1 - \frac{1}{\xmax^4} \right)^{-1}
 			&\text{$x \leq \xmax$} \\
			\left( 1 - \frac{1}{x^2} \right) \left( 1 + \frac{1}{x^2} \right)
			&= \left( 1 - \frac{1}{x^4} \right)^{-1}
 			&\text{$x \geq \xmax$}
	\end{align*}
\end{minipage}
\end{lrbox}
\begin{equation}
\mu (x) = \left[ \left( 1 - \frac{m(x)}{x^2} \right) \left( 1 + \frac{m(x)}{x^2} - 2\kappa(x) \right) \right]^{-1} 
\end{equation}
Making the same first-order approximation as before:
\begin{equation}
\mu (x) \approx \left\{ \usebox{\casebox} \right.
\end{equation}
How does this result reinforce our intuition?  The magnification arises from the relativistic distortion of spacetime from the Minkowski metric, which alters the  infinitesimal area elements along the geodesics.  Thus, far from the lens, the distortion is minimal and the \mf{} is small.  We recover the $\mu \approx 1$ \swz{} result outside the lens: indeed the $\Tobs = 25\,\text{yr}$ simulations in \cref{fig:single-lens:a,fig:single-lens:b,fig:single-lens:c,fig:single-lens:d} show precisely this behaviour.   Inside the lens, the $\bmax = 10$ simulations in \cref{fig:single-lens:c,fig:single-lens:d,fig:single-lens:g,fig:single-lens:h} also show $\mu \approx 1$ regardless of emission time.  These results confirm that the magnification effected by the presence of the lens mass is inversely related to the separation between lens and source.  The remaining simulations in \cref{fig:single-lens:a,fig:single-lens:b,fig:single-lens:e,fig:single-lens:f} exhibit magnification which is not insignificant.  There is a clearly-defined peak which varies smoothly with time.  The maximum, as in our first-order approximation, does not depend on the lens mass.  Since we have fixed the physical radius of the lens, we cannot tell whether the $\xmax$-dependence of the \hdisc{} approximation is reflected in the \nfw{} lens.  

In contrast, a comparison of \cref{fig:single-lens:a} to \cref{fig:single-lens:b} shows that the width of the peak does vary with mass.  This is emphasised in the shorter simulations \cref{fig:single-lens:e} and \cref{fig:single-lens:f}, in which the lens does not transit as far from the source.  A geodesic at the same distance from the lens will have a tangent bundle which deviates more from the un-lensed tangent bundle, if the lens is more massive.  This is reflected in the \mf{} for \cref{fig:single-lens:e} and \cref{fig:single-lens:f}: the $M = 10^5 \, M_{\oplus}$ halo shows the same magnification at times $T \in \pm 0.3 \times 10^7\,\text{s}$ as the $M = 10^6 \, M_{oplus}$ exhibits for $T \in \pm 3 \times 10^7\,\text{s}$.  Inverting this logic, the \mf{} will be the same for geodesics close to a low-mass lens and further from a high-mass one.  This generates the narrower peak in \cref{fig:single-lens:a} compared to \cref{fig:single-lens:b}.  Overall, the results demonstrate that, provided the lens and source are separated on the order of the lens radius, the \mf{} is a useful indicator of the presence of the lens.

The time delay, while itself unobservable, creates variation in the pulsar period.  To see that it is only the \emph{change} in the delay that is measurable, recall that the time delay is defined up to a constant of integration.  (This constant is the light-travel-time relative to the light-travel-time in the absence of the lens.)  The period $P$ of the pulsar absorbs all zeroth-order time delay terms: a constant time delay would contribute equally to every period and never be detected.  Similarly, first-order terms $\dot{\tau} = \tau_1$ are absorbed into $\dot{P}$, second-order $\ddot{\tau} = \tau_2$ into $\ddot{P}$ \etc, where the period derivatives themselves have an intrinsic uncertainty due to the physics of the pulsar and to instrumental and signal processing limitations.  There are further complications due to the processing pipelines (discussed in \cite{ipta}): simulating these is beyond the scope of this project, so the results show all first-and-higher-order terms in the period changes: $\Delta P = P(t) - \bar{P}$.  

Which component --- relativistic or geometric --- of the time delay is dominant is determined by the separation between lens and source.  When the lens-source distance is large, \ie $x \gg \xmax$ (which from Lemma \cref{thm:1} corresponds to $y \geq \xmax$), the time delay is largely geometric.  Far from the lens, the convolution $\kappa(x) \ast \ln(x)$ means that the relativistic time delay is small because $\ln(x)$ is small.  This confirms our intuition that the time delay surface should be asymptotic to the geometric time delay surface.  When the lens-source distance is small, the convergence causes the relativistic component of the time delay to dominate the geometric one, resulting in a total time delay surface which is decidedly not quadratic.  Indeed, for sufficiently small turnover radius (not covered in these results owing to limits on the grid fineness) the time delay surface is sufficiently distorted to produce multiple extrema.  The extreme case of this is the \swz{} lens, which produces two images at all $x$ (although one is highly demagnified): we discuss this further in \cref{sec:discussion}.  Moreover, comparing the low-mass results to the high-mass ones demonstrates that the time delay scales proportionally to the lens mass.  Given that there is no closed form for the relativistic time delay in the \nfw{} case, we cannot quantitatively extrapolate   the observable --- the pulsar period variation --- from the lensing effect, \ie{} the time delay.  Nevertheless, the time delay is worth considering despite the fact that it is not observable.

\begin{figure}
\includegraphics[width=0.7\textwidth]{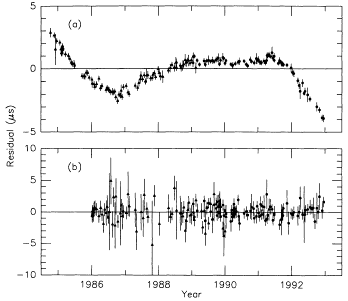}
\caption{Timing residuals for two \msp{s}: \textsc{(Top:)} \textsc{PSR~B1937+21} observed at $2\,380\,\text{MHz}$ and \textsc{(Bottom:)} \textsc{PSR~B1855+09} observed at at $1\,408\,\text{MHz}$. [Fig.~5 in \cite{1994ApJ...428..713K}]
\label{fig:psr-timing}}
\end{figure}

\begin{figure}
\includegraphics[width=0.7\textwidth]{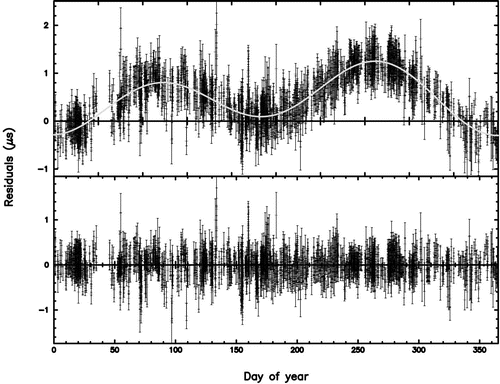}
\caption{Timing residuals for \textsc{PSR~J0437–4715}: \textsc{(Top:)} without parallax but including all remaining parameters at their best-fit values and \textsc{(Bottom:)} with a parallax fit of 6.65 mas (solid line in top figure). [Fig.~4 in \cite{15102}]
\label{fig:psr-accurate}}
\end{figure}

The remaining observable is the timing residual.  This is the gradient of the relative time delay, so it is dimensionless.  In \cref{fig:single-lens:a,fig:single-lens:b} the lens completely transits the source and we can see the geometric and relativistic influence on the variations.  The low-mass result in \cref{fig:single-lens:a} is the only instance in which the mass $M = 10^5 \, M_{\oplus}$ is sufficiently small and the observation period $\Tobs = 25\,\text{yr}$ sufficiently large that we see the lensing effects asymptote to zero.  In \cref{fig:single-lens:b} the larger mass $M = 10^6 \, M_{\oplus}$ influences lenses at larger distances; conversely, at the same distance, the time delay variation is larger.  The corresponding short observations \cref{fig:single-lens:e,fig:single-lens:f} illustrate the effects of the relativistic time delay: $\dot{\tau}$ decreases monotonically when the lens is sufficiently close, then becomes nearly constant.  While \cref{fig:single-lens:c,fig:single-lens:d,fig:single-lens:g,fig:single-lens:h} also exhibit monotonic, nearly-linear variations in the time delay, these are caused by a different process.  From the preceding paragraph, we deduce that the geometric time delay dominates because of the large impact factor: $\bmax = 10$.  This quadratic dependence in the time delay is equivalent to the linear behaviour of the variation.  There are small deviations from linear behaviour due to the relativistic term of the time delay, which has small but non-negligible influence at these distances.  Comparison to measured values of the timing residual determines whether these variations are observable in practice.  While many pulsars exhibit variations in the residuals, these fluctuations are caused by a variety of phenomena (a detailed list is given in \cite{1994ApJ...428..713K}), including the (poorly-understood) physics of the pulsar itself \cite{psr-summary}.  However, this noise is orders of magnitude greater in \lq\lq{}normal\rq\rq{} pulsars ($\sim 1 - 10^{2}\,\text{ms}$) than in millisecond ones ($\sim 1 - 10\,\mu\text{s}$).  \cref{fig:psr-timing} shows timing residuals of two characteristic \msp{s}, while \cref{fig:psr-accurate} shows one of the most stable pulsars to date.  A necessary condition for the gravitational lensing to be detected is that the residuals are large compared to the inherent fluctuations.  Thus we see that \cref{fig:single-lens:a,fig:single-lens:b,fig:single-lens:d} are easily detectable with current data.  Detection of \cref{fig:single-lens:c,fig:single-lens:f} are possible depending upon the stability of the source pulsar.  In contrast, \cref{fig:single-lens:g,fig:single-lens:h} require a decrease in the noise of two orders of magnitude.  Ultimately their detection depends upon the amplitude of inherent noise (caused by \eg{} superfluid behaviour in the neutron star) and the evolution of more sophisticated data reduction processes.  Thus we find that over short timescales, halos of $\sim 10^{6}\,M_{\oplus}$ are detectable regardless of impact parameter, whereas those of $\sim 10^{5}\,M_{\oplus}$ are detectable only when transiting close to the \los{}.  Taking longer observations removes this problem: lensing of a sufficiently stable pulsar produces a measurable signal independent of mass or impact parameter.
Therefore, variation in the pulsar times-of-arrival is a notable signature of gravitational lensing.

\begin{figure}
\includegraphics[width=\textwidth]{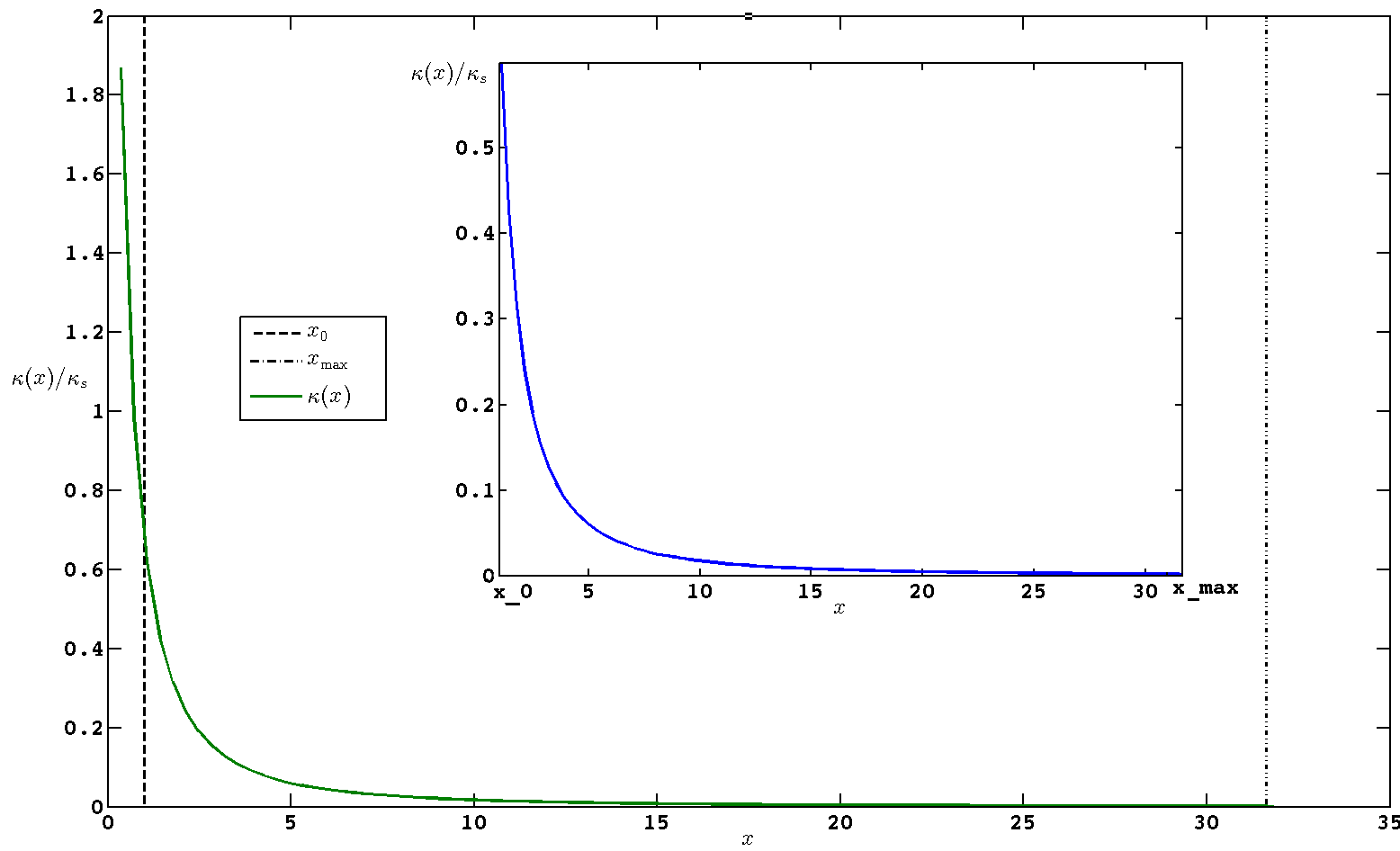}
\caption{Convergence for the \nfw{} lens.  The solid line shows the convergence as a function of radius.  The dashed and dash-dotted lines indicate the turnover radius $x_0 = 1$ and the physical radius $x_{\text{max}}$ respectively.  The inset shows the small values of $\kappa(x)$ after the turnover radius.
\label{fig:nfw-density}}
\end{figure}

While the single lens simulations provide unconvincing lensing evidence when examining the magnification of the pulsar signal, they also show easily detectable signatures in the times-of-arrival of the pulsar signal.  This suggests that the presence of even a single \dmh{} between the Earth and a nearby (Galactic) pulsar can be detected within a human lifetime.

\begin{landscape}
\begin{figure}
\noindent
\makebox[\paperwidth]{
\subfloat[$M=10^5 M_{\oplus}$, $b=1$, $T_{\text{obs}}=25\,\text{yr}$]{
    \label{fig:single-lens:a} 
    \includegraphics[height=\textwidth]{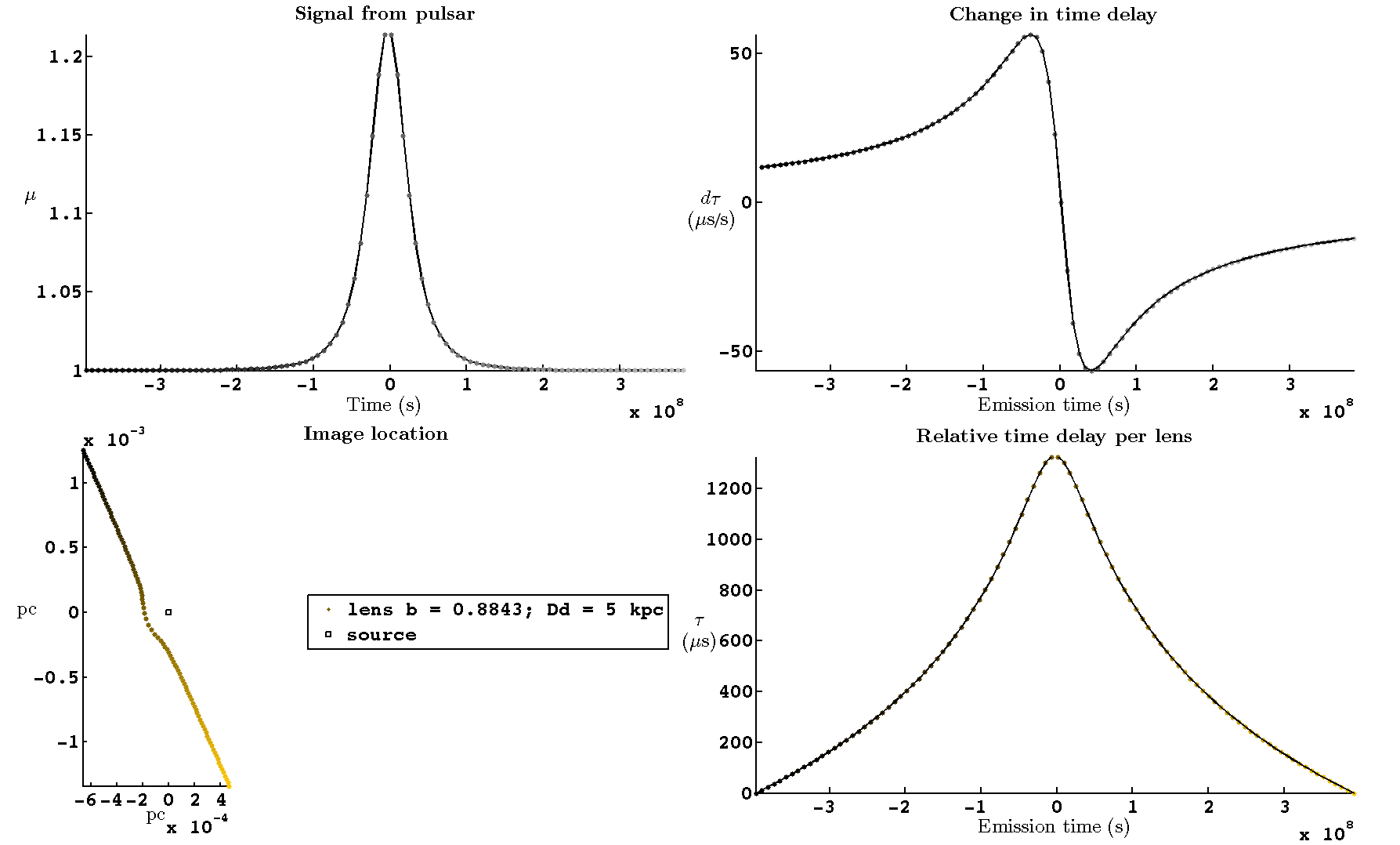}
}
}
\end{figure}

\begin{figure}
\ContinuedFloat
\noindent
\makebox[\paperwidth]{
\subfloat[$M=10^6 M_{\oplus}$, $b=1$, $T_{\text{obs}}=25\,\text{yr}$]{
    \label{fig:single-lens:b} 
    \includegraphics[height=\textwidth]{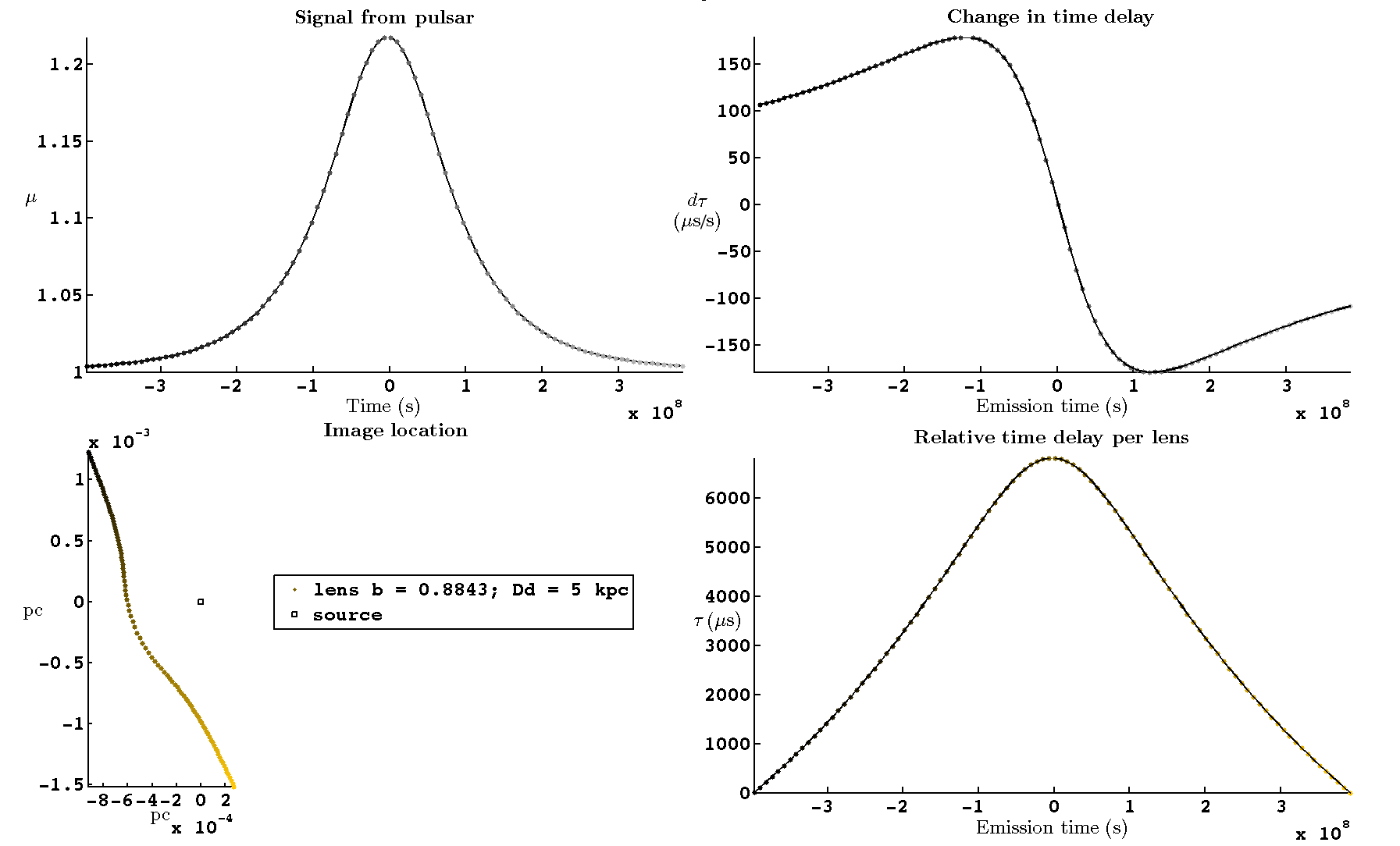}
}
}
\end{figure}

\begin{figure}
\ContinuedFloat
\noindent
\makebox[\paperwidth]{
\subfloat[$M=10^5 M_{\oplus}$, $b=10$, $T_{\text{obs}}=25\,\text{yr}$]{
    \label{fig:single-lens:c} 
    \includegraphics[height=\textwidth]{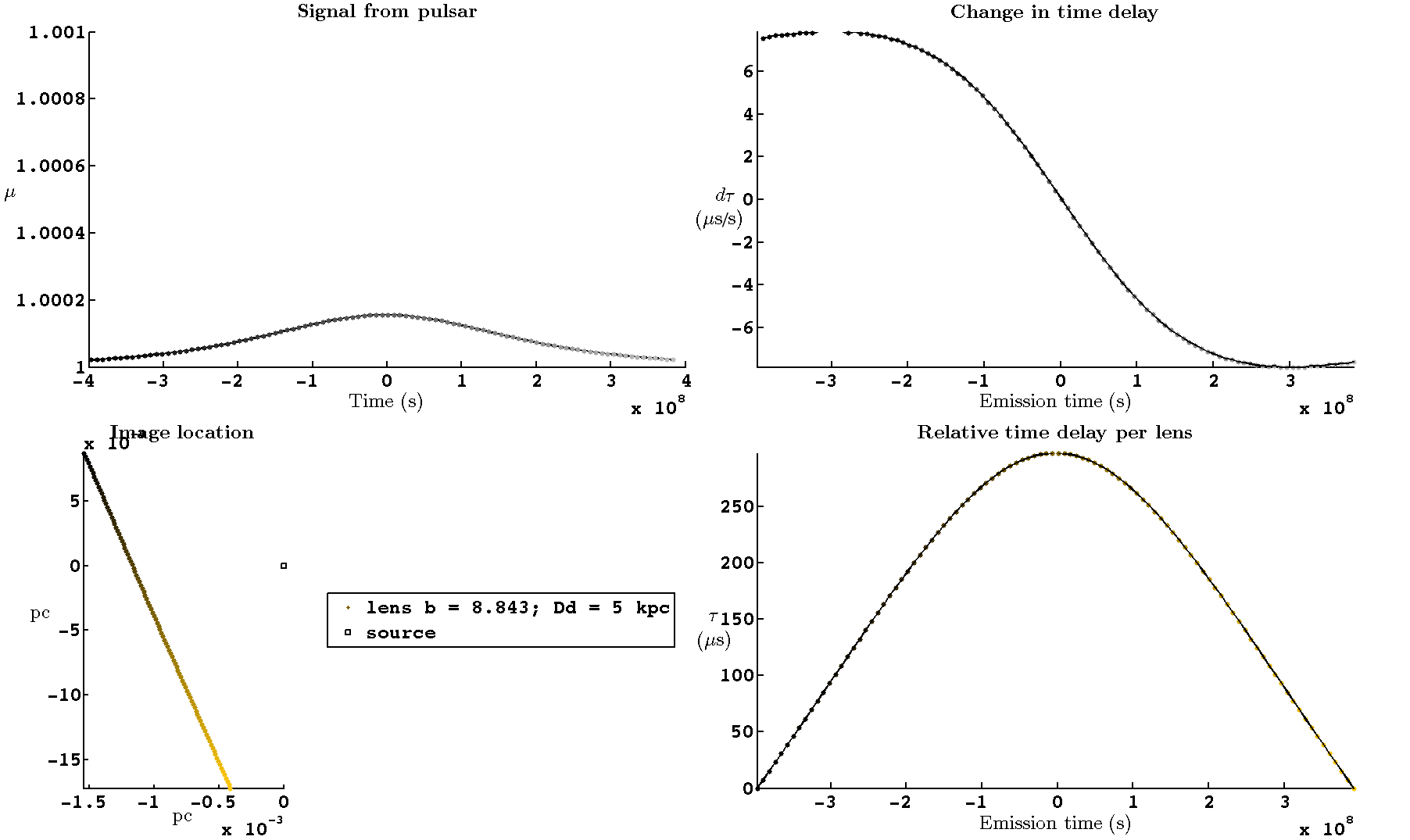}
}
}
\end{figure}

\begin{figure}
\ContinuedFloat
\noindent
\makebox[\paperwidth]{
\subfloat[$M=10^6 M_{\oplus}$, $b=10$, $T_{\text{obs}}=25\,\text{yr}$]{
    \label{fig:single-lens:d} 
    \includegraphics[height=\textwidth]{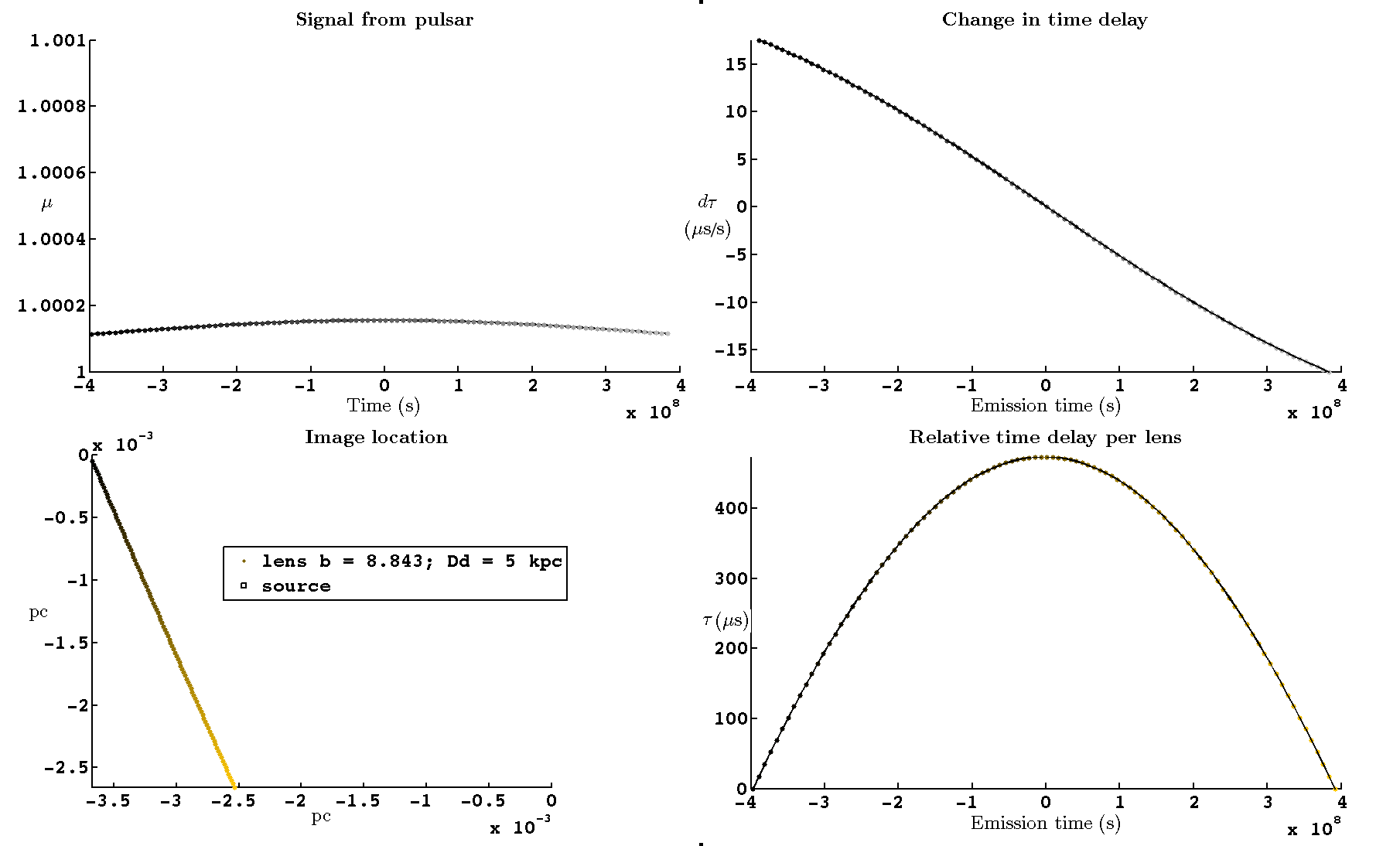}
}
}
\end{figure}

\begin{figure}
\ContinuedFloat
\noindent
\makebox[\paperwidth]{
\subfloat[$M=10^5 M_{\oplus}$, $b=1$, $T_{\text{obs}}=1\,\text{yr}$]{
    \label{fig:single-lens:e} 
    \includegraphics[height=\textwidth]{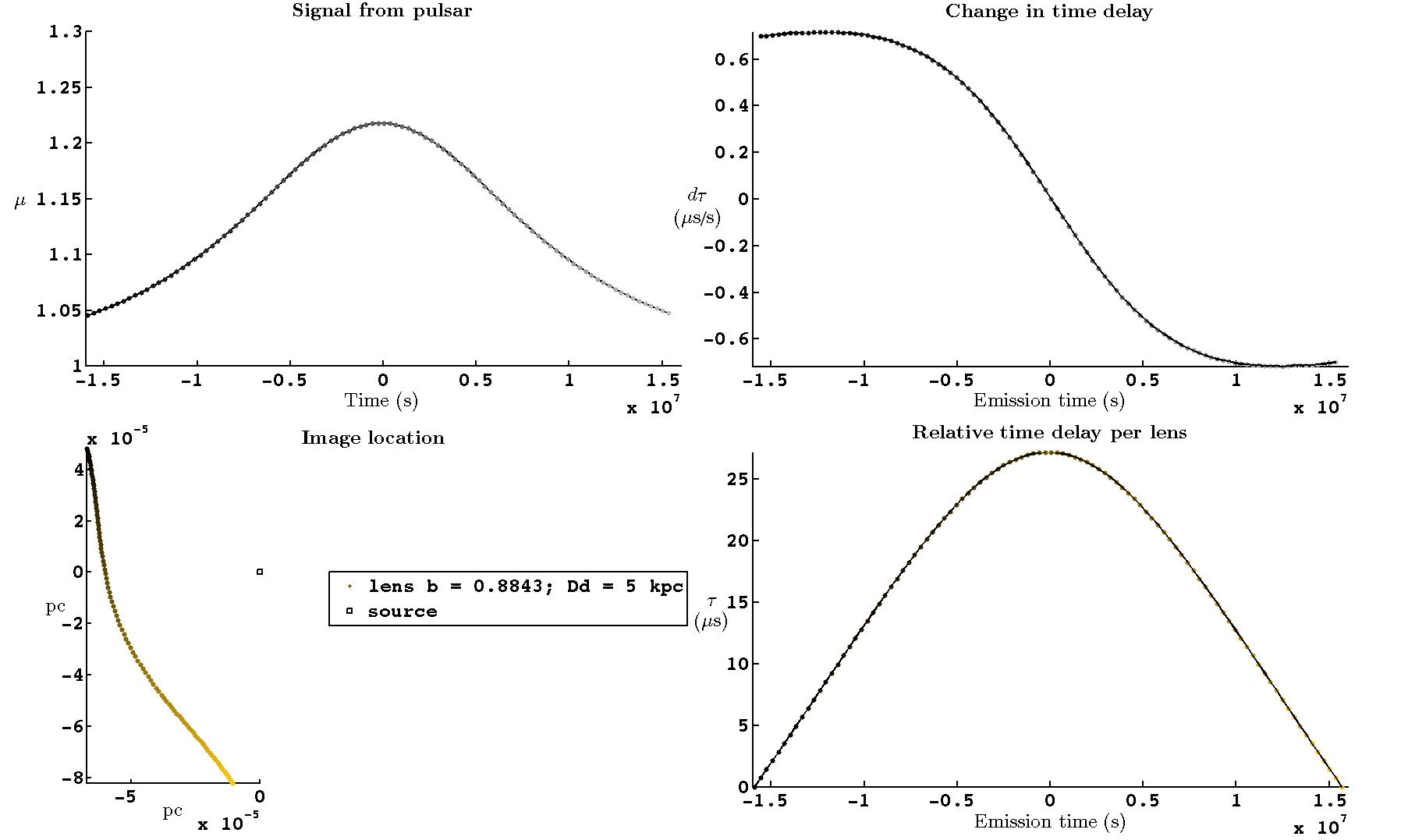}
}
}
\end{figure}

\begin{figure}
\ContinuedFloat
\noindent
\makebox[\paperwidth]{
\subfloat[$M=10^6 M_{\oplus}$, $b=1$, $T_{\text{obs}}=1\,\text{yr}$]{
    \label{fig:single-lens:f} 
    \includegraphics[height=\textwidth]{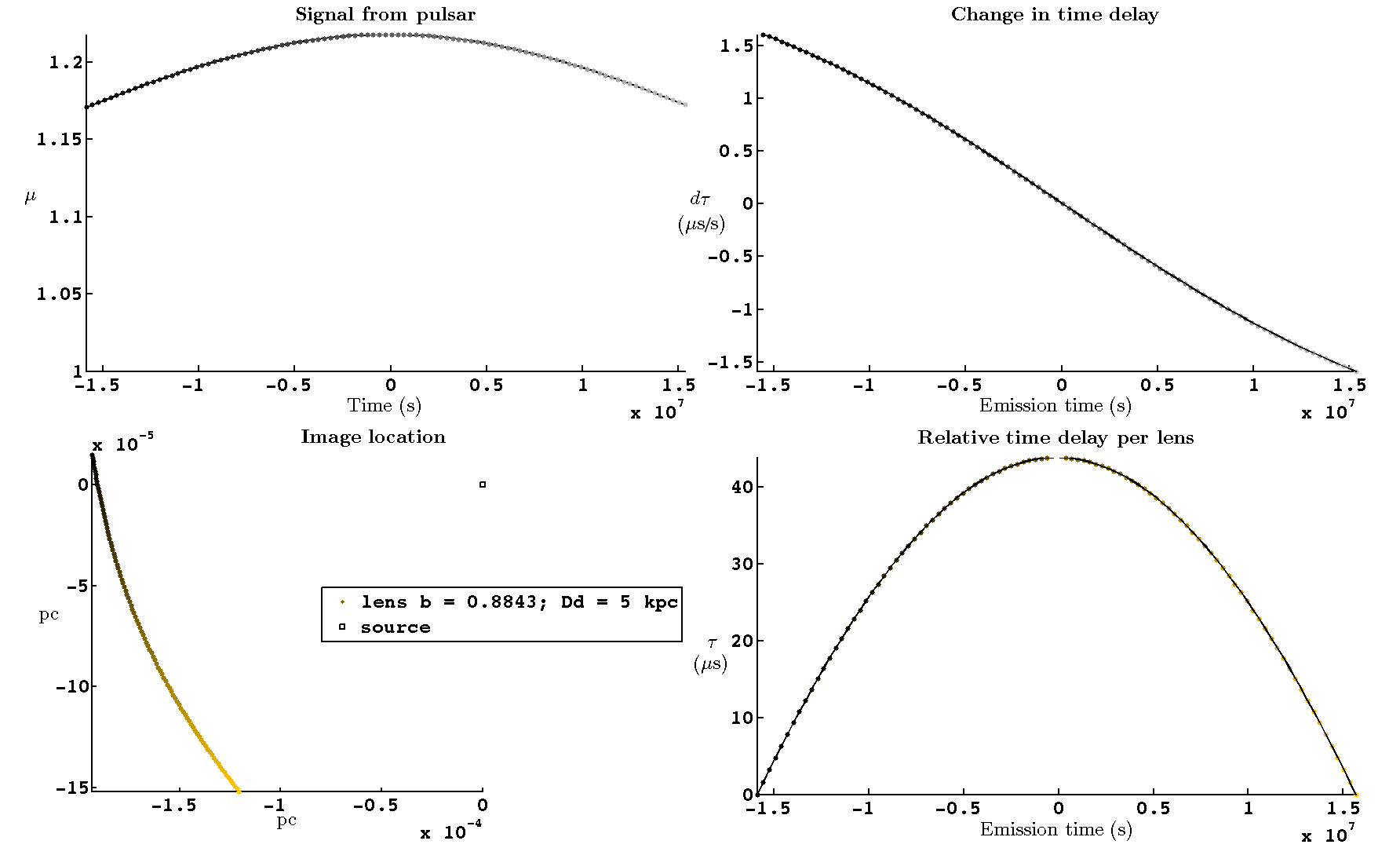}
}
}
\end{figure}

\begin{figure}
\ContinuedFloat
\noindent
\makebox[\paperwidth]{
\subfloat[$M=10^5 M_{\oplus}$, $b=1$, $T_{\text{obs}}=1\,\text{yr}$]{
    \label{fig:single-lens:g} 
    \includegraphics[height=\textwidth]{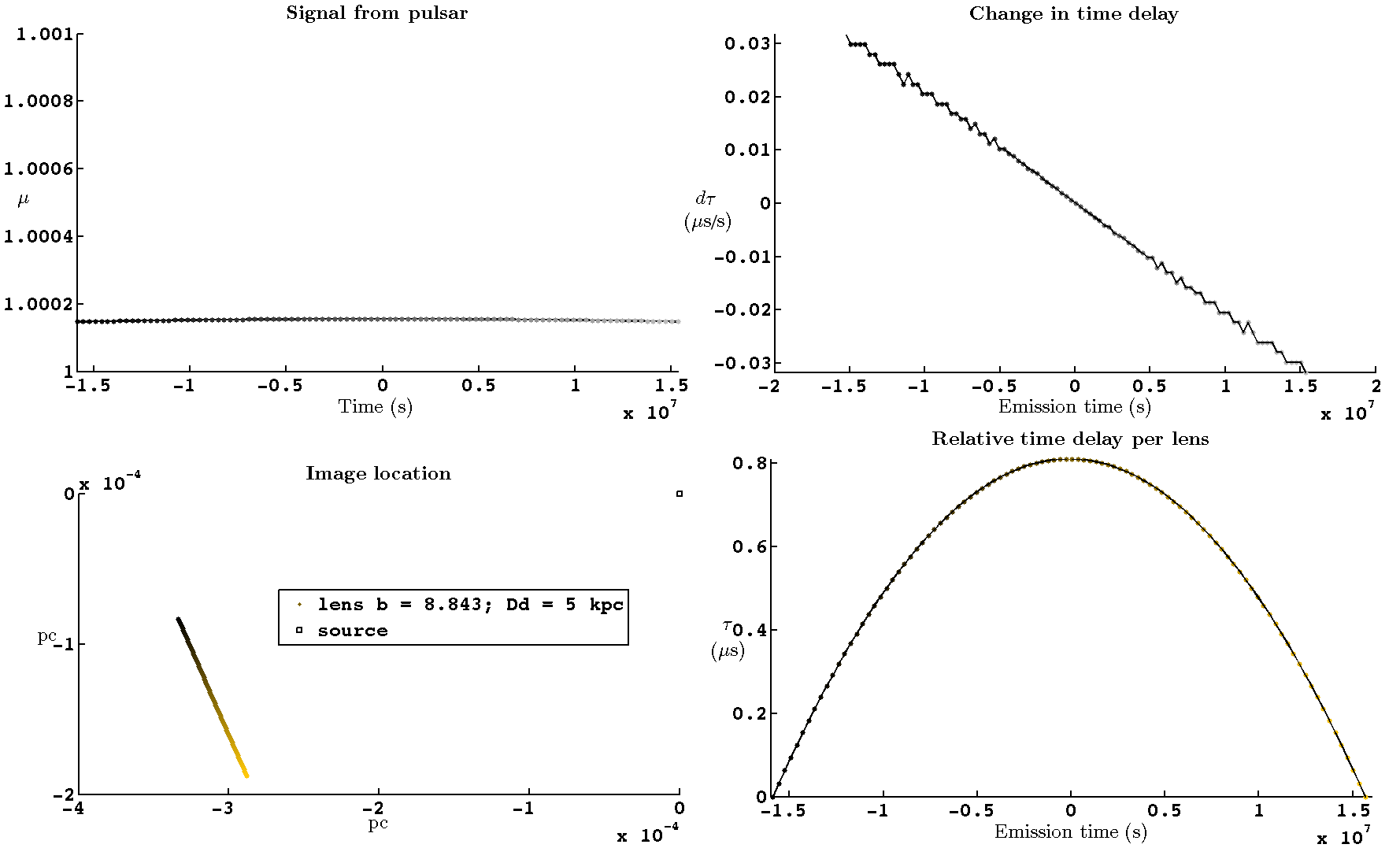}
}
}
\end{figure}

\begin{figure}
\ContinuedFloat
\noindent
\makebox[\paperwidth]{
\subfloat[$M=10^6 M_{\oplus}$, $b=10$, $T_{\text{obs}}=1\,\text{yr}$]{
    \label{fig:single-lens:h} 
    \includegraphics[height=.9\textwidth]{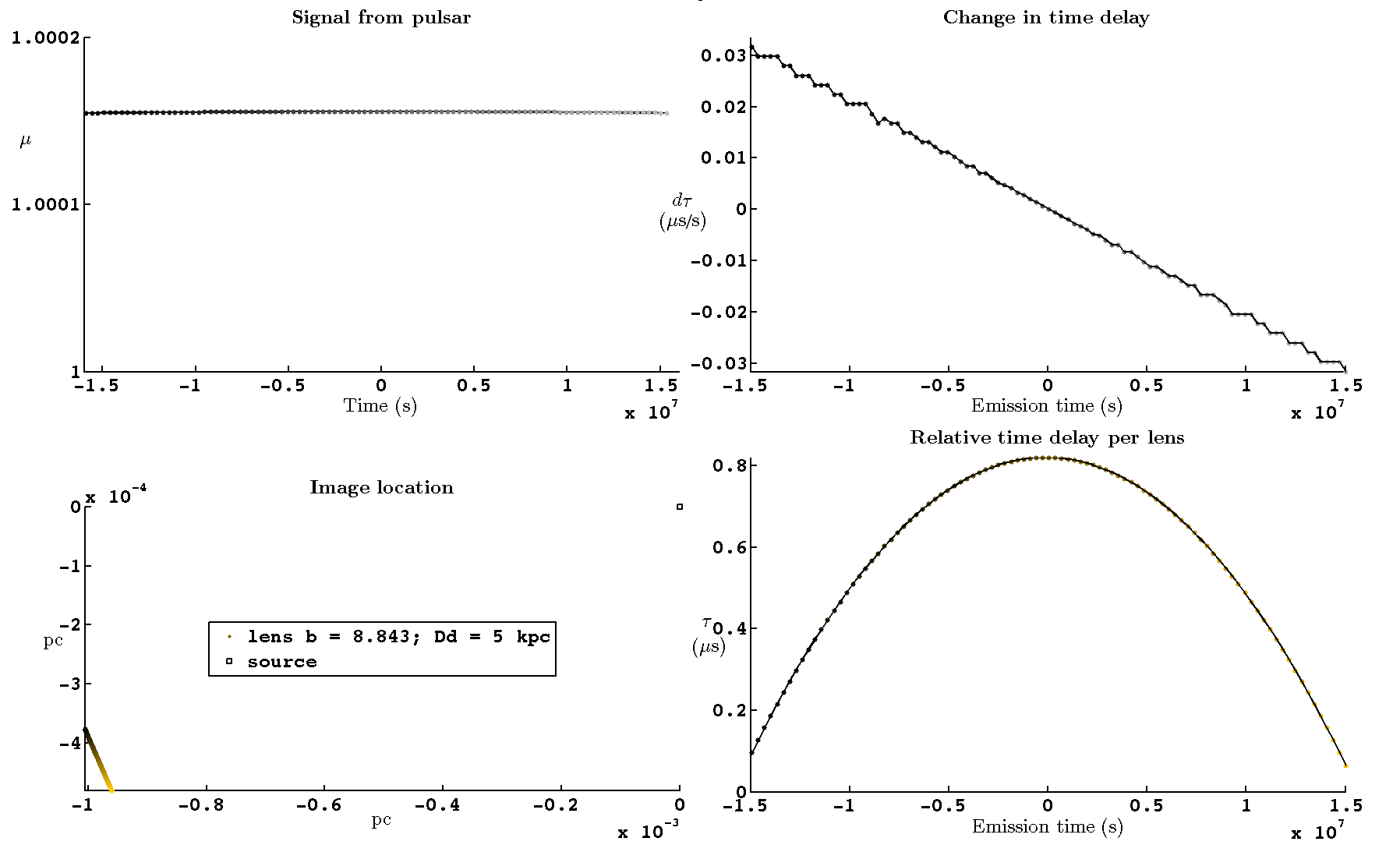}
}
}
\caption{Example of a single halo with scale radius $10^{-3}\,\textrm{pc}$ at $5\,\textrm{kpc}$ transiting between source at $10\,\textrm{kpc}$ and observer.  The observations are \textsc{(top left)}: the amplitude of the signal relative to that from the pulsar; \textsc{(top right)} the change in times-of-arrival of the signal.  \textsc{(bottom left)}: the image locations; \textsc{(bottom right)}: the relative time delay.  Later times are indicated by lighter colours.  The lens parameters are given in the sub-captions.}
\label{fig:single-lens} 
\end{figure}
\end{landscape}

\section{Fixed-distance model }\label{sec:multi-lens}

The multiple-lens model distributes the total mass of the dark matter between source and observer into a discrete number of halos.  The integrated mass remains the same: whereas the single halos have mass $10^6 M_{\oplus}$, the ten halos each have mass $10^5 M_{\oplus}$.  The impact factors are set at $b_\text{max} \in \left[1, 10 \right]$ and drawn from a uniform distribution $\left( 0, b_{\textrm{max}} \right)$.  The observation periods are set at $\left[1,\, 25\right]\,\text{yr}$ as before.  The Earth-lens distance is fixed at $D_{\text{d}} = 5\,\text{kpc}$.  (A further $10^2$ halo simulation was also run, but the graphical output is too complex to be illuminating.)

The primary difference between the single and multiple lenses is that the \lq\lq{}mapping\rq\rq{} from the actual effects of each lens to the resulting observations is now surjective.  The observations are a superposition of the effects of each lens: the time delay is the sum of those from each lens and the \mf{} is the product, just as in the \mpl{} formulae of \cref{ch:mpl}.\footnote{A key difference between this result and the full \mpl{} scenario is that the time delay surface is calculated for each lens separately.  A more accurate process would be to sum the convergences of each lens (similarly to a microlensing simulation) and calculate a single time delay by convolving the total convergence with the transform kernel.  A significant disadvantage to this method is that it removes the symmetry which we have used to minimise the computational expense.}
Unlike the full \mpl{} case, the cause of the surjectivity is not a recurrence relation, but the limits of angular resolution.  The individual images from each lens are unresolved because there are $\sim 6$ decades of length scale between the angular diameter distance to the lenses (and therefore the images) and the image separations.  This superposition removes any \lq\lq{}typical\rq\rq{} lensing characteristics from the signal because the total mass is distributed between \dmh{s}.  

The \mf{}, unlike the single lens case, is rapidly-varying and larger than unity.  This makes it practical to use as evidence of a lensing detection.  While an image with constant magnification $\mu$ is indistinguishable from a source with $\mu$-fold larger flux, an image whose magnification fluctuates is likely to be affected by external physics.  In comparison to the single-lens case, we can clearly see that the total magnification factor results from the superposition of individual signals with the same characteristics as \cref{fig:single-lens}.  The long observations \cref{fig:mult-lens:b,fig:single-lens:d} show artificially narrow peaks, an artefact of the scaling on the $x$-axis compared to those in the short period simulations \cref{fig:mult-lens:a,fig:mult-lens:c}.  The short observations \cref{fig:mult-lens:a,fig:mult-lens:c} are more useful for demonstrating the effect of the impact factor.  The $\bmax = 10$ result (\cref{fig:mult-lens:a}) is dominated by the effect of two lenses while the other eight have slowly-varying, smaller amplitudes.  The dominant lenses are similar to those in \cref{fig:single-lens:e} whereas the other eight closely resemble \cref{fig:single-lens:g}.  It is not implausible that the closest lenses with $b \approx 1.5$  and $b \approx 2$ make the largest contribution to the magnification factor.  By correlating the time of the peaks in $\mu$ with the individual time delays, this is reinforced; when those two lenses are at conjunction corresponds to the maxima in the \mf{}.  The $\bmax = 1$ result has a similar envelope behaviour, with larger, narrower peaks due to the smaller lens-source distance.  This smaller separation increases the relativistic effects on the geodesics, as described in \cref{sec:single-lens}.  Only \cref{fig:single-lens:a} displays a magnification substantially greater than unity, \ie one which is readily observable.  The remaining plots show oscillations of only a few percent, which suggests that the \mf{} may not be a useful indicator of the presence of multiple lenses.

\begin{figure}
\includegraphics[width=.8\textwidth]{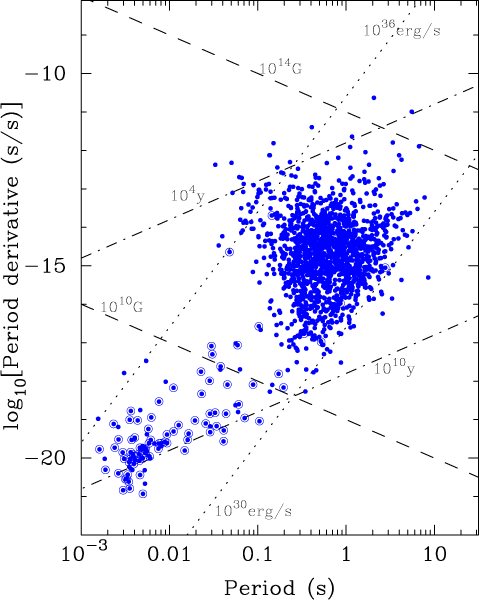}
\caption{Diagram of the $P-\dot{P}$ plane with the sample of radio pulsars as of 2008. Binary pulsars are highlighted by open circles. Lines of constant magnetic field (dashed), characteristic age (dash-dotted) and spin-down energy loss rate (dotted) are also shown. [Fig. 3 from \cite{psr-summary}]
\label{fig:ppdot}
}
\end{figure}

The superposition generates oscillations in the times-of-arrival of the pulsar signal.  Like the bell-shape of the \swz{} lens, the \nfw{} model produces period changes which are smoothly-varying over the observation time.  The summation process creates a result which is neither a continuous, nor easily-fitted function.  Under these circumstances, one may be forgiven for concluding that the lensing variations may be mistaken for noise.  Comparison to typical uncertainties in $\dot{P}$ (\cref{fig:ppdot}) show that the lensing effects are far greater: $\dot{\tau} \approx 1$ compared to $\dot{P} \approx 10^{-20}$ for a \msp{} \cite{psr-summary}, and the discontinuity in the delays as a function of time makes them difficult to attribute to natural properties of the pulsar (\eg spin-down or possible binary interaction) or gravitational waves \cite{ppta,ipta,psr-summary}.  Even on (relatively) short observational timescales, the time delay variations in \cref{fig:mult-lens:a,fig:mult-lens:c} leave a detectable and highly unusual signature.  The long measurements \cref{fig:mult-lens:b,fig:mult-lens:d} display similar behaviour.  The time delay changes appear sharper than the short-observing case, with each peak corresponding to the variations from a single lens.  Each individual lens dominates when it is close to conjunction, creating fluctuations on a much shorter timescale than the smooth variations seen when no other lenses are present.  These fluctuations also determine the amplitude of the variations.  Comparison between the $M = 10^5 \, M_{\oplus}$ simulations in \cref{fig:single-lens} and \cref{fig:mult-lens} shows that the extrema of the variations in the times-of-arrival are much reduced in the multiple-lens case.  This behaviour is best explained by the time delay plots for the individual lenses.  A set of lenses with similar period variations interferes destructively, dampening the magnitude of the total variation; those with greatly differing delays interfere constructively to emphasise the variation.  The latter case occurs most frequently when lenses are near conjunction, at which time the gradient of the time delay variations is very steep.   The oscillatory behaviour of the times-of-arrival of the pulsar signal form strong evidence of a lensing detection with multiple halos.

The multiple-lens simulations demonstrate that the observability of the lensing events is improved by the new distribution of mass.  The period measurements $\Delta P$ display comparatively smaller but far more irregular fluctuations than their single-lens counterparts.  The \mf{} is still near unity in three of the four cases  for most of the observing time, apart from short periods.  These short-lived peaks are too small to be definite indicators of lensing.  Nevertheless, the multiple-lens case is significantly easier to detect than the single-lens case and the two can be readily distinguished.

\begin{landscape}
\begin{figure}
\noindent\makebox[\paperwidth]{
    \subfloat[$b=1$, $N=10$, $T_{\text{obs}}=1\,\text{yr}$]{
        \label{fig:mult-lens:a} 
        \includegraphics[height=\textwidth]{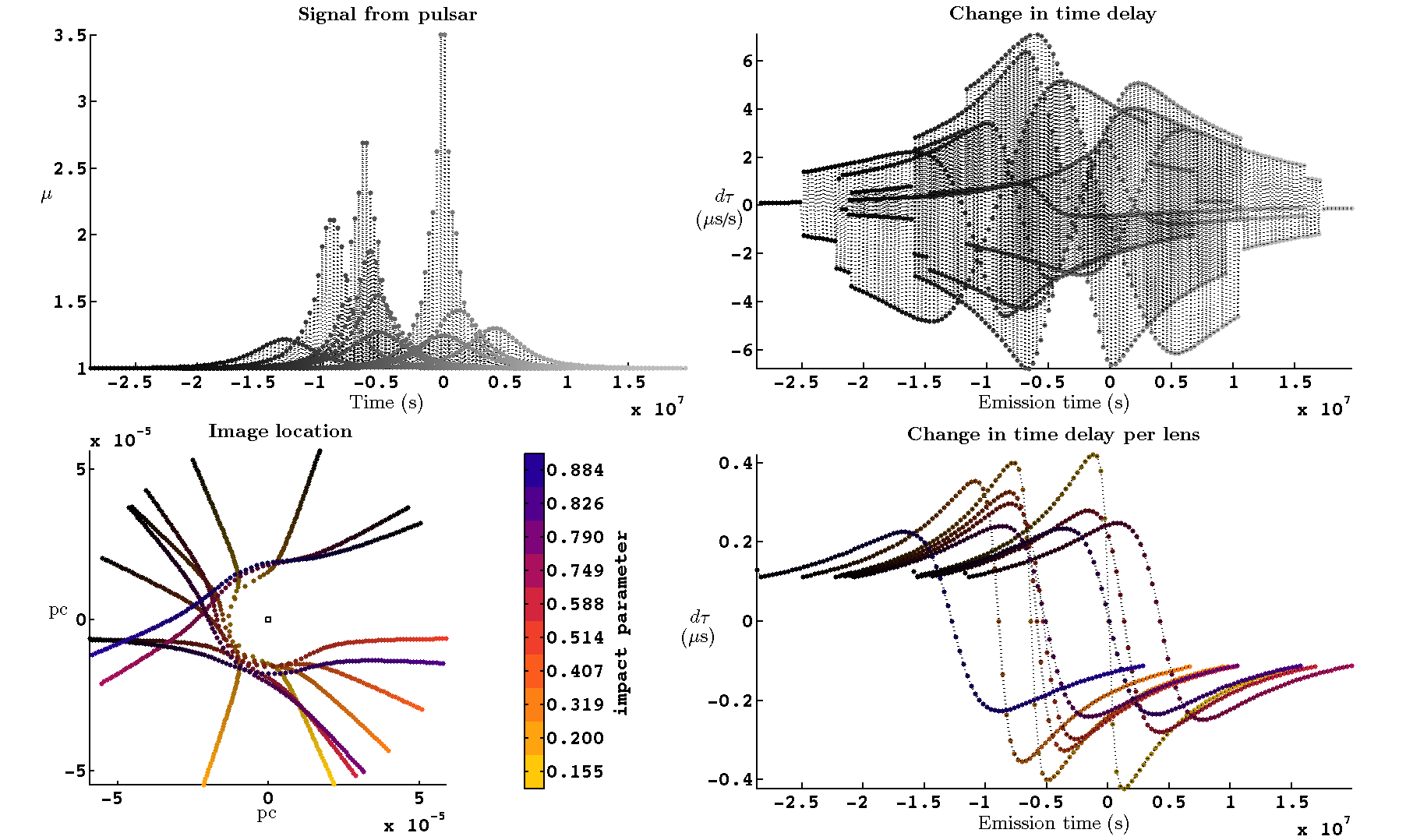}
    } 
}
\end{figure}

\begin{figure}
\ContinuedFloat
\noindent
\makebox[\paperwidth]{
  \subfloat[$b=1$, $N=10$, $T_{\text{obs}}=25\,\text{yr}$]{
      \label{fig:mult-lens:b} 
      \includegraphics[height=\textwidth]{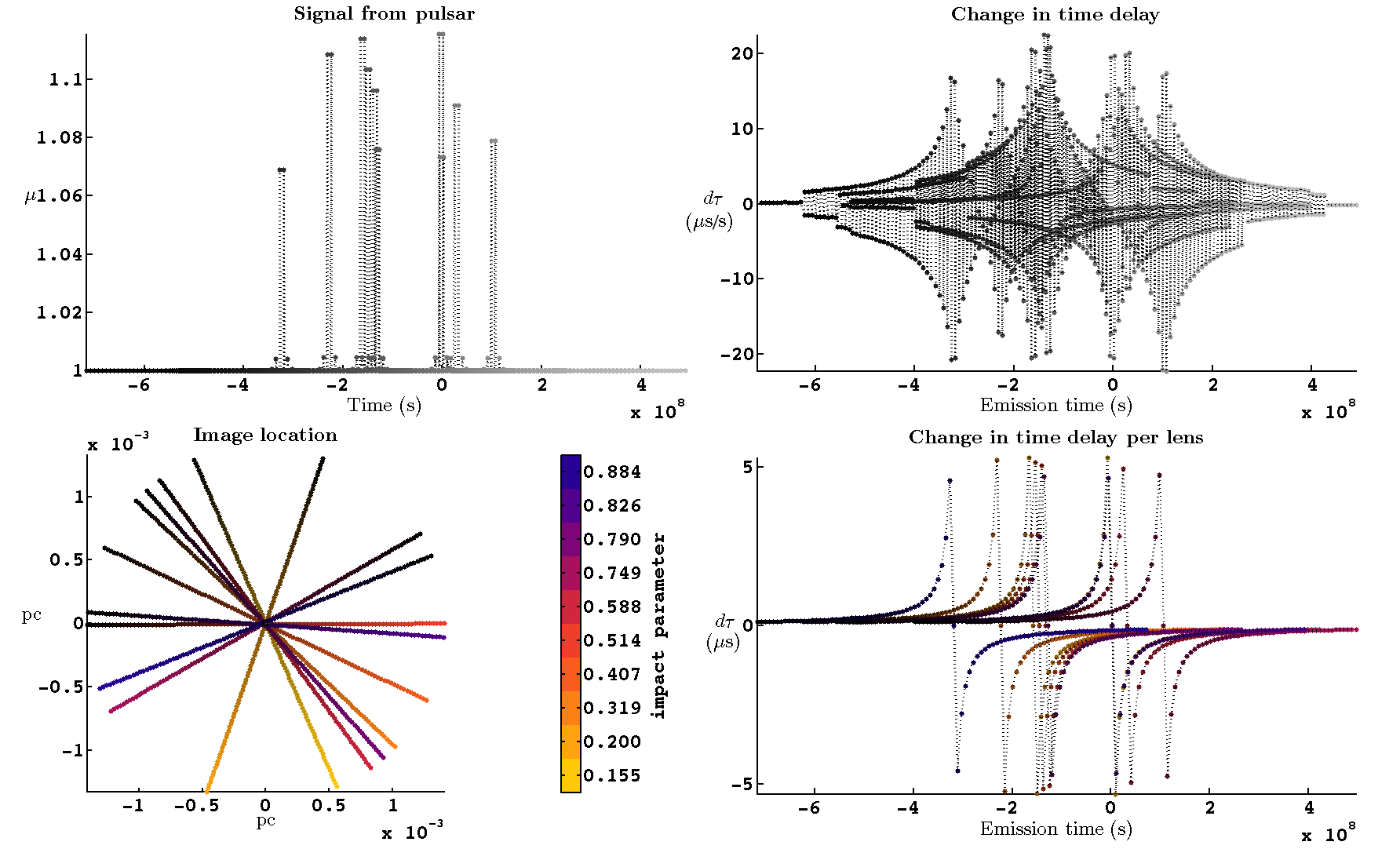}
} 
}
\end{figure}

\begin{figure}
\ContinuedFloat
\noindent
\makebox[\paperwidth]{
  \subfloat[$b=10$, $N=10$, $T_{\text{obs}}=1\,\text{yr}$]{
      \label{fig:mult-lens:c} 
      \includegraphics[height=\textwidth]{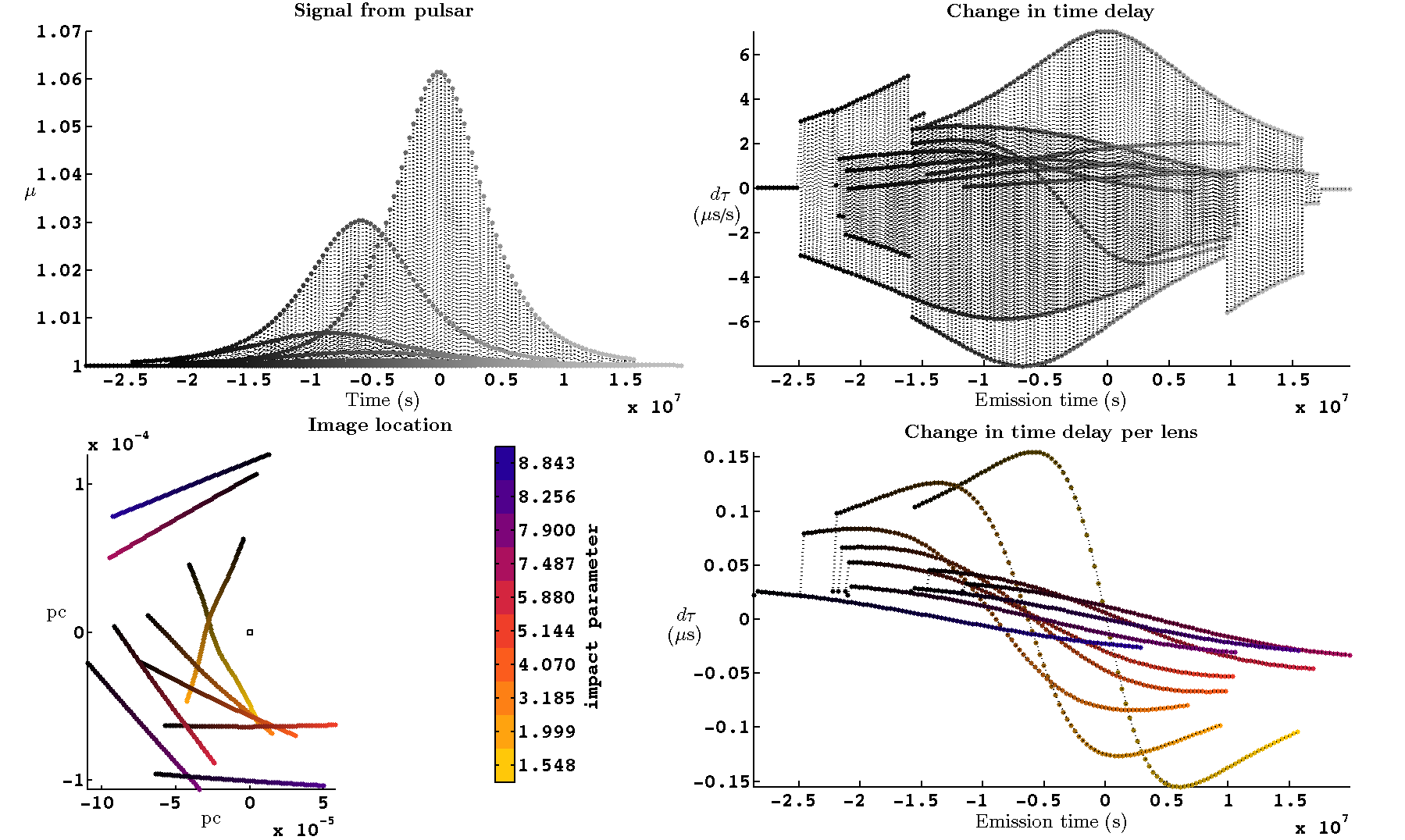}
  }
}
\end{figure}

\begin{figure}
\ContinuedFloat
\noindent
\makebox[\paperwidth]{
    \subfloat[$b=10$, $N=10$, $T_{\text{obs}}=25\,\text{yr}$]{
        \label{fig:mult-lens:d} 
        \includegraphics[height=.9\textwidth]{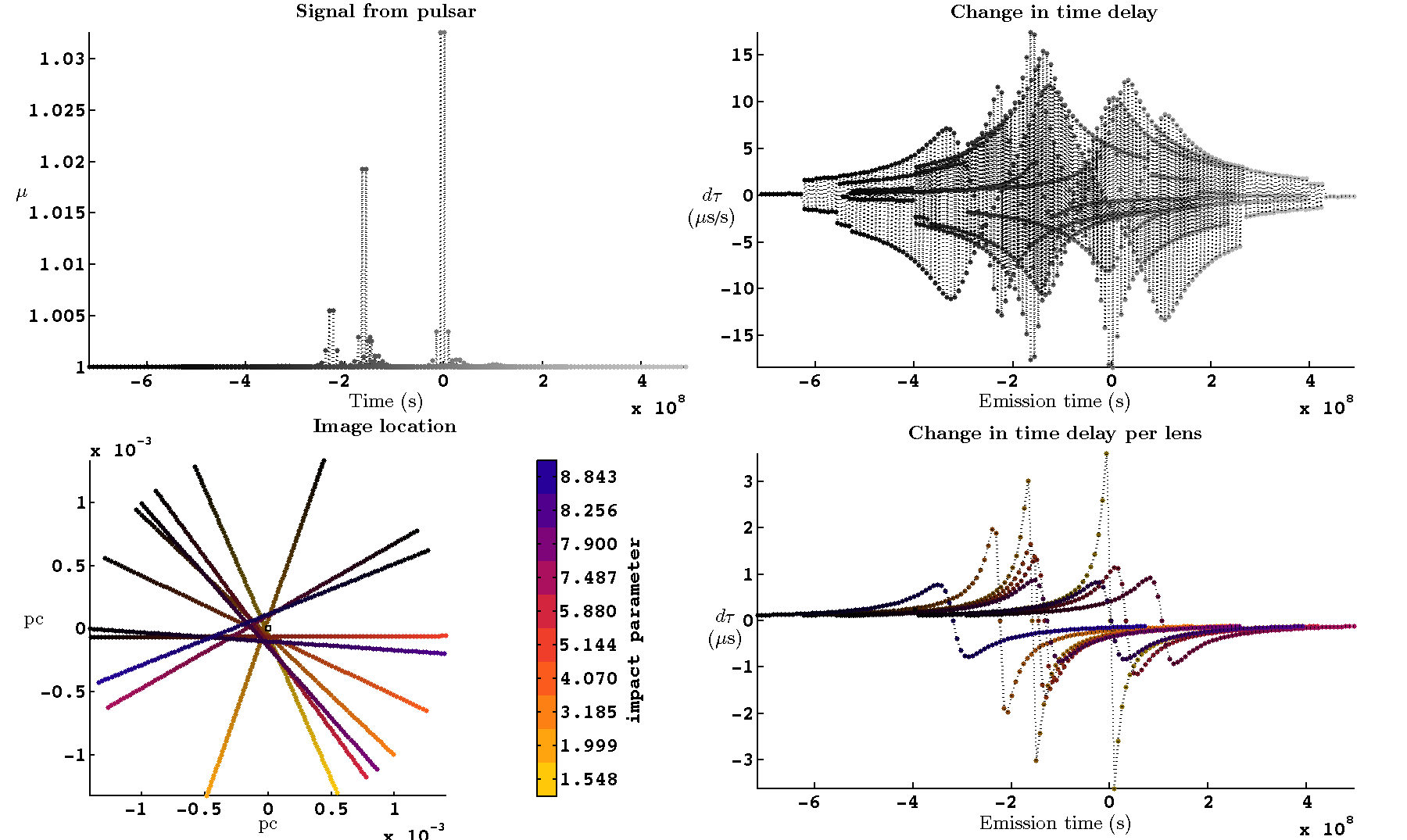}
    }
}
\caption{Example of multiple halos with scale radius $10^{-3}\,\textrm{pc}$ at $5\,\textrm{kpc}$ transiting between source at $10\,\textrm{kpc}$ and observer.  The observations are \textsc{(top left)}: the amplitude of the signal relative to that from the pulsar; \textsc{(top right)} the change in times-of-arrival of the signal.  \textsc{(bottom left)}: the image locations; \textsc{(bottom right)}: the relative time delay.  Later times are indicated by lighter colours.  The lens parameters are given in the sub-captions.}
\label{fig:mult-lens} 
\end{figure}
\end{landscape}

\section{Lenses distributed along the line-of-sight }\label{sec:vardist-lens}

This section introduces another free parameter, allowing the lenses to be distributed between source and observer.  The different distances involved factor into the conversion of the time delay from lens-plane units to physical units.  In order to maximise the effect of variation, the other parameters were kept precisely the same as in the fixed-distance case.

This additional degree of freedom creates two competing effects.  The sub-galactic scale of the pulsar-halo-observer system admits the use of Euclidean distances.  The angular diameter distances are then linear, \ie $D_{\textrm{ds}} = D_{\textrm{s}} - D_{\textrm{d}}$: having fixed $D_{\text{s}}$, we can then introduce a reduced parameter $d \equiv \sfrac{D_{\textrm{d}}}{D_{\textrm{s}}}$.  The time delay scaling is symmetric and non-linear in $d$, whereas the image location scaling is linear in $d$.  These two scaling mechanisms counteract one another in two of the three characteristics of strong lensing.  

\begin{figure}
\includegraphics[width=0.6\textwidth,clip]{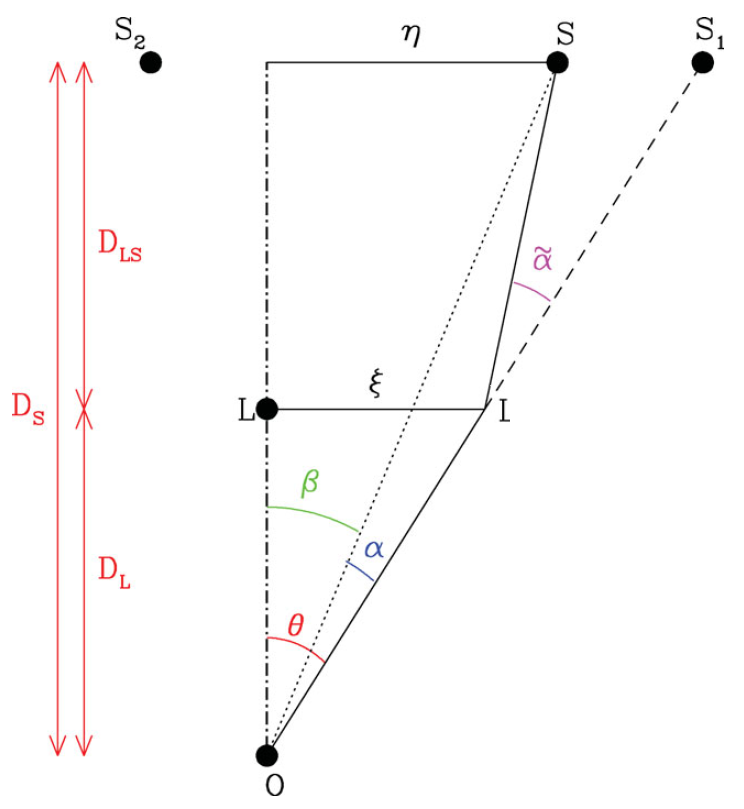}
\caption{Geometry of a typical gravitational lens system. The positions of the observer, source and lens are represented by \lq{}$O$\rq{},  \lq{}$L$\rq{} and \lq{}$S$\rq{} respectively.  The two apparent image locations are denoted \lq{}$S_1$\rq{} and \lq{}$S_2$.\rq{}  The angular diameter distances $D_{\text{L}}$, $D_{\text{S}}$ and $D_{\text{LS}}$ are between observer-lens, observer-source, and source-lens.  
Image credit: Fig.~3 in \cite{gl-bartelmann}
\label{fig:lens-geometry}} 
\end{figure}

The image locations are scaled according to the angular geometry of the lensing system.  Recall from \cref{sec:parameters} that we have set the relative velocity between pulsar and halos to be a constant.  Thus, independent of their angular diameter distance, each \dmh{} moves a fixed transverse distance along the sky.  However, the important quantity in the lensing calculations is not their linear motion, but rather their angular motion.  Returning to the lens geometry (\cref{fig:lens-geometry}), we are interested in the deflection angle and the angular diameter distance between source and image.  The deflection angle between the \lq\lq{}true\rq\rq{} position of the source and its observed image is calculated in the lens plane, not the source plane.  Thus the (transverse) distance corresponding to the deflection angle depends upon the (radial) angular diameter distance.  A cursory examination of \cref{fig:lens-geometry} demonstrates that when a source moves a distance $\eta$ across the sky, the corresponding distance moved by its image is $\xi = d\eta$.  This is reflected in \cref{fig:vardist-lens}, which shows that the transverse motion of the images is no longer equal for all lenses.  Compared to their fixed-distance counterpart, lenses with $d > \sfrac{1}{2}$ produce images which transit more of the sky in the same time period, whereas those with $d < \sfrac{1}{2}$ are compressed.  Images in the $d \approx 1$ limit trace (approximately) the motion of the pulsar, which is not observable directly.  If the individual images were resolvable, comparison of the proper motion of the images would allow estimation of $d$.  In practice, the individual images are separated by at most $\sfrac{a}{D_{\text{s}}} \approx \sfrac{10^{-2} \, \text{pc}}{10^{4} \, \text{pc}} = \mu \text{as}$, well below the resolution of modern radio telescopes.  The image scaling, while observable in principle, is not a useful indicator of lensing in practice.  

The similarities between the magnification in the variable- and fixed-distance models arise from the individual lens behaviour.  A \mf{} approximate to unity, with sharp peaks corresponding to single lenses appears both in \cref{fig:mult-lens} and \cref{fig:vardist-lens}.  Individual images do not incur scaling of their amplitudes.  This is because the intervening lens does not change the surface brightness of the signal (\ie flux per unit area), but merely the area over which the flux is distributed.  The ratio \cref{eq:mu-defn} of infinitesimal areas in the lensed and un-lensed cases is dimensionless.  The  \lq\lq{}numerator\rq\rq{} and \lq\lq{}denominator\rq\rq{} of the Jacobean are the angular distances $\vect{\beta}$ from the optical axis to the source in the source plane and $\vect{\theta}$.  The resulting matrix (and its determinant $\sfrac{1}{\mu}$) is independent of distance.  The overall form of the \mf{} is largely unaffected by the variation in distance, displaying approximately the same magnitude and shape as the fixed-distance case.

The differences between the magnification in the variable- and fixed-distance models is a consequence of the interaction between lenses.  Since the time delays are no longer equally weighted for each lens, the conversion of time delays from lensing to physical units is different for each lens.  This scaling increases the spread in the time delays, altering the probability that successive signals will be superimposed.  When the difference between successive times-of-arrival is on a longer timescale than the timing residual, the signals from different lenses do not superimpose.  Conversely, signals arriving within the timing residual are superimposed into a single signal with amplitude equal to the product of the \mf{}s of each component.  
We see from the time delay plot of the combined lenses that these delays are highly oscillatory functions, so the scaling reduces the likelihood of superposition.  This results in fewer instances of superposition and consequently a lower maximum.  In particular, the two simulations with detectable effects have their maxima greatly reduced: $\mu \approx 1.8$ (\cref{fig:vardist-lens:a}) rather than $\mu \approx 3.5$ (\cref{fig:mult-lens:a}) and $\mu \approx 1.06$ (\cref{fig:vardist-lens:b}) rather than $\mu \approx 1.1$ (\cref{fig:mult-lens:b}).  The remaining simulations, in which the source is never eclipsed, produce small variations in both models, $\mu \approx 1.05$, which are not strong evidence of magnification by a \dmh{}.  Thus, although the magnification for individual lenses is unaffected by the variation in distance, the magnification for the unresolved signal is damped compared to the fixed-distance case.  This behaviour renders the varying magnifications too small to be useful as evidence of lensing, except for a lens which is observed occluding the source.

\begin{figure}
 \includegraphics[width=\textwidth]{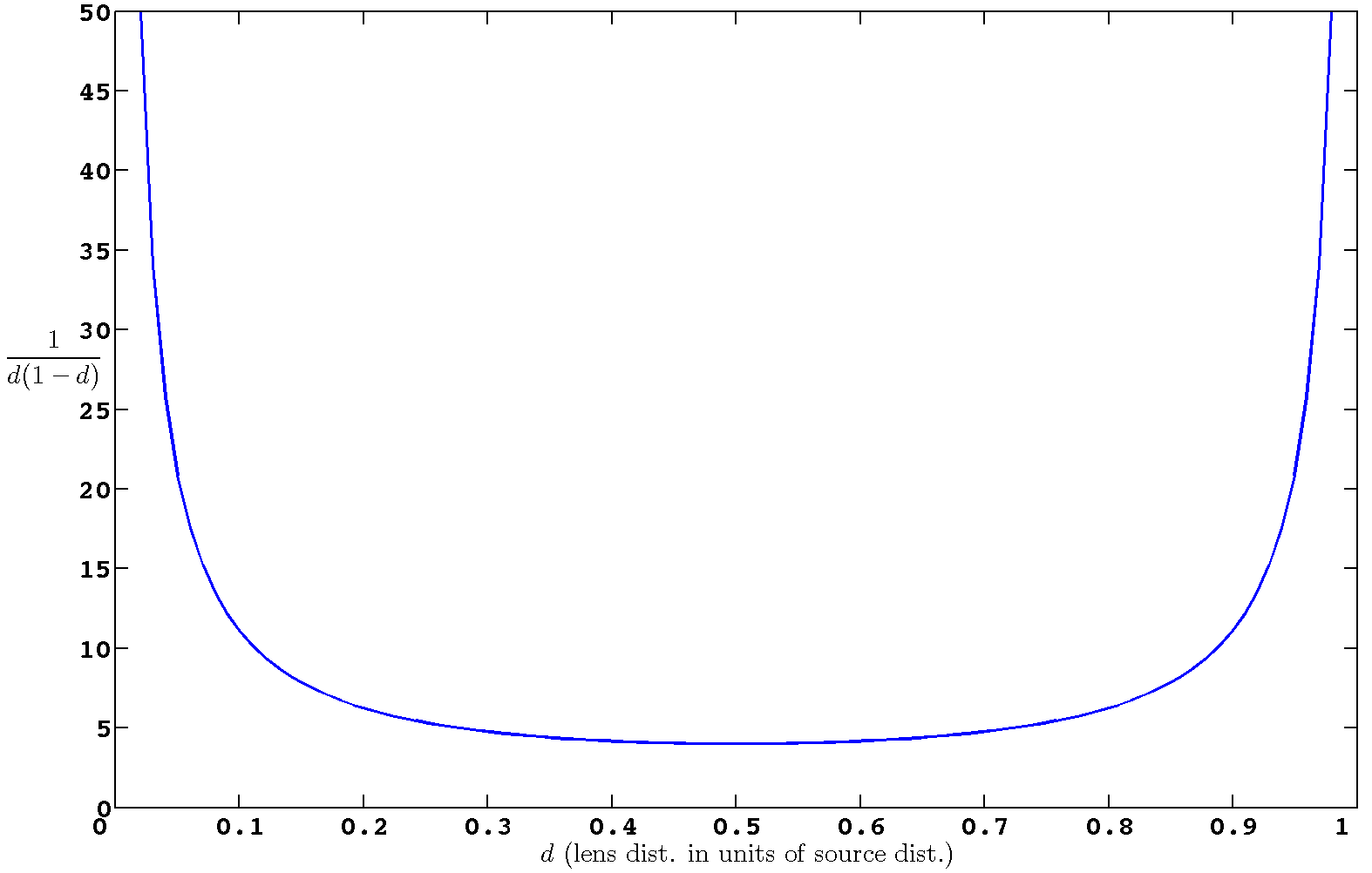}
 \caption{The scaling of a time delay of unity in lensing units according to the fractional distance of the source from the lens which produced the delay.}
\label{fig:scaling} 
 \end{figure}

The variation in distance has two competing consequences for the time delays.  A key difference between the fixed-distance and variable-distance models is the magnitude of the delays for different images.  The physical time delay is calculated from the Fermat potential by:
 \begin{equation}\label{eq:td-scaling}
 \tau = (1 + z_d)\frac{D_{\textrm{s}} \xi_0^2}{D_{\textrm{ds}}D_{\textrm{d}}} \phi
 = \frac{r_0^2}{D_{\textrm{s}}} \frac{1}{(1-d)d} \phi 
 \end{equation}
 where the latter equality holds for the parameters chosen here.  The distribution is shown in \cref{fig:scaling}.  The slowly-varying scaling for $d \mathrel{\substack{\textstyle\in\\[-0.1ex]\sim}} [0.2, 0.8]$ generates similar time delays for lenses with a wide range of distances.  The very steep gradient at the extremes of $d$ generates a very large difference in scaling in even the shortest of differences in distance.  Thus we expect scaling of different orders of magnitude for $d \mathrel{\substack{\textstyle\in\\[-0.1ex]\sim}} [0, 0.1] \cup [0.9, 1]$.  Therefore the lens distribution $N(d) \propto d^2$ produces a single lens at $d_{\text{max}}$ with a far larger scaling than the others and a cluster of lenses over a range of $d$ whose scaling is roughly equivalent to the scaling if the lenses were fixed at $d = \sfrac{1}{2}$.  If all the lenses produced the same time delay in lensing units, their physical time delay would be dominated by one lens.  

That this is not the case is a consequence of the image locations.  The images are produced at the extrema of the time delay surface.  Thus, a different image location caused by a change in $d$ corresponds to a new time delay surface, even when the other lens parameters are the same.  The separation between source and lens alters the value of the geometric time delay $\sfrac{1}{2} \left(\vect{x} - \vect{y} \right)^2$ and the relativistic delay $\kappa(\vect{x}) \ast \norm{\vect{x}}$.  In the limit where the lens becomes infinitely distant, the image location $\vect{x}$ converges to that of the source $\vect{y}$ and the convolution $\kappa(\vect{x}) \ast \norm{\vect{x}}$ must approach zero.  Accordingly, we expect that the $d\approx 1$ images which have a large transverse motion, produce a small time delay in lensing units.  The scaling to physical units amplifies a small quantity, which does not dominate the total delay from all lenses.  Instead, it is comparable to the delays from lenses with $d \lessapprox \sfrac{1}{2}$, which have a large time delay in lensing units (because their images are closest to the source locations), but are not scaled significantly by the conversion \cref{eq:td-scaling}.  

The variations in times-of-arrival $\Delta P$ suggest that the lensing is observable.  This follows from comparison of $\dot{\tau}$ to $\dot{P}$, as in the other simulations.  

The results suggest that the variable-distance model produces the strongest evidence for lensing.  The time-of-arrival variations, when taken in conjunction with the oscillations in the signal amplitude, demonstrate two of the three characteristics of \gl{ing}.  The argument for a lensing detection (as opposed to other causes for the observations) is enhanced by the observational timescales, which are sufficiently large that it is difficult to attribute the behaviour to changes in the pulsar itself, especially when different mechanisms might be required to explain the two phenomena.  The similarity between \cref{fig:mult-lens} and \cref{fig:vardist-lens} indicates that it is difficult to distinguish between multiple lenses at a fixed and varying radial (\ie non-transverse) distance.  Nevertheless, the presence of multiple lenses distributed --- as is likely --- over a range of angular diameter distances, is easily distinguishable from the presence of a single \dmh{}.

\begin{landscape}
\begin{figure}
\noindent
\makebox[\paperwidth]{
    \subfloat[$b=1$, $N=10$, $T_{\text{obs}}=1\,\text{yr}$]{
        \label{fig:vardist-lens:a} 
        \includegraphics[height=\textwidth]{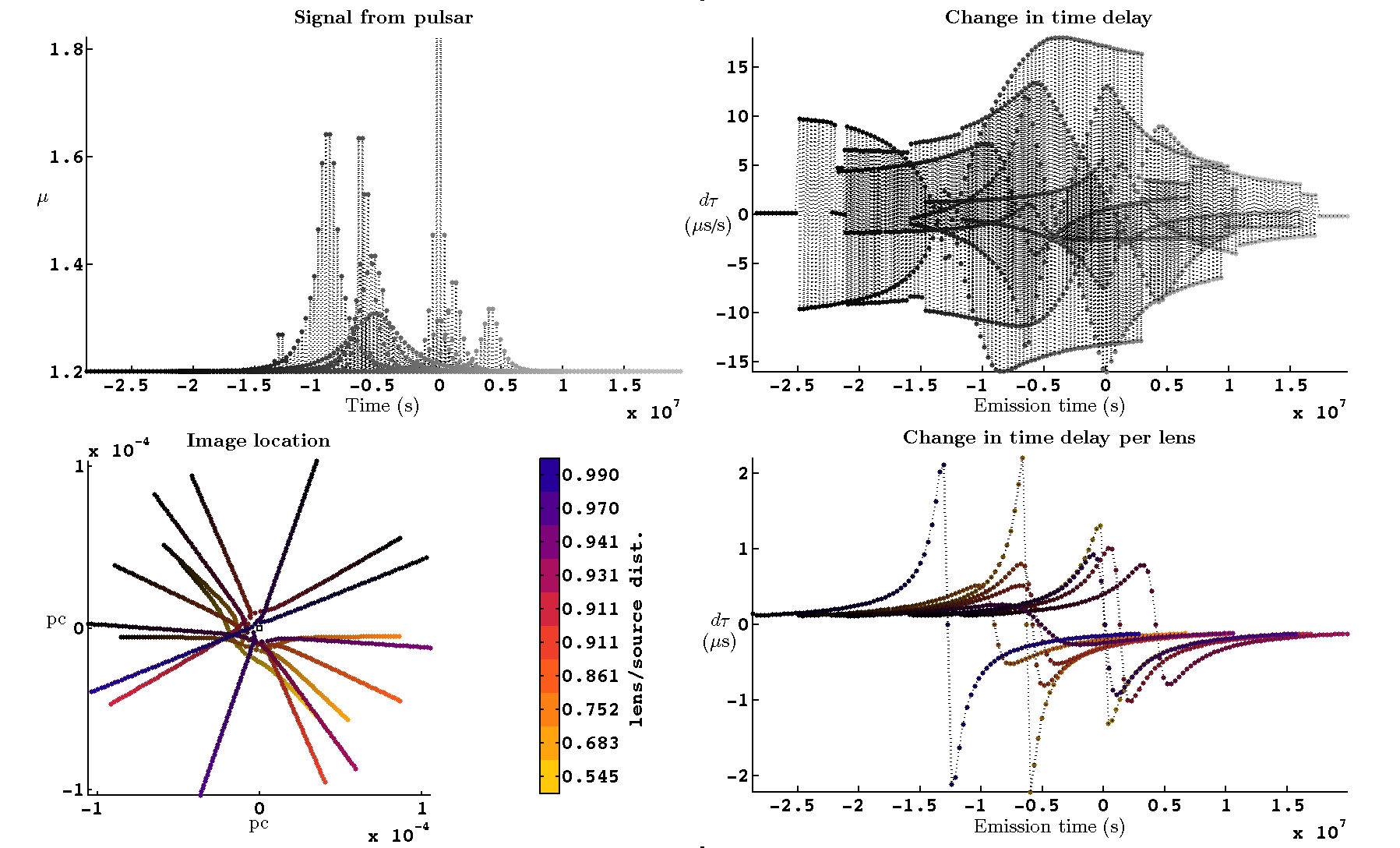}
    }
}
\end{figure}

\begin{figure}
\ContinuedFloat
\noindent
\makebox[\paperwidth]{
    \subfloat[$b=1$, $N=10$, $T_{\text{obs}}=25\,\text{yr}$]{
        \label{fig:vardist-lens:b} 
        \includegraphics[height=\textwidth]{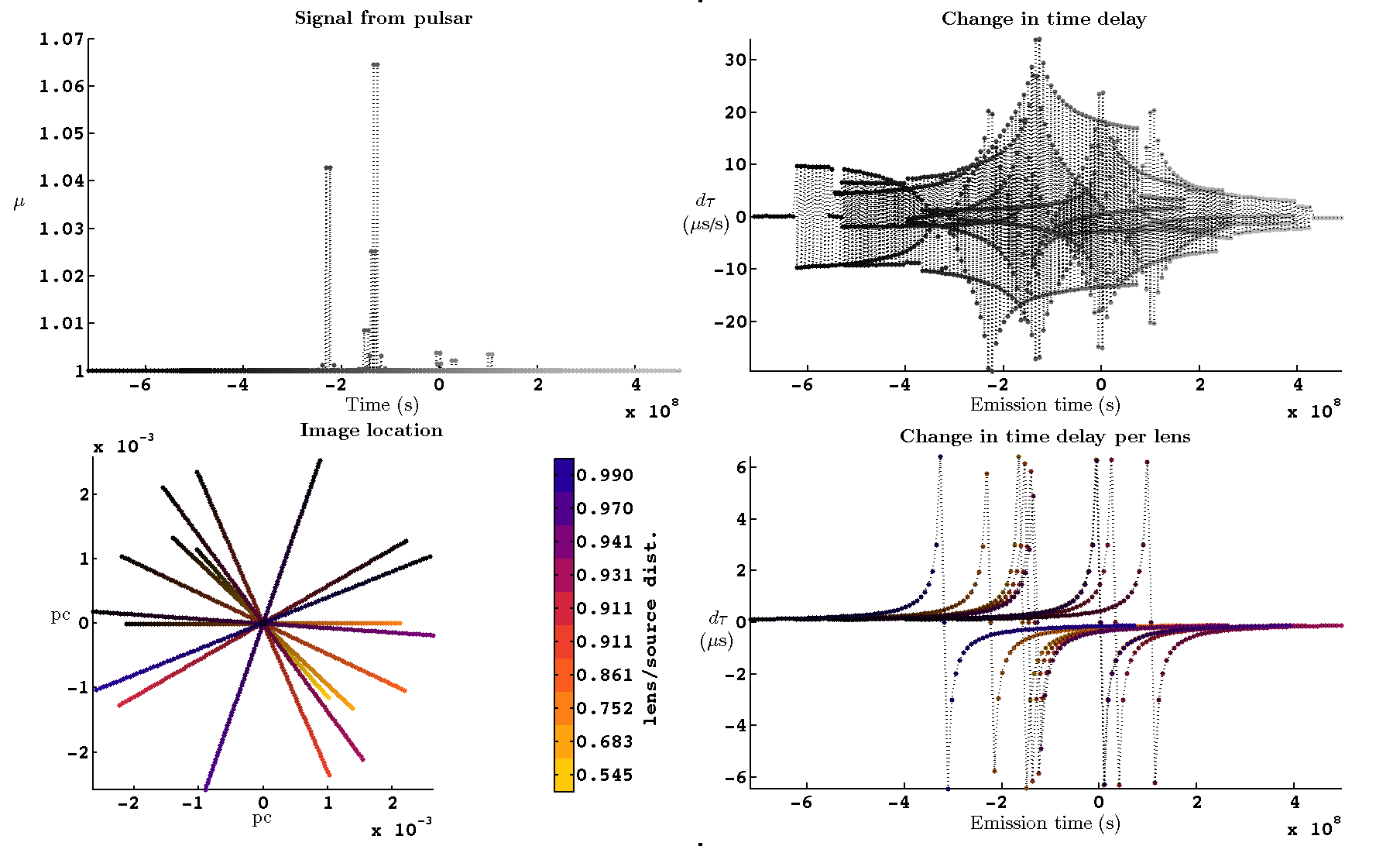}
    } 
}
\end{figure}

\begin{figure}
\ContinuedFloat
\noindent
\makebox[\paperwidth]{
    \subfloat[$b=10$, $N=10$, $T_{\text{obs}}=1\,\text{yr}$]{
        \label{fig:vardist-lens:c} 
        \includegraphics[height=\textwidth]{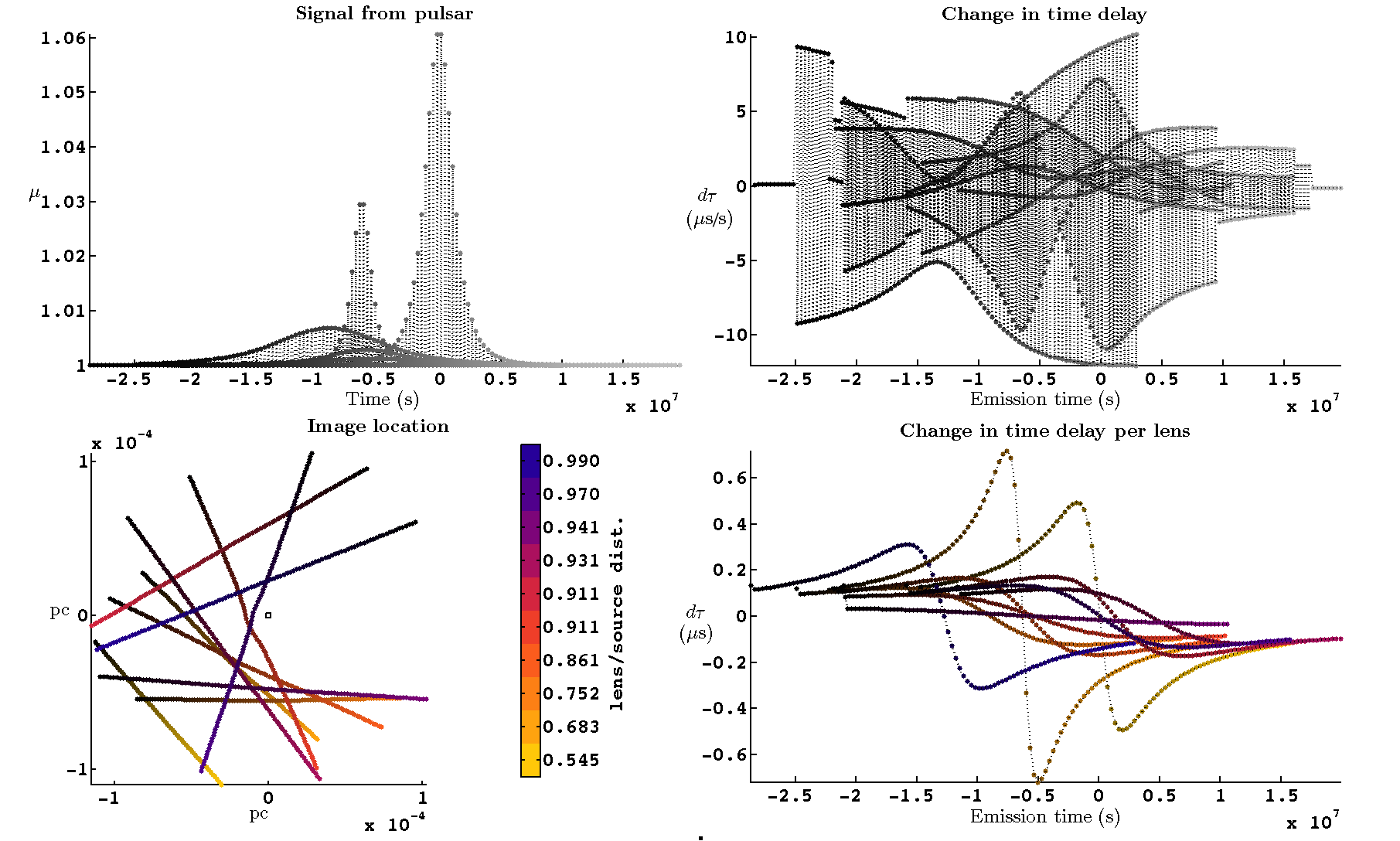}
    }
}
\end{figure}

\begin{figure}
\ContinuedFloat
\noindent
\makebox[\paperwidth]{
    \subfloat[$b=10$, $N=10$, $T_{\text{obs}}=25\,\text{yr}$]{
        \label{fig:vardist-lens:d} 
        \includegraphics[height=.9\textwidth]{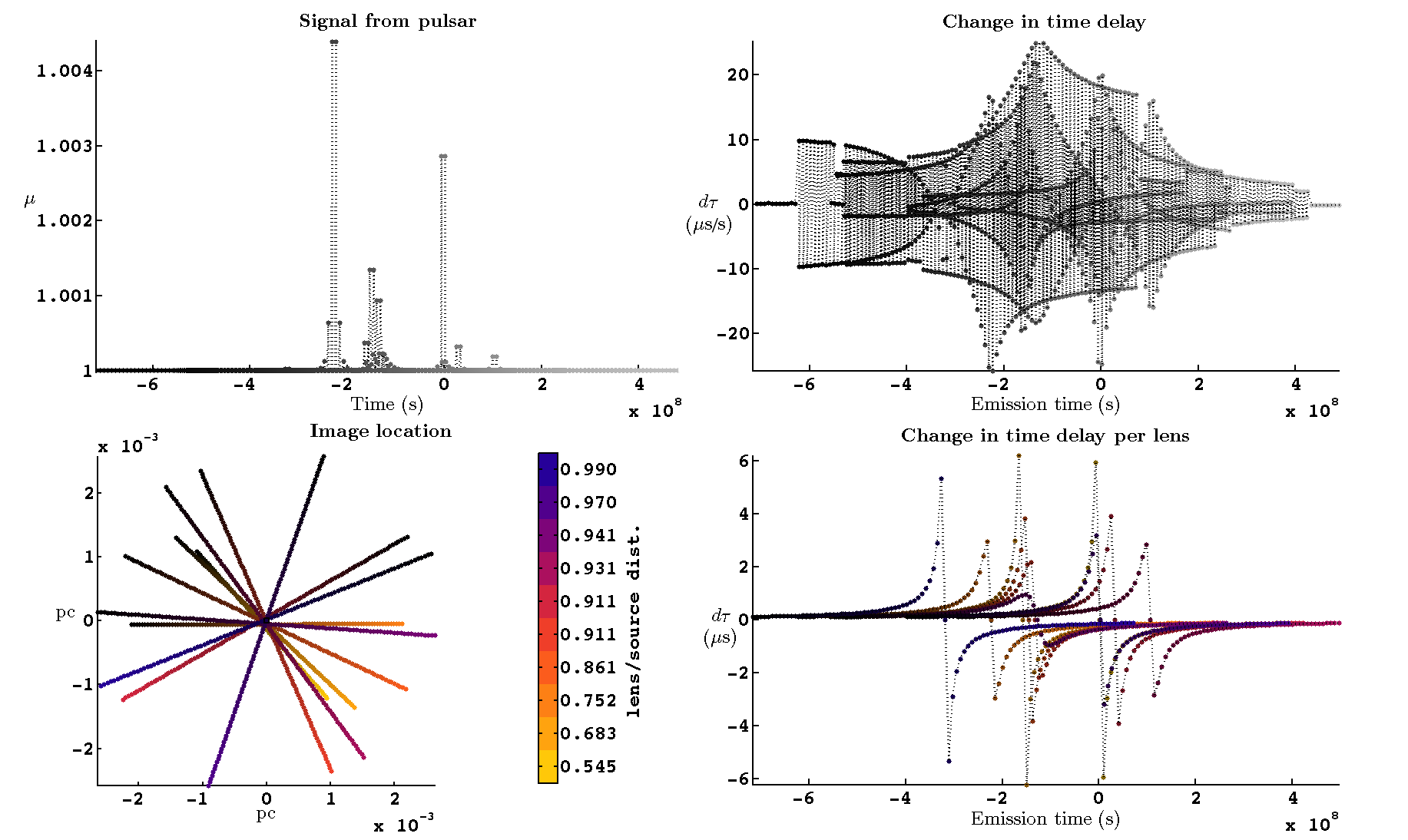}
    }
}
\caption{Example of multiple halos with scale radius $10^{-3}\,\textrm{pc}$ transiting at various distances between source at $10\,\textrm{kpc}$ and observer.  The observations are \textsc{(top left)}: the amplitude of the signal relative to that from the pulsar; \textsc{(top right)} the change in times-of-arrival of the signal.  \textsc{(bottom left)}: the image locations; \textsc{(bottom right)}: the relative time delay.  Later times are indicated by lighter colours.  The lens parameters are given in the sub-captions.}
\label{fig:vardist-lens} 
\end{figure}
\end{landscape}

\section{Discussion}\label{sec:discussion}
In this section we compare these results to those from the \swz{} lens most commonly used in the literature and consider the probability of observing a geometrically-lensed pulsar signal.

Comparison of the \nfw{} and \swz{} results is of interest because the \swz{} lens is prevalent in the literature.  In the case of papers with \mpl{} (\eg \cite{multiple-einstein-rings}, the point-mass lens highlights effects due to the presence of multiple lenses rather than effects of the individual lens geometries.  Alternatively, pulsar papers involving only a single lens (\eg \cite{psr-detection-unlikely,siegel2,psr-krauss-small} are motivated by relatively simple expressions for the period (and period derivative) contributions which can then be compared to the data to an order of magnitude.  In addition, any lens-source interaction with $x_{\textrm{max}} \ll b$ should behave asymptotically towards the point-mass lens case, despite the lens having a radial extent $x_{\textrm{max}}$ which is finite.

The most significant difference is in the number of images.  The \swz{} lens always produces two images \cite{multiple-einstein-rings,catastrophe,petters,sef}, whereas in our results the \nfw{} lens always produces a single image.  Hence we are faced with several questions:
\begin{samepage}
\begin{center}
\begin{minipage}[][][c]{.8\textwidth}
\begin{enumerate}
\item{Why is there always only a single image in our results?}\label{item:image}
\item{How can this be reconciled with the two images produced by the \swz{} lens?}\label{item:swz}
\item{Does this contradict the assumption that the point-mass lens is a practical approximation to a radially-extended lens?}\label{item:approx}
\end{enumerate}
\end{minipage}
\end{center}
\end{samepage}

Having already considered \cref{item:image} in \cref{sec:single-lens}, we turn to \cref{item:swz}.  Recall \cref{tab:lens-models} that the lens equation in the \swz{} case is invertible:
\begin{align}
y 
&= x - \frac{1}{x} 
\implies x_{+,-} = \frac{1}{2} \left( y \pm \sqrt{y^2 + 4} \right)
\intertext{which produces one image on each side of the lens \cref{fig:plot-summary}, with \mf{}}
\mu(x) 
&= \left( \left( 1 - \frac{m(x)}{x^2} \right) \left( 1 + \frac{m(x)}{x^2} - 2\kappa(x) \right)  \right)^{-1} 
= \left( 1 - \dfrac{1}{x^4} \right)^{-1}
\intertext{Substituting the image locations:}
\mu(x_{+,-})
&= \pm \frac{1}{4} \left[ \frac{y}{\sqrt{y^2 + 4} } + \frac{\sqrt{y^2 + 4} }{y} \pm 2 \right]
\end{align}
The two images are a positive-parity image $x_{+}$ near the lens and a negative-parity image $x_{-}$ near the source.  At the limit as the impact factor approaches infinity, $x_{+}$ approaches the true lens position with magnification $\mu_{+} = 1$; the other image is demagnified $\mu_{-} = 0$ as $x_{-}$ approaches the source.  Taking the opposite limit, when lens and source are aligned (\ie $b=0$), the two images are equidistant and form an Einstein ring.  The \mf{} theoretically approaches infinity, but in fact has a maximum of $\sfrac{\sqrt{4+R^2}}{R}$ for a source of radius $R$ \cite{sef}.  Accordingly, at some intermediate impact factor the second image becomes negligible due to demagnification, with the exact details depending on the sensitivity of the observing telescope.  The production of an even number of images is a result of its convergence being a delta function.  If the lens becomes an \hdisc{} lens of finite radius, an odd number of images are formed.  Thus, the \swz{} lens is an exception to the Odd Number Theorem.

The approximation \cref{item:approx} is acceptable, having resolved the apparent contradiction between \cref{item:image} and \cref{item:swz}.  In practice, it is only the positive-parity image which is resolved, unless the image and source are close to alignment (for $y \lesssim 1$ the images are approximately equal in brightness).  Indeed, it is precisely this argument which leads \cite{psr-detection-unlikely} to ignore the pulsar signal produced by the negative parity image throughout the paper.  This suggests that the trade-off for having an invertible lens mapping with analytical time delay is the assumption that the lens is extremely compact.  While Einstein rings have been observed (\eg \cite{walsh} and indeed multiple Einstein rings, \cf \cite{multiple-einstein-rings}), they are in an extragalactic context with either $D_{\textrm{d}} \ll D_{\textrm{ds}} \approx D_{\textrm{s}}$ or $D_{\textrm{ds}} \ll D_{\textrm{d}} \approx D_{\textrm{s}}$.  In such a context, the physical radius of the lens is several orders of magnitude less than the scaling lengths in the source and lens planes, so it is scaled to zero in the lensing geometry.  In contrast, the galactic lensing geometries have far less severe scaling of the physical radius of the lens.  Thus the only possibility that suits the point-mass approximation is a large impact factor.

The image-counting using the \nfw{} lens suggests that it is superior to the \swz{} model even at large impact factors, as the latter forces one to explicitly discount a root of the lens equation, whereas the former naturally produces realistic behaviour.

To date there is only a single pulsar observation attributed to \gl{ing}: \cite{psr-detection} propose that time-of-arrival distortions in the pulsar PSR~B0525+21 from 1968 to 1983 were caused by lensing from a $330 \, M_{\odot}$ black hole.  The original paper \cite{psr-detection} claims that the timing residuals have significant behaviour similar to that of a Shapiro time delay, which leads them to estimate the mass of a \swz{} lens which generates the best fit to the data.  In contrast, \cite{psr-fargion} suggest that the optical depth is far too small and \cite{psr-detection-unlikely} states (without proof) that the signals are not well-fitted by the expected delay curve.  Instead, he argues that the density of intermediate mass black holes is too low (using stellar matter as a proxy, $\sim 0.1 M_{\odot}\,\text{pc}^{-3}$) for such an observation to be probable on human timescales (a probability of $\sim 3 \times 10^{-5}$).  Thus, a detection of lensing has not been confirmed: while there is little uncertainty over the timing residuals themselves, their appropriate interpretation remains unresolved.

The probability of observing a gravitationally-lensed pulsar is not a well-constrained estimate.  The initial proposal by \cite{psr-krauss-small} found that a \lq\lq{}non-negligible probability\rq\rq{} of discovering a lensing event was possible with only $\sim 10^3$ pulsars catalogued within the Galaxy \cite{psr-krauss-small}.  (In fact, the authors note that the finite length of the time delay signal increases the probabilities from the \lq\lq{}raw\rq\rq{} estimates.)  Furthermore, \cite{psr-fargion} have an even more optimistic value of $\sim 500$ pulsars required for the lensing observation probability to approach unity.  These differing \lq\lq{}optimistic\rq\rq{} estimations are caused by different models for the distribution of matter within the Galaxy, namely that derived from the \lq\lq{}Bahcall-Soneira\rq\rq{} luminosity function  and a double exponential model respectively.  The \lq\lq{}pessimistic\rq\rq{} prediction of \cite{psr-detection-unlikely} is generated by simulation of a pulsar with velocity $1\,000\,\text{km}{s}^{-1}$ and $10^6$ solar-mass stars in a $0.1 \times 0.1 \times 1\,\text{kpc}$ box: it does not include any dark matter and uses a large relative velocity for the pulsar, in contrast to the other papers.  

\begin{figure}
\centering
\includegraphics[width=0.75\textwidth]{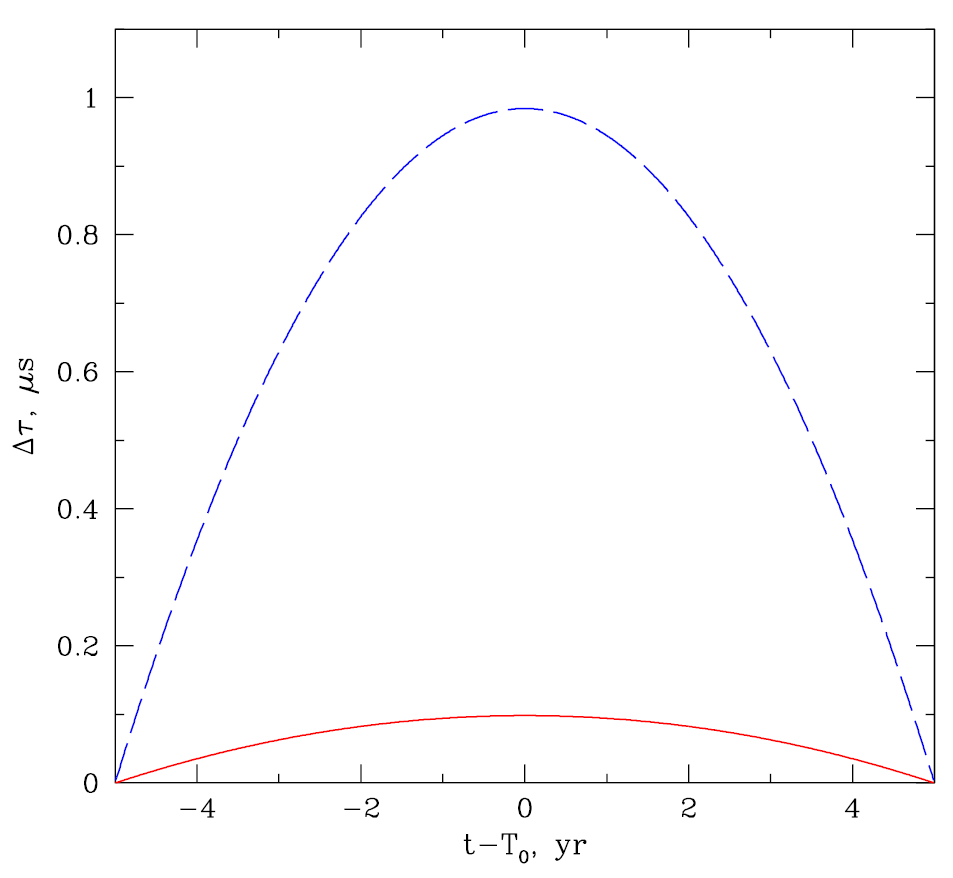}
\caption{Time delay curves for a pulsar lensed by an intermediate mass black hole of mass $10^3 M_{\odot}$ (dashed line) and $10^4 M_{\odot}$ (solid line). [Fig. 3 from \cite{psr-bh-detection}]}
\label{fig:imbh-lensing}
\end{figure}

The situation is even more different within globular clusters.  Given the high concentration of pulsars within globular clusters (as of 2006, 129 pulsars have been catalogued within 25 globular clusters \cite{psr-globular-clusters}), they are an ideal location to begin searching for lensed pulsar signals.  Following the calculations in \cite{psr-krauss-small}, \cite{psr-globular-clusters} obtain the probability for a Shapiro-like time delay detection for a pulsar at the centre of various globular clusters.  These estimates are more optimistic still, compared to those for a pulsar in the galaxy itself \cref{tab:gc-events}.  However, a follow-up paper \cite{psr-bh-detection} concludes that even the (proposed) intermediate-mass black holes at the centre of the globular clusters will not produce a detectable lensing event \cref{fig:imbh-lensing}.  This is primarily caused by the greater impact parameters involved, which are significantly larger than the Einstein radii of the black holes.  This demonstrates how the lensing geometry dominates the observation probability.  Even in the simplest lensing scenario, the observational predictions for \gl{ing} of \msp{s} vary greatly.

\begin{figure}
\centering
\begin{tabular}{*{4}{l}}
\toprule
       & \multicolumn{2}{c}{Probability ($\text{yr}^{-1}$)} & Events\\
       \cmidrule(r){2-3}
       & Galaxy & Cluster & $(5\,\text{yr})^{-1}$ \\
\midrule
M 15   & $1.15\times10^{-3}$    &  $3.4\times10^{-3}$  &  0.18 \\
47 Tuc & $5.44\times10^{-4}$    &  $7.6\times10^{-4}$  &  0.14 \\
Ter 5  & $1.05\times10^{-2}$    &  $4.8\times10^{-3}$  &  2.45 \\
\bottomrule
\end{tabular}
\captionof{table}{Lensing detection rates for a pulsar in various globular clusters.  The lens is \textsc{(left)} in the galactic disc, bulge or halo, or \textsc{(centre)} within the cluster.  \textsc{(right)}: the number of events observed over a five-year period. \cite{psr-globular-clusters}}
\label{tab:gc-events}
\end{figure}

Fortunately, observation habits need not be changed to improve the likelihood of a detection.  Typically, pulsars are surveyed such that observations of the same pulsar are a few weeks apart \cite{ipta}.  Given the assertion in \cite{psr-uncertainty} that potentially-lensed pulsars require constant observing due to the transience of lensing events, then the lack of (firm) lensing detections is inevitable.  However, even for the \swz{} lens the overall time delay signal (the characteristic bell shape) occurs over a matter of years.  Adopting the \nfw{} model, we have seen in \cref{sec:single-lens} that the gaps between observations facilitate the lensing detection.  Therefore, not only is it possible to examine already-reduced data for lensing signals, but also future data recorded for other purposes --- particularly gravitational wave detection --- will be easily analysed for lensing signals.  This maximises the possibility of detecting lensing events.

\chapter{Conclusions}\label{ch:conclusion}
This thesis illustrates a method of \dmh{} detection on astrophysical scales via the halos' gravitational lensing effects on \msp{s}.  While the use of gravitational lensing phenomena --- namely time delays, multiple images and image magnification --- in the detection of dark matter is not new, this thesis combines a number of previously disparate elements.  The halo model uses a realistic \nfw{} profile rather than idealising the halos as point masses, the relativistic time delays are calculated using Hankel transforms to take full advantage of the spherical symmetry of the problem and this efficiency allows multiple lenses ($\sim 10^2$) to be included in a single simulation. 

\section{Summary of results}\label{sec:conclusion}

The results discussed in \cref{sec:single-lens}--\cref{sec:vardist-lens} suggest that \dmh{s} have a characteristic lensing signature.  Although no multiple images were produced, the single image is in accordance with the Odd Number Theorem.

The other characteristics of lensing --- namely magnification of the images and time delays --- are seen in the simulations.  These effects do not always generate observational signatures.  All of the simulations suggest that the strongest evidence for a lensing detection is variation in the signal times-of-arrival.  The time delay imposes a far larger variation on the pulsar period than the astrophysical properties of the pulsar.  Furthermore, transient effects can be discarded as an alternative explanation because the variations occur over the entire observing period, on the order of years.  The magnification effects due to lensing are not a useful indicator unless a single lens is observed for $\sim 25\,\text{yr}$ or there are multiple lenses with impact factors comparable to the lens radius.  In the remaining cases, namely a single lens observed for $\sim 1\,\text{yr}$ or multiple lenses with $\bmax \approx 10$, the magnitude of the variations are too small to be decisive.  The simulations show that the presence of \dmh{s} can be inferred from their lensing effects. 

It is possible to distinguish between the presence of a single halo and multiple halos using the smoothness of the time-of-arrival variations.  However, using either the \mf{} or time delays, it is difficult to determine whether the halos are at a fixed distance or distributed along the \los{}.

Current observation routines are sufficient to produce a detection.  Continuous monitoring (as suggested by some authors) is unnecessary: maintaining the current dictum of observing every few weeks produces detectable results.  While a longer observation period of $\sim 25 \, \textrm{yr}$ is preferable due to the increased proper motion of the lens (relative to its impact factor and radius), a shorter period of a year produces useful results.  Thus we have demonstrated that lensing from \dmh{s} produces observational signatures difficult to attribute to other causes and does so over human timescales.

\section{Generalisations of the method}\label{sec:general}
The method can be generalised in three major ways: the properties of the lenses can be complexified, the distribution of the lenses can be altered or the processing of the lensing simulations can be altered.

Change to the lens profile is motivated by the fact that a suitable density profile for low-mass \dmh{s} ($M \lesssim 10^{6} \, \textrm{M}_{\oplus}$) remains unknown.  Computational power has evolved sufficiently that it is no longer necessary to use analytically tractable models such as the \swz{} or \hdisc{} lenses in most lensing situations (notable exclusions being \mpl{} and microlensing simulations).  Rather, it may be more useful to examine modifications to the \nfw{} model.  Currently these are proposed for galaxy and cluster halos as these were the scales upon which the \nfw{} model was originally generated.  Considering its now-ubiquitous use, it is not unreasonable to hypothesise that extensions to the profile may be useful at the low-mass end of the \dmh{} spectrum.  It is also possible, as discussed in \cite{sef}, to approximate elliptical lensing profiles by multipole expansion of radially symmetric terms.  Three-dimensional lens models (\ie those not adhering to the thin-lens approximation) can be modelled by projecting the radial density distribution $\rho(\vect{r})$ onto the lens plane $\vect{\xi}$ to obtain the convergence $\kappa\left(\sfrac{\vect{\xi}}{\xi_0}\right)$, as was performed for the \nfw{} profile.  These three possibilities for expanding the lens model reflect the lack of an empirical density profile for \textsc{macho}s.

The lens distribution can be drawn from a different probability distribution function.  The fractional lens-observer distance $d = \sfrac{D_d}{D_s}$ is a crucial component of the image behaviour and the time delay scaling.  Since the scaling factor of the images is $d$ and the time delays is $\sfrac{1}{d(1-d)}$, it may be desirable to emphasise these conflicting effects by re-distributing the \dmh{s}.  Alternatively, the distribution can be altered to reflect a change in the mass profile of the Galaxy.  (The reason for drawing the lens distribution from $\textrm{pdf}(d) \propto d^2$ was the assumption that the Milky Way followed an \nfw{} profile.)  This extension is a minor one which would only be of interest once a comparison to observation could be made.

The most significant improvement in the method would be a refinement of the signal-producing code.  Modern pulsar observations absorb linear and quadratic time delay terms into the uncertainties for the period and its differential respectively \cite{psr-uncertainty}.  Consequently, this should be reflected in the simulations before any firm conclusions can be drawn on whether or not this effect can be practically observed.  Using a point-mass lens, Siegel concludes that \msp{s} are useful probes of the dark matter present in the Galaxy \cite{siegel2}.  It is natural to ask whether a different lens profile or multiple lenses would alter this forecast.  However, this cannot be done rigorously without the subtraction of the best-fit quadratic from the times-of-arrival.  (The alternative is to develop new techniques for pulsar analysis when specifically searching for gravitational lensing effects.)  Such a modification is the most important further work arising from this thesis.

\section{Open questions}\label{sec:open}
There are three open questions which also arise from this thesis: can it be extended to extragalactic sources; on what grounds is the omission of \mpl{} justified, or even necessary; and whether the realism of the \nfw{} profile warrants the additional complexity.

The extragalactic application of this method is unlikely at the present time.  The reason for this is that a small percentage of detected pulsars are \msp{s}.  The total pulsar population within the Galaxy is estimated to be $2 \times 10^{5}$, comprising $\sim 40\,000$ \msp{s} and $\sim 160\,000$ normal pulsars \cite{psr-summary}.  Only a tiny fraction of this expected amount have been found due to technological limitations \cite{ipta,ppta} and selection effects (some general, \eg Malmquist bias and others specific to pulsar surveys \cite{psr-summary}).  Figures from the \textsc{psrcat} Pulsar Catalogue\footnote{version 1.59 can be found at: \texttt{http://www.atnf.csiro.au/people/pulsar/psrcat/}}
show that 12 of the $2\,193$ listed puslars fall into the millisecond category, \ie periods of $P \leq 2 \,\textrm{ms}$ and period derivatives of $\dot{P} \leq 1\,\mu\textrm{s}$.  Comparatively, 21 are extragalactic (\ie $D \geq 50\,\text{kpc}$) \cite{psrcat}.  Assuming that the two properties are uncorrelated, $19\,084$ pulsars would have to be surveyed before one might expect an extragalactic \msp{} to be discovered.  Such large surveys require next-generation radio arrays such as the \textsc{ska}, which will also have sufficient sensitivity to probe the Large and Small Magellanic Clouds.  Current pulsar surveys can also be examined for signals of gravitational lensing.  In particular, pulsar surveys optimised to detect gravitational waves \eg the Parkes Pulsar Timing Array, facilitate this by providing high-precision data on the times-of-arrival of pulsars distributed over the sky \cite{ppta}.  The gravitational waves and gravitational lensing effects are quite distinct \cite{siegel1}, particularly in the quadrupole effect induced by a gravitational wave \cite{ipta}, so the two can be distinguished from one another.  Thus, the chance of probing the \dmh{} structure of nearby extragalactic objects is unlikely now but highly probable in the near future.

The justification to avoid multi-plane lensing in a lensing geometry with multiple halos is somewhat contentious.  This is a pragmatic rather than a scientific  simplification, motivated by the assumption that the additional computing time and memory requirements outweigh the benefits of a more accurate simulation.  Unfortunately, this cannot be confirmed without directly implementing the recursive  \mpl{} equations \cref{tab:mpl}.  The lensing of images by other images raises the possibility of more multiple images than are detected in the straightforward case (\cf \cite{multiple-einstein-rings} for a two-lens example).  While there are mathematical possibilities to place limits on the number of images produced in \mpl{} (\eg via Morse\rq{}s theorem), this remains a complicated problem \cite{petters}.  Currently, application of \mpl{} to the method demonstrated herein would be better done using a \swz{} profile, which has closed forms for the key formulae \cref{tab:lens-models}, than the \nfw{} model. 

The most promising open question is whether the \nfw{} lenses can be differentiated from the \swz{} results.  Were this false, the simulations could be greatly improved because the point-mass time delay has an analytical form \cref{tab:lens-models}.  Were this true, it would enable sample observations to be compared over a spectrum of \nfw{} parameters, to determine where in the parameter space $(M, \rho_s, r_s)$ the \dmh{s} would lie.  (The \swz{} profile can be considered as the limiting case of a \nfw{} profile as the scale radius approaches zero.)  A key factor in this comparison is the demagnification of secondary images, which may cause one of the two images produced by the \swz{} lens profile to be demagnified below the observational threshold.   Investigating whether the realistic model is observably different to the maximally-simplified model is the most useful further work in this thesis.

In conclusion, I have demonstrated that it is possible to simulate the effect of multiple \dmh{s} transiting between Earth and a \msp{} in an efficient manner.  I reviewed the motivation for dark matter and summarised the plausible candidates, the breadth of which necessitates the use of \gl{ing} to detect all possibilities.  Subsequently, I presented the principles of \gl{ing} in the case of a single lens and how axial symmetry facilitates the computations.  In particular, I showed that the relativistic portion of the time delay simplifies from a two-dimensional integral into a one-dimensional Hankel transform.  I developed a simulation for multiple lenses with realistic properties for both source and lens and their distribution within the Galaxy.  The results suggest that pulsar timing can be used to detect \dmh{s} using current telescopes.  Therefore, the method illustrated by this thesis is an efficient and practical way in which to probe that dark matter content of the Galaxy. 

\printbibliography[]

\appendix
\cleardoublepage\chapter{Multiplane lensing}\label{ch:mpl}

The major simplification in this thesis was the assumption that the effects of each lens were independent of the others.  This was necessary to reduce the computational requirements.  A brief explanation of \mpl{} is necessary to appreciate the full complexity of the problem.  

\begin{figure}
\centering
\includegraphics[width=0.5\textwidth]{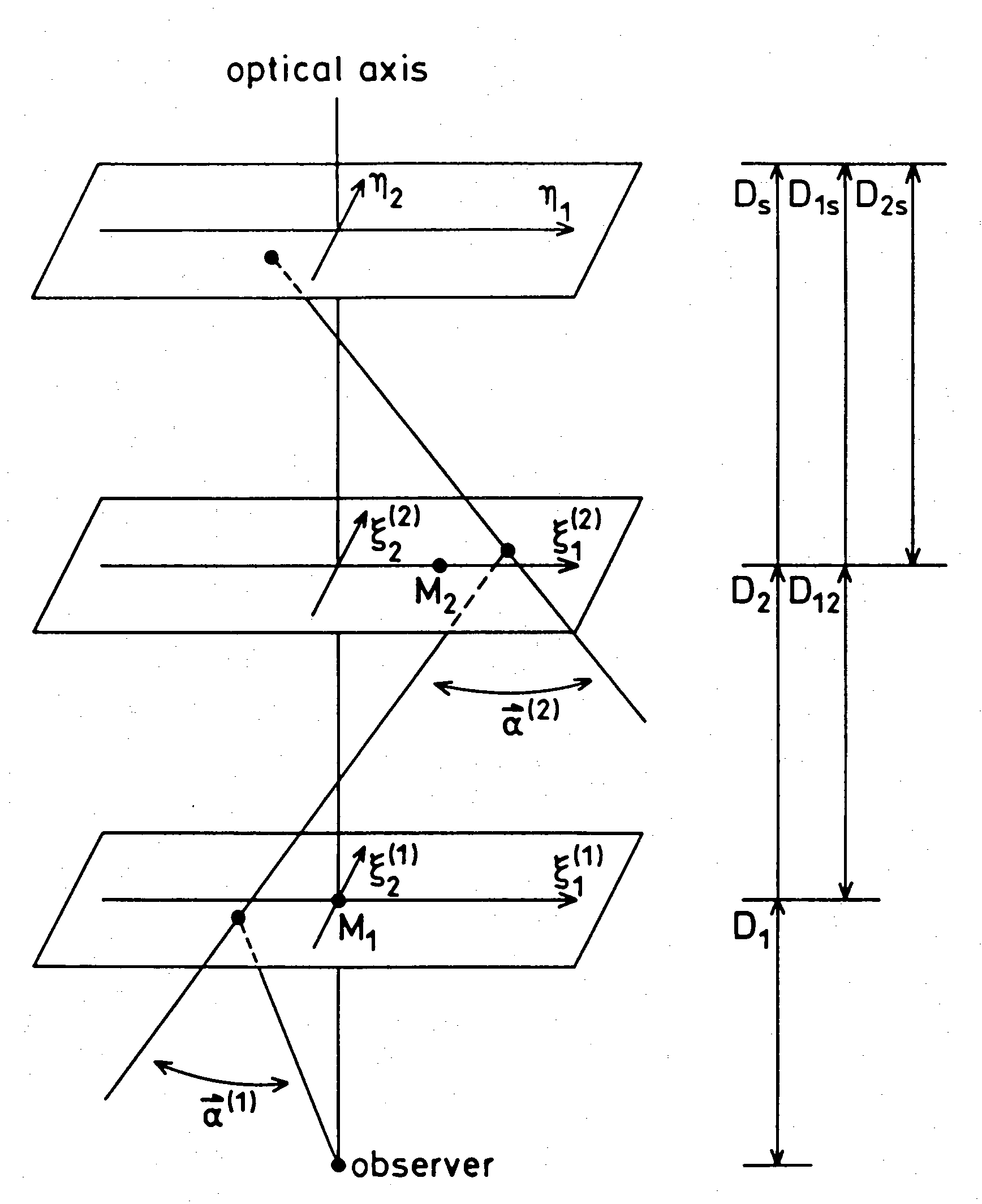} 
\caption{Diagram of the geometry of a multiplane lensing scenario.  (Fig.~1 in \cite{catastrophe})
\label{fig:mpl}}
\end{figure}

\begin{figure}
\centering
\includegraphics[width=0.75\textwidth]{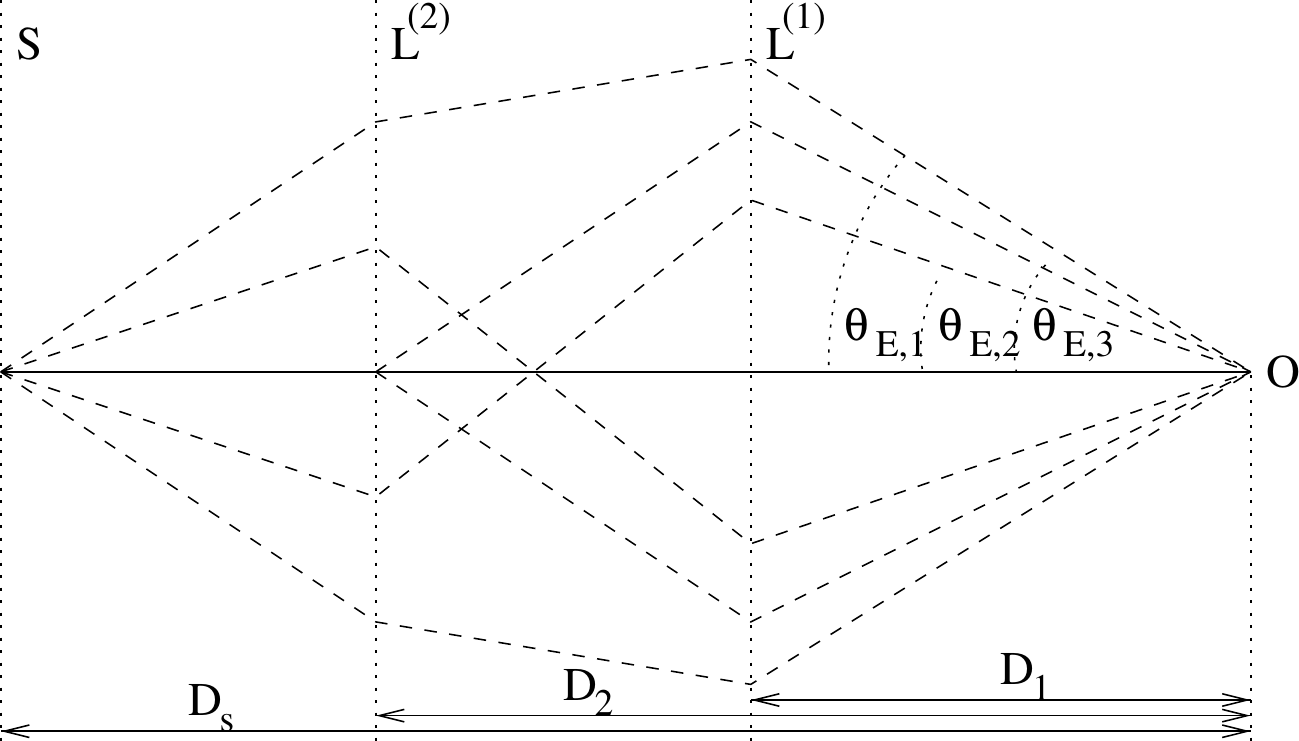} 
\caption{Diagram of Einstein rings produced by \swz{} lenses $L^{(1)}$ and $L^{(2)}$ in two different lens planes. (Fig.~1 in \cite{multiple-einstein-rings})
\label{fig:multiple-einstein-rings}}
\end{figure}

\section{Concept}
The central purpose of \mpl{} is to quantitatively determine the effect of the presence of more than one lens between source and observer.  The geometric setup is illustrated in \cref{fig:mpl}: it remains essentially unchanged from \cref{fig:lens-setup} (for simplicity only two lenses are shown).  There are $N$ lenses at distances $D_1 < \ldots D_i < \ldots D_N$, with the source at $D_s > D_N$.  The new lenses follow the same geometry as the single lens case: the photons emitted by the source have their geodesics perturbed by the presence of the lens, which introduces the same phenomena of time delays and magnification effects as discussed in \cref{ch:gl}.  The effect of the additional lens is shown in \cref{fig:multiple-einstein-rings}.    The dotted lines show the deflection of the light rays from the source $S$ to the observer $O$ via the lens planes $L^{(1)}$ and $L^{(2)}$.  This geometry produces not two, but three Einstein rings with different deflection angles $\theta$.  The rings $\theta_{E,1}$ and $\theta_{E,2}$ are the images produced by geodesics from $S$ to $O$, which are lensed by both lenses. The third ring $\theta_{E,3}$ is an image produced by the image of $L^{(1)}$ being lensed by $L^{(2)}$ and would not be present without use of the multiplane lensing algorithm.  Hence we see that the time delay surfaces can be so deformed by the presence of more than one lens that additional extrema appear, which correspond to extra images.

We introduce dimensionless parameters analagous to those in \cref{eq:scaling}:
\begin{align}
\vect{x}_i &= \frac{\vect{\xi}_i}{D_{i}} &&
\vect{x^{\mathsmaller\prime}}_i &= \frac{\vect{\xi^{\mathsmaller\prime}}_i}{D_{i}} &&
\vect{y} &= \frac{\vect{\eta}}{D_{s}} &&
\beta_{i,j} &= \frac{D_{ij} D_{s}}{D_{j} D_{is}} &&
\vartheta_{i} &= (1+z_{i}) \frac{D_i D_{i+1}}{D_{i,i+1}}&&
\end{align}
The $i-$th lens is located at $\vect{x}_i$ in the lens plane and the source at $\vect{y}$ in the source plane.  The distances are expressed in pc: $D_{i,j}$ is the angular diameter distance between the $i-$th and $j-$th (lens) plane, a subscript $s$ refers to the source plane and the second subscript is dropped when it refers to the observer.  We now have a set of dimensionless parameters with which to determine the recurrence relations in \cref{sec:formulae}.

\section{Recurrence relation formulae}\label{sec:formulae}
The formulae shown in \cref{ch:gl} are modified via use of a recurrence relation.  The structure of the equations are preserved, but the complexity is increased by contributions to the $j-$th lens from all $(j+1) \leq N$ lenses between it and the source.  In the specific case of \dmh{} lenses, this is simplified by the fact that the lenses themselves are not imaged, as they are not luminous themselves.  The resulting formulae are shown in \cref{tab:mpl}.  
The hindrance to numerical use of the \mpl{} formalism is the recurrence relations in the formulae.  They are neither vectorisable nor parallelisable readily, since the output of the previous lenses forms the input for the next.  Nevertheless, recent efforts in \cite{hilbert-lensing-millenium} demonstrate an effective use of multiplane lensing by dark halos using the Millenium simulation. 
In the galactic regime, which is the focus of this thesis, \cite{multiple-einstein-rings} illustrates a simple model for multiple Einstein rings in a two-lens system using the \swz{} (point mass) lens.  Ultimately \mpl{} is problematic to simulate due to the inherent numerical inefficiency of the recurrence formulae.

\clearpage{\newsavebox{\zerothbox}
\sbox{\zerothbox}{
\begin{minipage}{0.25\textwidth}
\begin{equation*}
\vect{y} = \vect{x}_1 - \sum_{i=1}^{N} \vect{\alpha}_i(\vect{x}_i)
\end{equation*}
\end{minipage}
} 

\newsavebox{\firstbox}
\sbox{\firstbox}{
\begin{minipage}{0.35\textwidth}
\begin{equation*}
\tau(\vect{x};\vect{y}) = 
\vartheta \left[ \frac{1}{2} \left( \vect{y} - \vect{x} \right)^2 - \beta\psi(\vect{x}) \right]
\end{equation*}
\end{minipage}
} 

\newsavebox{\secondbox}
\sbox{\secondbox}{
\begin{minipage}{0.5\textwidth}
\begin{multline*}
\tau(\vect{x}_1 \ldots \vect{x}_N;\vect{y}) = \\
\sum_{i=1}^{N-1} \vartheta_{i,i+1} \left[ \frac{1}{2} \left( \vect{x}_{i+1} - \vect{x}_i \right)^2 - \beta_{i,i+1} \psi_i(\vect{x}_i) \right]
\end{multline*}
\end{minipage}
} 

\newsavebox{\thirdbox}
\sbox{\thirdbox}{
\begin{minipage}{0.5\textwidth}
\begin{equation*}
    \mu(\vect{x};\vect{y}) = 
    \dfrac{1}{
        \mathrm{det} \Biggl(I - \dfrac{\partial \vect{\alpha}(\vect{x})} {\partial \vect{x}} \Biggr)  
    }
\end{equation*}
\end{minipage}
} 

\newsavebox{\fourthbox}
\sbox{\fourthbox}{
\begin{minipage}{0.5\textwidth}
\begin{multline*}
    \mu(\vect{x}_1 \ldots \vect{x}_N;\vect{y}) = \\
    \dfrac{1}{ 
        \mathrm{det}
        \left(I - \sum_{i=1}^{N} \dfrac{\partial \vect{\alpha}_i(\vect{x}_i)} {\partial \vect{x}_i} \, 
        \dfrac{\partial \vect{x}_i} {\partial \vect{x}_1} ) \right) 
    }
\end{multline*}
\end{minipage}
} 

\setlength{\LTcapwidth}{\textwidth}
\renewcommand{\arraystretch}{2}
\begin{landscape}
\begin{longtable}{ l l l } 
\toprule 
Property
& Single lens
& Multiple lenses
 \\
\midrule
\endfirsthead 

\bottomrule
\caption{Comparison of the key effects of \gl{ing} in the case of single and multiple lenses \cite{sef,petters,singularity-theory,practical-model}. The $i-$th of $N$ lenses is represented by a subscript $i$, except when $N=1$, when it is dropped.  The source is denoted by $s$.
\label{tab:mpl}}
\endlastfoot 

Lens equation 
& $\vect{y} = \vect{x} - \vect{\alpha}(\vect{x})$
& $\vect{y} = \vect{x}_1 - \displaystyle\sum_{i=1}^{N} \vect{\alpha}_i(\vect{x}_i)$ 
\\ \midrule 
Time delay
& $\tau(\vect{x};\vect{y}) = \vartheta \left[ \frac{1}{2} \left( \vect{y} - \vect{x} \right)^2 - \beta\psi(\vect{x}) \right]$
& \usebox{\secondbox}
\\ \midrule 
\begin{minipage}{2.7cm}Magnification \\ factor \end{minipage}
& \usebox{\thirdbox} 
& \usebox{\fourthbox}
\end{longtable}
\end{landscape}

\renewcommand{\arraystretch}{1.0}} 

\cleardoublepage\chapter{Numerical routines}\label{ch:code}
This appendix shows the complete schemes discussed in \cref{ch:gl}, written in \textsc{matlab}.  All routines (rather than the key ones listed here) can be found at the repository \cite{vonbraunbates-pulsar-lensing}.

\section{Roots of the lens equation}\label{code:rootsearch}
\begin{lstlisting}
function x_0 = rootsearch(f,df,d2f,a,b)

eps_ = double(eps('single'));
d2x0 = []; dx0 = []; x0 = []; x_0 = [];
options = optimset('FunValCheck','on', ... % f(x0) finite
    'TolFun',eps_/1e1); % tolerance f(x)
%{ % Plot functions to check:
xx = linspace(a,b,1e4);
figure(6); hold all
plot(xx,zeros(1e4,1),'k-','DisplayName','f = 0','MarkerSize',4)
plot(xx,d2f(xx),'.','DisplayName','d^2f/dx^2','MarkerSize',4)
plot(xx,df(xx),'.','DisplayName','df/dx','MarkerSize',4)
plot(xx,f(xx),'.','DisplayName','f(x)','MarkerSize',4)
axis([a b -10 10]); legend('show','Location','Best');%}
% find where d2f/dx2 changes sign:
if (sign(d2f(a)) ~= sign(d2f(b)))
    d2x0 = fzero(d2f,[a b],options);
else
    d2x0 = NaN;
end

d2x0 = d2x0(isfinite(d2x0)); % remove NaN
d2x0(abs(d2x0) < eps_) = 0;

% between each root look for roots of df/dx
rangeint = unique([a;d2x0;b]); % sort

for i = 2:length(rangeint)
    if(~isfinite(df(rangeint(i-1)))) % df(a) = +/- Inf
        if (sign(df(rangeint(i-1) - eps_)) ~= sign(df(rangeint(i))))
            int = [rangeint(i-1) - eps_ rangeint(i)];
            dx0(i-1) = fzero(df,int,options);
        elseif (sign(df(rangeint(i-1) + eps_)) ~= sign(df(rangeint(i))))
            int = [rangeint(i-1) + eps_ rangeint(i)];
            dx0(i-1) = fzero(df,int,options);
        end
    elseif(~isfinite(df(rangeint(i)))) % df(b) = +/-Inf
        if (sign(df(rangeint(i-1))) ~= sign(df(rangeint(i) - eps_)))
            int = [rangeint(i-1) rangeint(i) - eps_];
            dx0(i-1) = fzero(df,int,options);
        elseif (sign(df(rangeint(i-1))) ~= sign(df(rangeint(i) + eps_)))
            int = [rangeint(i-1) rangeint(i) + eps_];
            dx0(i-1) = fzero(df,int,options);
        end
    elseif(sign(df(rangeint(i-1))) ~= sign(df(rangeint(i)))) % f continuous over int
        int = rangeint(i-1:i);
        dx0(i-1) = fzero(df,int,options);
    else % df is undefined at a or b
        dx0(i-1) = NaN;
    end
end

dx0(abs(dx0) < eps_) = 0;
dx0 = dx0(isfinite(dx0)); % remove NaN

% between those roots look for roots of f
rangeint = unique([a;d2x0';dx0';b]); % sort

for i = 2:length(rangeint)
    % f = +/- Inf at a or b breaks fzero(f,[a b])
    if(~isfinite(f(rangeint(i-1)))) % f(a) = +/- Inf
        if (sign(f(rangeint(i-1) - eps_)) ~= sign(f(rangeint(i)))) 
            int = [rangeint(i-1) - eps_ rangeint(i)];
        elseif (sign(f(rangeint(i-1) + eps_)) ~= sign(f(rangeint(i)))) 
            int = [rangeint(i-1) + eps_ rangeint(i)];
        end
    elseif(~isfinite(f(rangeint(i)))) % f(b) = +/- Inf
        if (sign(f(rangeint(i-1))) ~= sign(f(rangeint(i) - eps_))) 
            int = [rangeint(i-1) rangeint(i) - eps_];
        elseif (sign(f(rangeint(i-1))) ~= sign(f(rangeint(i) + eps_))) 
            int = [rangeint(i-1) rangeint(i) + eps_];
        end
    elseif(sign(f(rangeint(i-1))) ~= sign(f(rangeint(i)))) % f continuous over int
        int = rangeint(i-1:i);
    end
    % Now find root within modified interval
    try
        [x0(i-1),~,exitflag,~] = fzero(f,int,options);
        if(exitflag==1) 
        x_0 = [x_0;x0(i-1)]; 
        else
        % f is undefined over [a,b]
        x0(i-1) = NaN;
        end
    catch
        disp('FZERO error.')
    end; % try
end

% concatenate zeros
xvals = [d2x0 dx0];
fvals = f(xvals); % check roots of f', f" zeros of f
x_0 = sort([x_0; xvals(abs(fvals) < eps_)']); % keep true zeros
if(isempty(x_0)); 
    disp('No roots!'); x_0 = NaN;
else
    x_0 = x_0(logical([1,(diff(x_0) > eps_)'])); % remove elements equal within tol
end; % if
%{% The value x returned by fzero is near 
% a point where fun changes sign, 
% or NaN if the search fails.
% ONLY a zero if fun is continuous
% Otherwise a divergent discontinuity.

%plot(x_0,zeros(size(x_0)),'x','DisplayName','zeros')
%hold off;%}
end % function
\end{lstlisting}

\newpage
\section{Calculation of the Hankel transform}\label{code:hankel}
\begin{lstlisting}
function H = hankel_matrix(ord, R, N, varargin)
%{HANKEL_MATRIX: Generates data to use for Hankel Transforms
The algorithm used is that from:
		"Computation of quasi-discrete Hankel transforms of the integer
		order for propagating optical wave fields"
		Manuel Guizar-Sicairos and Julio C. Guitierrez-Vega
		J. Opt. Soc. Am. A 21(1) 53-58 (2004)
paper defn: (eqn 1)
H[f(r)]   \equiv 2*pi \int dr f(r)J_p(2*pi*kr)r 
H-1[F(k)] \equiv 2*pi \int dk F(k)J_p(2*pi*kr)k 
RHB defn.:
H[f(r)]   \equiv \int dr f(r)J_p(kr)r 
H-1[F(k)] \equiv \int dk F(k)J_p(kr)k 
scaling:
forward:   H[f] = (T * (f.*s_HT.JR) ) ./ s_HT.JV   (eqn 6a)
backward: ~H[F] = (T * (F.*s_HT.JV) ) ./ s_HT.JR   (eqn 6b)%}
%{ if(~isempty(varargin))
    fhandle1 = figure('visible','on');
     %fhandle2 = figure('visible','on');
    % set figure size, visibility
    set(0,'DefaultFigureVisible','off');
    scrsz = get(0,'ScreenSize');
    set(0,'DefaultFigurePosition',[1 .9*scrsz(4) .99*scrsz(3) .9*scrsz(4)]);
    % axes in plot
    ncols = 2;     nrows = 2;     len=nrows*ncols;
    ax.min = 0.05; ax.max = 0.95; ax.gap = 0.05;
    ax.size = (ax.max - ax.min)./[ncols nrows];
    ax.box = ax.size - ax.gap;
    ax.coord(:,1) = ax.min + ax.size(1).*mod([1:len]-1,ncols); % x starting co-ord
    ax.coord(:,2) = ax.max - ax.size(2).*ceil([1:len]./ncols); % y starting co-ord
    ax.coord(:,3) = ax.box(1); % width
    ax.coord(:,4) = ax.box(2); % height
    ax.coord(end+1,:) = [0 0 1 1]; % figure axes
    
    % Remove warnings in legend
    warning('off','MATLAB:legend:UnsupportedFaceColor');
    warning('off','MATLAB:legend:PlotEmpty');
    warning('off','MATLAB:legend:IgnoringExtraEntries');
    
    clear len scrsz ax.min ax.max ax.size ax.box
 end % if %}
%% Transformation matrix
if(~isinteger(N)); N = floor(N); end; % int nr of Bessel roots
%	Calculate N+1 roots:
c = bessel_zeros('J',ord,N+1);
% [jn,jm] = meshgrid(c(1:N),c(1:N)); % alpha_{p,1:N}
% Jn = besselj(ord+1,jn); Jm = Jn';
% But meshgrid runs out of memory!
J = besselj(ord+1,c(1:N)'); 
Jn = abs(repmat(J,N,1)); % rows of Jn are copies of J
% Calculate hankel matrix
C = (2/c(N+1))*besselj(ord,(c(1:N)*c(1:N)')/c(N+1))./(Jn.*Jn'); %c*c' = jn.*jm
clear Jn 

% Co-ordinate vectors: f_n = f(j_n/V); F_m = F(j_m/R); 
V = c(N+1)/R;           % Maximum frequency
r = c(1:N)/V;           % /V instead of *R/c(N+1);   % Radius vector
v = c(1:N)/R;           % Frequency vector

% Scaling: f_qdht = f(x)/m1; F_qdht = F(k)/m2
% F(k) =  ht[f_qdht] * m2 = (C * (f(x)/m1)) * m2;
% f(x) = iht[F_qdht] * m1 = (C * (F(k)/m2)) * m1;
m1 = abs(J')/R;         %% m1 prepares input vector for transformation
m2 = abs(J')/V;         %% m2 prepares output vector for display

%% Analytical soln if necessary
if(~isempty(varargin))
    % input
    f = [];
    
    % transform and inverse transform
    ht  = @(f) (C*(f(:)./m1)).*m2;
    iht = @(F) (C*(F(:)./m2)).*m1;
    for j=1:2
        f2(:,j)   = ht ( f(:,j));  % forward
        fiht(:,j) = iht(f2(:,j));  % backward
    end % for
    clear j
    f2(:,3)   = 2*pi*f2(:,1).*f2(:,2); % convolution thm.
    fiht(:,3) = iht(f2(:,3));
    %{ 
	%% Plotting
    title_str = sprintf('N = %8.0g, R_{max} = %8.0g, N/R = %8.0g',[N,R,N/R])
    
    % actual plots
    figure(fhandle1), subplot(1,2,1), hold all,
    plot(r,fiht,'o'), axis tight;
    xlabel('r'), ylabel('f(r)'); 
    
    subplot(1,2,2), hold all, 
    plot(v,f2,'o'), axis tight;
    xlabel('v'), ylabel('F(v)'); 
    
    mtit(title_str); %}
end % if

%% assign to struct
H = struct('C',C,'r',r,'v',v,'m1',m1,'m2',m2);
clear C r v m1 m2 f f2 fiht

end % function
\end{lstlisting}

\newpage
\section{Generation of observations from the lensing results}\label{code:tdelay}
\begin{lstlisting}
function t_struct = plot_tdelay(t_delay,mu,t_lens,T_res)
    
    % get time delays and magnification
    N_lens    = numel(t_delay);         % t is a cell: t{i} = dt(ith lens)
    N_pulse   = size(t_delay{1},1);
    N_images  = max(cellfun(@(x)size(x,2), t_delay));
    tau_array = cellextract(t_delay)';  % t_arr(:,i)  = t{ith lens}(:,:)
    mu_array  = cellextract(mu)';       % mu_arr(:,i) = mu{ith lens}(:,:)
    
    % setup signal
    t_struct.T_res = T_res;               % timing residual (s)
    s_source = ones(size(t_lens));        % original pulsar signal
    
    %% Create signals
    % add lensing effects to signal
    t_images = t_lens + tau_array;
    t_images = reshape(t_images,[N_pulse N_lens*N_images]);
    s_images = s_source .* mu_array;
    s_images = reshape(s_images,[N_pulse N_lens*N_images]);
    clear mu_array
    
    % bin signals by time
    t_images = t_images(:); s_images = s_images(:);
    [temp1,ind] = sort(t_images);  % sort t
    temp2 = s_images(ind);         % sort mu by t value
    temp3 = t_lens(ind);           % sort emission time by t value
    clear ind;
    k = 1; l = 1;
    while k < length(temp1);
        % find signals close together
        dt = temp1 - temp1(k);
        t = temp1((dt >= 0) & (dt < 10*T_res));
        s = temp2((dt >= 0) & (dt < 10*T_res));
        d = temp3((dt >= 0) & (dt < 10*T_res));
        k = find(dt > 10*T_res,1,'first');
        % bin only those signals
        [counts, bin] = histc(t, [min(t) : T_res : max(t)+T_res]);
        max_counts = max(counts); % largest nr of superposed signals
        m=0; % number of non-empty bins
        % Non-empty bins contain signals which will be superimposed
        for i = 1:length(counts)
            if(~isempty(t(bin==i)))
                m = m+1;
                t_mat(m,:) = vec2mat(t(bin==i),max_counts,NaN);
                s_mat(m,:) = vec2mat(s(bin==i),max_counts,0);
                d_mat(m,:) = vec2mat(d(bin==i),max_counts,NaN);
            end % if
        end; % for
        clear bin i max_counts nbins
        % get time, signal for composite
        T{l} = t_mat(:,1);          % only need unique t
        S{l} = sum(s_mat,2);        % sum s with same i
        D{l} = d_mat(:,1);
        % increment search
        clear *_mat dt
        l = l+1;
    end; % while
    
    clear i k l m s t temp1 temp2
    
    % extract composite signal
    T = cellextract(T); T = T(:); [t_sorted,ind] = sort(T);
    S = cellextract(S); S = S(:); s_sorted = S(ind);
    D = cellextract(D); D = D(:); d_sorted = D(ind);
    clear ind D S T
    
    %% Plotting
    % colourbar shows impact parameter
    thermal_map = ...
        [1.0000   0.7857    0.0357
        1.0000    0.5714    0.0714
        0.9857    0.3643    0.1143
        0.9143    0.1857    0.1857
        0.6714    0.0643    0.3714
        0.4000         0    0.5286
        0.1500         0    0.6000];
    thermal_map = colormap_helper(thermal_map, N_lens);
    for j=1:N_lens
        colour_hsv = rgb2hsv(thermal_map(j,:));
        map_hsv = [repmat(colour_hsv(1:2),[N_pulse 1]) linspace(0,colour_hsv(3),N_pulse)'];
        colour{j} = hsv2rgb(map_hsv);
        cbar_map(j,:) = colour{j}(N_pulse,:);
    end % for
    grey_map = repmat([0 .25 .5 .75]',[1 3]);
    colour{j+1} = colormap_helper(grey_map, numel(t_sorted));
    cmap = vertcat(colour{:});              % concatenated maps
    clen = cellfun(@(x)(size(x,1)),colour); % length of each map
    csum = cumsum(clen) - clen;             % starting index of each map
    
    % get axis handles for subplots
    ax = plot_axes((N_lens > 1),2,2,{1,2,3,4}); % {[1 2],3,4});
    
    % Line style default for single lens, dotted for multiple
    if(N_lens==1); linespec = 'k-'; else linespec = 'k:'; end; % if
    
    % plot radio signal from each image
    set(ax.figure,'CurrentAxes',ax.handle(1)); hold on; colormap(cmap);
    ctemp = csum(end) + [1:length(t_sorted)]'; % end was j
    hLine = plot(t_sorted(:), s_sorted(:), linespec);
    set(get(get(hLine,'Annotation'),'LegendInformation'),...
        'IconDisplayStyle','off'); % Exclude line from legend
    scatter(t_sorted(:), s_sorted(:), 360, ctemp, 'Marker','.');
    set(gca,'CLim',[1 sum(clen)]); clear ctemp; freezeColours(gca);
    xlabel('Time (s)','interpreter','latex');
    ylabel('$\mu$','interpreter','latex','rotation',0);
    title('{\bf Signal from pulsar}','interpreter','latex');
    axis tight; clear h*
    ticklabelformat(gca,'xy','%2.6g');
    set(gca,'XTickLabel',get(gca,'Xticklabel'),'FontName','Courier 10 Pitch','FontSize',20,'fontweight','bold');
    set(gca,'XTickMode','auto','XTickLabelMode','auto');
    
    % plot change in time delay vs observation time
    set(ax.figure,'CurrentAxes',ax.handle(2)); colormap(cmap); hold on;
    dt = nan([1 length(t_sorted)]);
    ctemp = csum(end) + [1:length(t_sorted)]';
    dt(2:end) = 1e6*diff(t_sorted - d_sorted, 1, 1);
    scatter(t_sorted, dt, 360, ctemp, 'Marker','.');
    plot(t_sorted, dt,linespec);
    set(gca,'CLim',[1 sum(clen)]); freezeColours(gca);
    xlabel('Emission time (s)','interpreter','latex');
    ylabel('$d\tau$','interpreter','latex','rotation',0);
    title('{\bf Change in time delay}','interpreter','latex');
    axis tight; clear h*
    ticklabelformat(gca,'xy','%2.6g');
    set(gca,'XTickLabel',get(gca,'Xticklabel'),'FontName','Courier 10 Pitch','FontSize',20,'fontweight','bold');
    set(gca,'XTickMode','auto','XTickLabelMode','auto');
    
    % plot time delay per lens
    set(ax.figure,'CurrentAxes',ax.handle(4)); hold on; colormap(cmap);
    for j=1:N_lens
        ctemp = csum(j) + [1:clen(j)]'; % end was j
        if(N_lens~=1) % change in time delay if > 1 lenses
            dt(2:N_pulse,j) = 1e6*diff(tau_array(:,j) - min(tau_array(:,j)), 1, 1);
            scatter(t_lens(:,j), dt(:,j), 360, ctemp, 'Marker','.');
            plot(t_lens(:,j), dt(:,j),linespec);
            ylabel('$d\tau$','interpreter','latex','rotation',0);
            title('{\bf Change in time delay per lens}','interpreter','latex');
        else % time delay if 1 lens
            ttemp = (tau_array - min(tau_array))*1e6;
            scatter(t_lens, ttemp, 360, ctemp, 'Marker','.');
            plot(t_lens, ttemp,linespec);
            ylabel('$\tau$','interpreter','latex','rotation',0);
            title('{\bf Relative time delay per lens}','interpreter','latex');
        end; % if
    end; clear j % for
    set(gca,'CLim',[1 sum(clen)]); freezeColours(gca);
    xlabel('Emission time (s)','interpreter','latex');
    axis tight; clear h*
    ticklabelformat(gca,'xy','%2.6g');
    set(gca,'XTickLabel',get(gca,'Xticklabel'),'FontName','Courier 10 Pitch','FontSize',20,'fontweight','bold');
    set(gca,'XTickMode','auto','XTickLabelMode','auto');
    
    %% Output
    t_struct.t = t_sorted; t_struct.mu = s_sorted;
    t_struct.ax = ax.handle; t_struct.fig = ax.figure;
    t_struct.colour = colour; t_struct.cbar_map = cbar_map;
    
end % function
\end{lstlisting}


\end{document}